\documentclass[
10pt,twocolumn,
 amsmath,amssymb,
 aps,
pre,
]{revtex4-2}

\usepackage{graphicx}
\usepackage{dcolumn}
\usepackage{bm}

\usepackage[export]{adjustbox}
\usepackage{bbold}
\usepackage[utf8]{inputenc} 
\usepackage[T1]{fontenc}    
\usepackage{graphicx}
\usepackage{hyperref}  
\usepackage{mathtools}
\usepackage{tikz}
\usepackage{times}
\usepackage{verbatim}
\usepackage{xfrac}
\usepackage[normalem]{ulem}
\usepackage{comment}
\usepackage{lineno}
\usepackage{cancel}
\usepackage{nicematrix}

\NiceMatrixOptions
{
    custom-line = 
    {
        command = dashedline , 
        letter = : , 
        tikz = { dashed } , 
        total-width = \pgflinewidth
     }
}
\allowdisplaybreaks

\usetikzlibrary{arrows.meta,automata,calc,decorations.markings}
\definecolor{purpura}{rgb}{0.5, 0.0, 0.5}
\definecolor{violeta}{rgb}{0.2, 0.0, 0.4}
\definecolor{azul}{rgb}{0, 0.0, 0.6}
\definecolor{rojo}{rgb}{0.6, 0, 0}
\definecolor{verde}{rgb}{0, 0.6, 0}
\definecolor{turquesa}{rgb}{0, 0.5, 0.5}
\definecolor{marron}{rgb}{0.6, 0.4, 0}
\definecolor{gris}{rgb}{0.4, 0.4, 0.4}
\definecolor{celeste}{rgb}{0., 0.3, 0.5}

\begin{document}

\title{Expansion of marginal correlations in terms of partial correlations}

\author{
Bautista Arenaza$^{12}$, Sebastián Risau-Gusmán$^{1}$, and Inés Samengo$^{12}$ }

\affiliation{
{\sl 1: Conicet and Department of Medical Physics, Centro Atómico Bariloche, Argentina \\
2: Instituto Balseiro, San Carlos de Bariloche, Argentina}}

\begin{abstract}
The marginal correlation between two variables is a measure of the linear dependence of one variable on the other. The two original variables need not interact directly, because  marginal correlation may arise from the mediation of other variables in the system. The underlying network of direct interactions can be captured by a weighted graphical model. The connection between two variables can be weighted by their partial correlation, defined as the residual correlation left after accounting for the linear effects of mediating variables. While matrix inversion can be used to obtain marginal correlations from partial correlations, in large systems this approach does not reveal how the former emerge from the latter. Here we present an expansion of marginal correlations in terms of partial correlations, which shows that the effect of mediating variables can be quantified by the weight of the paths in the graphical model that connect the original pair of variables. The expansion is proved to converge for arbitrary probability distributions.  The graphical interpretation reveals a close connection between the topology of the graph and the marginal correlations. Moreover, the expansion shows how marginal correlations change when some variables are severed from the graph, and how partial correlations change when some variables are marginalised out from the description. It also establishes the minimum number of latent variables required to replicate the exact effect of a collection of variables that are marginalised out, ensuring that the partial and marginal correlations of the remaining variables remain unchanged. Notably, the number of latent variables may be significantly smaller than the number of variables that they effectively replicate. Finally, for Gaussian variables, marginal correlations are shown to be related to the efficacy with which information propagates along the paths in the graph.

\end{abstract}

\maketitle


\section{Introduction} 
\label{sect:intro}

In statistics, it is important to distinguish between \emph{direct} and \emph{marginal} correlations between variables. A humorous example was offered by \citet{Matthews2000}, where the number of stork breeding pairs in a collection of countries was shown to be positively and significantly correlated with the number of human births in those same countries. If this correlation were to be interpreted as a \emph{direct} mechanistic interaction between the variables, storks might be concluded to deliver human babies. Or perhaps, human births might be concluded to incentivise storks to breed. The more likely conclusion, however, is that the variables ``storks'' and ``babies'' are only related because there are additional variables that modulate the two of them, for example, the size of the country where the two measurements were made: The larger the country the more of everything, in particular, the more of storks and the more of babies. If the size of country is fixed, then the induced interaction between the other two variables vanishes.

In more general terms, given a set of $d$ random variables $\{X_1, \dots, X_d\}$, the underlying mechanistic description of the system can be represented by a graphical model \citep{lauritzen1996}, where node $i$ stands for variable $X_i$, and the edge between nodes $i$ and $j$ represents the direct interaction between $X_i$ and $X_j$. A non-zero edge implies that $X_i$ and $X_j$ influence each other directly, that is,  without requiring the involvement of the remaining variables to propagate the interaction. The weight of the edge captures the strength of the influence, and different graphical models can be defined by varying the metric that quantifies the influence.

Among the many possible metrics, partial correlations are an easily computable measure of pairwise direct interactions   \citep{baba2004partial}, since they can be obtained from the fitted coefficients of a multiple linear regression model \citep[Chapter~23]{cramer1946mathematical}. They have been used to construct mechanistic descriptions of a broad spectrum of systems \citep{alves2019improved, millington2020partial, von2021exploratory, zhang2021partial, werme2022integrated}. The typical approach is to build a ``partial correlation network'', that is, a graphical model where the edge between nodes $i$ and $j$ is weighted by the partial correlation $r_{ij}$ between variables $X_i$ and $X_j$. 

Partial correlations can always be related to geometric properties of the best linear approximation to the system \citep[Chapter~23]{cramer1946mathematical}. In certain special cases, these geometric properties have an intuitive interpretation. For example, for multivariate Gaussian distributions, partial correlations coincide with conditional correlations. For some other widely-used distributions, partial correlations are averaged conditional correlations (Sect.~\ref{sec:notation}). We are not aware, however, of the existence of an intuitive interpretation of partial correlations for arbitrary distributions. 

In this paper we derive an explicit and intuitive relation between partial and marginal correlations. Our main result is that the value of the marginal correlation $\rho_{ij}$ between the variables $X_i$ and $X_j$ can be expanded as a sum of contributions from all the paths of the graph that connect the nodes $i$ and $j$ travelling through links with non-zero partial correlations. The expansion reveals how \emph{net} interactions (measured by marginal correlations) arise from \emph{direct} interactions (measured by partial correlations). 

The conditions of validity of the expansion impose no restrictions on the probability distribution. Therefore, the  role of partial correlations as the local building blocks from which non-local net pairwise interactions arise is completely general. The expansion itself, therefore, provides a way to visualise and interpret partial correlations for arbitrary probability distributions. 

In addition to this new insight, the expansion can be used to study the effect of modifying the structure of the graphical model on both the marginal and partial correlations. Finally, the expansion can also be employed as an algorithmic tool with which marginal correlations can be obtained from partial correlations.

The paper is organised as follows. In Sect.~\ref{sec:notation} we define the relevant quantities, and explore the meaning of partial correlations both for Gaussian and non-Gaussian variables. In Sect.~\ref{sec:correl}, we introduce the graphical model and expand the marginal correlation between two variables as a sum of contributions from paths in the graph. In Sect.~\ref{sect:ejemplos} four examples are discussed, the first two with the goal of developing intuition, and the last two, to illustrate the algorithmic utility of the expansion to perform calculations in more involved cases. Sections \ref{sec:separatingnode} and \ref{sec:alterations} use the series expansion to describe the strong influence imposed by the topology of the graphical model on the marginal correlations. First, in Sect.~\ref{sec:separatingnode}, the role of separating nodes is discussed. Then, Sect.~\ref{sec:alterations} analyses how additions and deletions of nodes and edges in the graph modify marginal and partial correlations. For example, the marginal correlation between two variables changes when the edges that connect some other variables to the rest of the graph are severed. Another example involves studying how partial correlations change when a subset of the variables is marginalised out. The section closes with a derivation of the minimum number of latent variables that need to be introduced, and the connectivity pattern that they need to establish with the rest of the network, to effectively reproduce the influence exerted by a set of variables eliminated from the description through marginalisation. Finally, since for Gaussian variables there is a one-to-one correspondence between the covariance matrix and the mutual information, in Sect.~\ref{sec:info} the mutual information between subsets of variables is also expanded as a sum over paths connecting the subsets. The paper closes with a brief summary and discussion of the results.

\section{Marginal and partial correlations}
\label{sec:notation}

We consider $d$ random variables $\{X_1, \dots, X_d\}$ with a positive-definite covariance matrix $C$, but an otherwise arbitrary probability distribution. The indices $i$ and $j$ are reserved for the variables $X_i$ and $X_j$ whose  interaction is being quantified. The index $\kappa$ runs over the rest of the variables. Matrices are denoted with capital letters (for example, $A$) and their entries, with the lowercase variant ($a_{ij}$). Sets are represented by a calligraphic font ($\mathcal{A}$).

The present section provides the definition and the interpretation of marginal and partial correlations both for Gaussian and non-Gaussian variables. Readers already familiar with these concepts may skip this section and proceed directly to the next one.

Given the covariance matrix $C$, the \emph{marginal correlation} (also called Pearson correlation coefficient) between variables $X_i$ and $X_j$ is 
\begin{equation} \label{eq:rho_C}
    \rho_{ij} = \frac{c_{ij}}{\sqrt{c_{ii} \, c_{jj}}} = \frac{{\rm Cov}(X_i, X_j)}{\sqrt{{\rm Var}(X_i) \, {\rm Var}(X_j)}}.
\end{equation}
The value of $\rho_{ij}$ is a statistical measure of the degree up to which fluctuations in the component $X_i$ are linearly related to fluctuations in $X_j$, irrespective of what happens with all other components. Only the linear component of the interaction is captured, since $\rho_{ij}$ can be computed from the fitted coefficients of the least square linear regressions of $X_i$ in terms of $X_j$, and of $X_j$ in terms of $X_i$ \citep[Chapter~21]{cramer1946mathematical}. The maximum absolute value $|\rho_{ij}| = 1$ is achieved if and only if a strictly linear relation between $X_i$ and $X_j$ exists. In other words, $\rho_{ij}$ represents the marginal dependency between $X_i$ and $X_j$, and the word ``marginal'' underscores the fact that the remaining variables have been integrated out, and vary in whichever way the joint probability distribution governs their behaviour. The marginal covariance $\rho_{ij}$, hence, measures the \emph{net} linear effect that $X_i$ has over $X_j$ (or vice-versa), possibly mediated by other variables. The normalising factor in the denominator of Eq.~\ref{eq:rho_C} makes $\rho_{ij}$ independent from the units in which $X_i$ and $X_j$ are expressed. Importantly, $\rho_{ij}$ remains unaltered if other variables in the system (different from $i$ and $j$) are marginalised out.

The \emph{partial correlation} $r_{ij}$ is a widely-used measure of the {\sl direct} influence that $X_i$ has over $X_j$ after discounting the effects mediated by other variables \citep{baba2004partial}. To discount such effects, it is useful to consider the auxiliary variables $\hat{X}_i$ and $\hat{X}_j$, defined as the best linear approximations of $X_i$ and $X_j$ as a function of the remaining $\boldsymbol{X}_{\mathcal{K}}$. An explicit expression for $\hat{X}_i$ and $\hat{X}_j$ can be derived by partitioning the set of indices $\mathcal{D} = \{1, \dots, d\}$ into $\mathcal{IJ} = \{i, j\}$ and $\mathcal{K} = \mathcal{D} - \mathcal{IJ}$, with which the  covariance matrix $C \in \mathbb{R}^{d \times d}$ can be re-written in block format as
\begin{equation*}
    C =
    \left(\begin{array}{cc}
        C_\mathcal{IJ} & C_\mathcal{IJ,K} \\
        C_\mathcal{K,IJ} & C_\mathcal{K}
    \end{array} \right),
\end{equation*}
where $C_\mathcal{IJ} \in \mathbb{R}^{2 \times 2}$ has the covariances between variables with indices in $\mathcal{IJ}$, $C_\mathcal{K} \in \mathbb{R}^{(d-2) \times (d-2)}$ contains the covariances of the variables in $\mathcal{K}$, and $C_\mathcal{IJ,K} = C_\mathcal{K,IJ}^T \in \mathbb{R}^{2 \times (d - 2)}$ and has the covariances between the two sets. The linear influence of the variables $\boldsymbol{X}_\mathcal{K}$ associated to $\mathcal{K}$ over $\boldsymbol{X}_\mathcal{IJ} = (X_i, X_j)^T$ is 
\begin{equation*}
    \begin{pmatrix}
        \hat{X}_i \\
        \hat{X}_j
    \end{pmatrix}
    = {\rm E}(\boldsymbol{X}_\mathcal{IJ}) + C_\mathcal{IJ,K}\,C_\mathcal{K}^{-1} \left[\boldsymbol{X}_\mathcal{K} - {\rm E}( \boldsymbol{X}_\mathcal{K} ) \right].
\end{equation*}
It may be shown \citep[Chapter~23]{cramer1946mathematical} that $\hat{X}_i$ and $\hat{X}_j$ thus defined constitute the best linear approximations of $X_i$ and $X_j$ as a function of $\boldsymbol{X}_\mathcal{K}$, in terms of minimising the mean squared error. The partial correlation between $X_i$ and $X_j$ is defined as
\begin{equation*}
    r_{ij} = \frac{{\rm Cov}(X_i-\hat{X}_i,X_j-\hat{X}_j)}{\sqrt{{\rm Var}(X_i-\hat{X}_i)\,{\rm Var}(X_j-\hat{X}_j)}}.
\end{equation*}
Similar to the marginal correlation $\rho_{ij}$, the partial correlation $r_{ij}$ measures the degree up to which $X_i$ and $X_j$ influence each other, the difference being that $r_{ij}$ only correlates the part of $X_i$ and $X_j$ that is not linearly predicted by the remaining variables. It turns out (see for example \citep{williams2020back}) that $r_{ij}$ can be alternatively computed from the precision matrix $\Omega = C^{-1}$ as
\begin{equation} \label{eq:rij_omega}
    r_{ij} = \frac{-\omega_{ij}} { \sqrt{\omega_{ii} \ \omega_{jj}}}.
\end{equation}

As mentioned in Sect.~\ref{sect:intro}, partial correlations are often interpreted as a measure of conditional dependence \citep{millington2020partial,von2021exploratory,werme2022integrated}. However, a measure that is more precisely related to conditional dependence is the conditional correlation
\begin{equation}  \label{eq:cond_corr}
    \rho_{ij|\mathcal{K}} = \frac{{\rm Cov}(X_i, X_j | \boldsymbol{X}_\mathcal{K} = \boldsymbol{x}_\mathcal{K})}{\sqrt{{\rm Var}(X_i | \boldsymbol{X}_\mathcal{K} = \boldsymbol{x}_\mathcal{K})\,{\rm Var}(X_j | \boldsymbol{X}_\mathcal{K} = \boldsymbol{x}_\mathcal{K})}}.
\end{equation}
If variables $X_i$ and $X_j$ are conditionally independent, then $\rho_{ij|\mathcal{K}} = 0$. The partial correlation $r_{ij}$, however, need not vanish \citep{baba2004partial}. Still, as far as we know, conditional correlations are rarely used in real-world applications, probably because the dependence they often bear on the conditioned values $\boldsymbol{x}_\mathcal{K}$ make them difficult to estimate and to work with.

Interestingly, some probability distributions exhibit a close relationship between partial and conditional correlations. If the conditional expectations ${\rm E}(X_i | \boldsymbol{X}_\mathcal{K} = \boldsymbol{x}_\mathcal{K})$ and ${\rm E}(X_j | \boldsymbol{X}_\mathcal{K} = \boldsymbol{x}_\mathcal{K})$ are linear functions of the conditioned values $\boldsymbol{x}_\mathcal{K}$, then \citep{baba2004partial}
\begin{equation} \label{eq:linear_exp_part_corr}
    r_{ij} = \frac{{\rm E}[{\rm Cov}(X_i, X_j | \boldsymbol{X}_\mathcal{K})]}{\sqrt{{\rm E}[{\rm Var}(X_i | \boldsymbol{X}_\mathcal{K})]\,{\rm E}[{\rm Var}(X_j | \boldsymbol{X}_\mathcal{K})]}},
\end{equation}
where the expectation values should be taken with respect to $\boldsymbol{X}_\mathcal{K}$.  Equation~\ref{eq:linear_exp_part_corr} only differs from Eq.~\ref{eq:cond_corr} in that all conditional covariances are averaged over the fixed values.

For some of the probability distributions satisfying the linear conditional expectation condition, Eq.~\ref{eq:linear_exp_part_corr} reduces to Eq.~\ref{eq:cond_corr}, and partial and conditional correlations coincide. For example, Gaussian random variables not only have linear conditional expectation values, but also have conditional covariances that are independent of the conditioned values. For other distributions (i.e. non-Gaussian), conditional covariances may depend on the conditioned values in such a way that the dependence is lost when dividing by the variances. This is the case, for example, of the multinomial, the multivariate hypergeometric, the Dirichlet and the family of elliptical distributions \citep{baba2005}.

Partial correlations may have an intuitive interpretation even if they differ from conditional correlations. Consider $d$ univariate distributions $f_1(x), \dots, f_d(x)$ that for simplicity are assumed to have equal variances. Assume also that the probability distribution of $\boldsymbol{X}$ can be written as
\begin{equation} \label{eq:factor_mod}
    p(\boldsymbol{x}) \propto \prod_{\ell=1}^d f_\ell(\boldsymbol{w}_\ell^T \boldsymbol{x}),
\end{equation}
where each $\boldsymbol{w}_\ell = (w_{\ell 1}, \dots, w_{\ell d})^T$ is a $d$-dimensional vector that mixes the components of $\boldsymbol{X}$, and the proportionality constant does not depend on $\boldsymbol{x}$. This condition implies that $\boldsymbol{X}$ is a linear transformation of a random vector of independent components. Then the partial correlation between $X_i$ and $X_j$ is
\begin{equation} \label{eq:interpretarij}
    r_{ij} = \frac{ - \sum_{\ell=1}^d w_{\ell i} \, w_{\ell j}}{\sqrt{\left[ \sum_{\ell=1}^d w_{\ell i}^2 \right] \left[ \sum_{\ell=1}^d w_{\ell j}^2 \right]}}.
\end{equation}
Thus, $r_{ij}$ quantifies the joint occurrence of $X_i$ and $X_j$ as arguments of the univariate distributions (numerator), relative to their individual occurrence (denominator). If $X_i$ and $X_j$ never appear together in any of the arguments, then $r_{ij} = 0$. The particular value and sign of $r_{ij}$ indicate how $X_i$ and $X_j$ tend to co-fluctuate around their means when the rest of the variables are fixed to their respective means. If the variances of the functions $f_\ell$ are not all equal, then each term in each sum in Eq.~\ref{eq:interpretarij} is weighted by the corresponding variance. A generalisation of Eq.~\ref{eq:interpretarij} to a number of factors larger than $d$ is possible by factorising the moment-generating function, rather than the probability distribution.

The interpretation of partial correlations provided by Eq.~\ref{eq:interpretarij} is only valid for distributions that factorise as in Eq.~\ref{eq:factor_mod}. The advantage is that the univariate distributions $f_1(x),$ $ \dots, f_d(x)$ can be arbitrary, and they can describe widely different types of random variables; some may be discrete, others continuous, some may have infinite support, others bounded, etc.


\section{Marginal correlations as a function of partial correlations}
\label{sec:correl}

Equations~\ref{eq:rho_C} and \ref{eq:rij_omega} define a precise relation between $\rho_{ij}$ and the set of all $r_{k\ell}$, implying that marginal correlations can be derived from partial correlations through matrix inversion and rescaling. Yet, in high-dimensional spaces, the relation obtained using standard matrix inversion techniques fails to yield conceptual insight, particularly when the matrix to be inverted is expressed in terms of indeterminate variables, as opposed to actual numbers. For example, it is by no means evident how a particular marginal correlation $\rho_{ij}$ changes when the partial correlation graph is modified. Modifications of interest include, for instance, setting one or a few partial correlations $r_{k\ell}$ to zero, marginalising out some variables from the model, or adding new ones. The framework developed here enables these questions to be answered in an intuitive way. The result of the calculation always coincides with that of matrix inversion and rescaling, but the approach developed here yields more compact and interpretable mathematical expressions.

To carry out the derivation, we introduce the matrix of partial correlations $R$ with vanishing diagonal elements $r_{ii} \equiv 0$, and off-diagonal entries equal to the partial correlations $r_{ij}$. Equation~\ref{eq:rij_omega} implies that the matrices $\Omega$ and $R$ can be related by defining a diagonal matrix $\Lambda$ with entries $\lambda_{ii} = \sqrt{\omega_{ii}}$, so
\begin{equation} \label{eq:omega_lambda_r}
    \Omega = \Lambda \, (\mathbb{1} - R) \, \Lambda,
\end{equation}
where $\mathbb{1}$ is the $d \times d$ identity matrix. Using Eqs.~\ref{eq:rho_C} and \ref{eq:omega_lambda_r} to write $\rho_{ij}$ in terms of $R$, we get
\begin{align} \label{eq:rho_of_R}
    \rho_{ij} &= \frac{\left(\Omega^{-1}\right)_{ij}}{\sqrt{\left(\Omega^{-1}\right)_{ii} \, \left(\Omega^{-1}\right)_{jj}}} \nonumber \\
    & = \frac{\big[\left(\mathbb{1} - R\right)^{-1}\big]_{ij}}{\sqrt{\big[\left(\mathbb{1} - R\right)^{-1}\big]_{ii} \, \big[\left(\mathbb{1} - R\right)^{-1}\big]_{jj}}}.
\end{align}
If a matrix $M$ has spectral radius $\nu(M) < 1$, i.e. all its eigenvalues belong to the interval $(-1, 1)$, then $\mathbb{1} - M$ is invertible, and \citep{Meyer2010}
\begin{equation} \label{eq:neumann}
    (\mathbb{1} - M)^{-1} = \sum_{n = 0}^{{\infty}} M^n,
\end{equation}
\noindent The series in the right-hand side is the Neumann series of $M$. If $\nu(R) < 1$, Eq.~\ref{eq:neumann} implies that the element $ij$ of $(\mathbb{1} - R)^{-1}$ can be written as
\begin{equation}
\begin{split}
    \big[\left(\mathbb{1} - R\right)^{-1}\big]_{ij} = \ & \delta_{ij} + r_{ij} + \sum_{k} r_{ik} r_{kj} \\ & + \sum_{k_1, k_2} r_{ik_1} r_{k_1 k_2} r_{k_2j} + \dots
\end{split}
\label{eq:r_neumann}
\end{equation}
\noindent where the indices of the sums vary in the set $\{1, \dots, d\}$. Since $\Omega$ is positive-definite, $\mathbb{1} - R$ has positive eigenvalues, and therefore, the eigenvalues of $R$ are smaller than unity. Yet, the condition $\nu(R) < 1$ is not guaranteed to hold, because the eigenvalues of $R$ may still be smaller than $-1$. In what follows, we first restrict the analysis to partial correlation matrices satisfying $\nu(R) < 1$. In fact, as explained below, we start by imposing an even stronger condition: $\nu(R_{+}) < 1$, where $R_+$ is the matrix with entries $(R_+)_{ij} = |r_{ij}|$. This condition is relaxed in Appendix~\ref{appendixB}, where $\nu(R) < 1$ is shown to suffice. In Appendix~\ref{app:gen_expansion}, the derivation is extended to arbitrary $R$ matrices.  

\subsection{Graphical model interpretation}

We now interpret the expansion of Eq.~\ref{eq:r_neumann} in terms of a graphical model. We associate a weighted graph $G$ to the set of random variables $\{ X_1, \dots, X_d\}$, where node $i$ represents variable $X_i$, and the weighted adjacency matrix is the partial correlation matrix $R$. Since the diagonal elements of $R$ vanish, $G$ does not contain self-loops. Figure~\ref{fig:corr_graph_example}a shows an example of a graph of 4 variables with weighted adjacency matrix
\begin{equation} \label{eq:example_R}
    R = \left(
    \begin{array}{cccc}
        0 & r_{12} & r_{13} & 0 \\
        r_{12} & 0 & r_{23} & r_{24} \\
        r_{13} & r_{23} & 0 & r_{34} \\
        0 & r_{24} & r_{34} & 0
    \end{array}
    \right).
\end{equation}
The absence of a link between nodes $1$ and $4$ stems from $r_{14} = 0$, and implies that these nodes do not interact directly.
\begin{figure*}
    \centering
    \begin{tikzpicture}[>=Stealth,auto,node distance=2cm,thick,baseline=(1a.base)]
                \tikzset{
                    mystate/.style={fill=black,circle,inner sep=0pt,minimum width=4pt},
                    myedge1/.style={opacity=0.2},
                    myedge2/.style={opacity=0.4}}

                \node[mystate,label={left:$1$}] (1a) at (0,0) {};
                \node[mystate,label={[label distance=0.05cm]above:$2$}] (2a) at ($(1a) + (2,1)$) {};
                \node[mystate,label={[label distance=0.05cm]below:$3$}] (3a) at ($(1a) + (2,-1)$) {};
                \node[mystate,label={right:$4$}] (4a) at ($(1a) + (4,0)$) {};
        
                \path   (1a) edge [myedge1,sloped,anchor=center,above,text opacity=0.5] node {$r_{12}$} (2a)
                        (2a) edge [myedge1,sloped,anchor=center,above,text opacity=0.5] node {$r_{24}$} (4a);
                \path   (1a) edge [myedge1,sloped,anchor=center,below,text opacity=0.5] node {$r_{13}$} (3a)
                        (3a) edge [myedge1,sloped,anchor=center,below,text opacity=0.5] node {$r_{34}$} (4a);
                \path   (2a) edge [myedge1,right,text opacity=0.5] node {$r_{23}$} (3a);

                \node[mystate,label={left:$1$}] (1b) at ($(1a) + (5.75,0)$) {};
                \node[mystate,label={[label distance=0.05cm]above:$2$}] (2b) at ($(1b) + (2,1)$) {};
                \node[mystate,label={[label distance=0.05cm]below:$3$}] (3b) at ($(1b) + (2,-1)$) {};
                \node[mystate,label={right:$4$}] (4b) at ($(1b) + (4,0)$) {};
        
                \path   (1b) edge [myedge1] node {} (2b)
                        (2b) edge [myedge1] node {} (4b);
                \path   (1b) edge [myedge1] node {} (3b)
                        (3b) edge [myedge1] node {} (4b);
                \path   (2b) edge [myedge1] node {} (3b);

                \draw[myedge2,postaction={decorate},decoration={markings,mark=at position 0.8 with {\arrow{>}}}] plot [smooth,tension=0.4] coordinates {($(1b) + (0.05,0.2)$) ($(1b) + (1.7,1.1)$) ($(1b) + (2.3,1.1)$) ($(1b) + (3.95, 0.2)$)};

                \node[mystate,label={left:$1$}] (1c) at ($(1b) + (5.75,0)$) {};
                \node[mystate,label={[label distance=0.05cm]above:$2$}] (2c) at ($(1c) + (2,1)$) {};
                \node[mystate,label={[label distance=0.05cm]below:$3$}] (3c) at ($(1c) + (2,-1)$) {};
                \node[mystate,label={right:$4$}] (4c) at ($(1c) + (4,0)$) {};

                \path   (1c) edge [myedge1] node {} (2c)
                        (2c) edge [myedge1] node {} (4c);
                \path   (1c) edge [myedge1] node {} (3c)
                        (3c) edge [myedge1] node {} (4c);
                \path   (2c) edge [myedge1] node {} (3c);

                \draw[myedge2,postaction={decorate},
            decoration={markings,mark=at position 0.26 with {\arrow{>}},mark=at position 0.9 with {\arrow{>}}}] plot [smooth,tension=0.4] coordinates {($(1c) + (0.2,0)$) ($(1c)+(1.75,0.7)$) ($(1c)+(1.75,-0.65)$) ($(1c)+(0.4,-0.05)$) ($(1c)+(0.4,-0.35)$) ($(1c)+(1.7,-1.1)$) ($(1c)+(2.3,-1.1)$) ($(1c)+(3.95, -0.2)$)};

            \node[label={\large (a)}] (a) at ($(1a) + (-0.4,1)$) {};
            \node[label={\large (b)}] (b) at ($(a) + (1b) - (1a)$) {};
            \node[label={\large (c)}] (c) at ($(b) + (1c) - (1b)$) {};
    \end{tikzpicture}
    \caption{(a) Graph representation of the partial correlations of Eq.~\ref{eq:example_R}. (b and c) Two example paths contributing to the marginal correlation $\rho_{14}$.}
    \label{fig:corr_graph_example}
\end{figure*}

We define an $ij$-path in graph $G$ as a sequence of vertices such that the first vertex is $i$, the last vertex is $j$, and two vertices $k$ and $\ell$ can be consecutive only if the weight of edge $k \ell$ is different from zero. Vertices and edges can be visited repeatedly in an $ij$-path. We define an $ij^{*}$-path as an $ij$-path where vertices $i$ and $j$ only appear in the initial and the final positions, respectively. Additionally, we define the {\em weight of a path} simply as the product of the weights of all edges joining consecutive vertices. Since $r_{ij}$ measures the \emph{direct} interaction between $X_i$ and $X_j$, the weight of an $ij$-path of length $\ge 2$ can be interpreted as a quantification of the \emph{indirect} interaction between $X_i$ and $X_j$ mediated by all the direct interactions along the path. Equation~\ref{eq:r_neumann} shows that the net interaction between $X_i$ and $X_j$ is the sum of their indirect interaction along all paths connecting them.

Figure~\ref{fig:corr_graph_example}b represents the term $r_{12} \, r_{24}$ as the contribution from the path $1 \to 2 \to 4$ to the net interaction between $X_1$ and $X_4$. Since paths can revisit nodes, as for example in Fig.~\ref{fig:corr_graph_example}c, the interaction from $i$ to $j$ can feed back on $i$, and can do so repeatedly. In the absence of restrictions, the enumeration of all the $ij$-paths is dauntingly laborious. Fortunately, as shown in Sect.~\ref{sec:cancellation}, the marginal correlation may be written as a function of weights from $ij^*$-, $ii^*$-, and $jj^*$-paths alone.

Equation~\ref{eq:r_neumann} can be rewritten diagrammatically as
\footnotesize
\begin{align*}
    \big[\left(\mathbb{1} - R\right)^{-1}\big]_{ij}
    &= \delta_{ij} +
    \begin{tikzpicture}[->,>=Stealth,auto,shorten >=0.5pt,node distance=1.5cm,semithick,baseline=(i.base)]
        \tikzset{
            mystate/.style={fill=black,circle,inner sep=0pt,minimum width=4pt},
            myedge/.style={text opacity=0.9,opacity=0.6}}
        \node[mystate,label={above:$i$}] (i)               {};
        \node[mystate,label={above:$j$}] (j)  [right of=i] {};
        \path (i) edge [myedge,below] node {$r_{ij}$} (j);
    \end{tikzpicture}
    + \\
    & + \sum_{k}
    \begin{tikzpicture}[->,>=Stealth,auto,shorten >=0.5pt,node distance=1.5cm,semithick,baseline=(i.base)]
        \tikzset{
            mystate/.style={fill=black,circle,inner sep=0pt,minimum width=4pt},
            myedge/.style={text opacity=0.9,opacity=0.6}}
        \node[mystate,label={above:$i$}] (i)               {};
        \node[mystate,label={above:$k$}] (k)  [right of=i] {};
        \node[mystate,label={above:$j$}] (j)  [right of=k] {};
        \path (i) edge [myedge,below] node {$r_{ik}$} (k);
        \path (k) edge [myedge,below] node {$r_{kj}$} (j);
    \end{tikzpicture}
    + \\
    &  + \sum_{k_1, k_2}
    \begin{tikzpicture}[->,>=Stealth,auto,shorten >=0.5pt,node distance=1.5cm,semithick,baseline=(i.base)]
        \tikzset{
            mystate/.style={fill=black,circle,inner sep=0pt,minimum width=4pt},
            myedge/.style={text opacity=0.9,opacity=0.6}}
        \node[mystate,label={above:$i$}]    (i)                 {};
        \node[mystate,label={above:$k_1$}] (k1) [right of=i]    {};
        \node[mystate,label={above:$k_2$}] (k2) [right of=k1]   {};
        \node[mystate,label={above:$j$}]    (j) [right of=k2]   {};
        \path (i) edge [myedge,below] node {$r_{ik_1}$} (k1);
        \path (k1) edge [myedge,below] node {$r_{k_1 k_2}$} (k2);
        \path (k2) edge [myedge,below] node {$r_{k_2j}$} (j);
    \end{tikzpicture} \nonumber \\
    & + \dots,
\end{align*}
\normalsize
or simply,
\begin{equation}
    \big[\left(\mathbb{1} - R\right)^{-1}\big]_{ij} = \delta_{ij} + \sum_{\mathtt{P}_{ij}}
    \begin{tikzpicture}[->,>=Stealth,auto,shorten >=0.5pt,node distance=1.5cm,semithick,baseline=(i.base)]
        \tikzset{
            mystate/.style={fill=black,circle,inner sep=0pt,minimum width=4pt},
            myedge/.style={text opacity=0.9,opacity=0.6}}
        \node[mystate,label={above:$i$}] (i)               {};
        \node[mystate,label={above:$j$}] (j)  [right of=i] {};
        \path (i) edge [myedge,bend left,above] node {$\scriptstyle \mathtt{p}_{ij}$} (j);
    \end{tikzpicture},
    \label{eq:r_neumann_diag}
\end{equation}
where the sum runs over all $ij$-paths, and 
\begin{tikzpicture}[->,>=Stealth,auto,shorten >=0.5pt,node distance=1.5cm,semithick,baseline=(i.base)]
    \tikzset{
        mystate/.style={fill=black,circle,inner sep=0pt,minimum width=4pt},
        myedge/.style={text opacity=0.9,opacity=0.6}}
    \node[mystate,label={above:$i$}] (i)               {};
    \node[mystate,label={above:$j$}] (j)  [right of=i] {};
    \path (i) edge [myedge,bend left,above] node {$\scriptstyle \mathtt{p}_{ij}$} (j);
\end{tikzpicture}
indicates the weight of a particular path $\mathtt{p}_{ij}$, i.e.
\begin{equation*}
    \begin{tikzpicture}[->,>=Stealth,auto,shorten >=0.5pt,node distance=1.5cm,semithick,baseline=(i.base)]
        \tikzset{
            mystate/.style={fill=black,circle,inner sep=0pt,minimum width=4pt},
            myedge/.style={text opacity=0.9,opacity=0.6}}
        \node[mystate,label={above:$i$}] (i)               {};
        \node[mystate,label={above:$j$}] (j)  [right of=i] {};
        \path (i) edge [myedge,bend left,above] node {$\scriptstyle \mathtt{p}_{ij}$} (j);
    \end{tikzpicture}
    = r_{i k_1} r_{k_1 k_2} \dots r_{k_{n-1} k_{n}} r_{k_n j}
\end{equation*}
with $n \in \mathbb{N}$ and $k_{\ell} \in \{1, \dots, d\}$. 

The prescription of ``summing over all paths'' only makes sense if any rearrangement of terms in the series yields the same result. In the Neumann series, rearrangements of different powers of the matrix $R$ are guaranteed not to change the result of the sum when $\nu(R) < 1$. The prescription, however, requires a stronger condition, since it not only relies on the freedom to rearrange the powers of $R$, but more generally, on that to  rearrange the paths arbitrarily, even by regrouping paths that belong to different powers of $R$. These rearrangements are inconsequential only if the series of paths converges absolutely, which is guaranteed if and only if $\nu(R_{+})<1$ \citep{Meyer2010}. If a particular correlation matrix satisfies $\nu(R) < 1 < \nu(R_+)$, then the sum in Eq.~\ref{eq:r_neumann_diag} yields different results for different orderings. The correct result (the one that makes Eq.~\ref{eq:r_neumann_diag} true) is obtained when paths are arranged as prescribed by the Neumann series, that is, when all those paths belonging to the same power of $R$ are grouped together. In short, paths must be ordered by length. In this case, Eq.~\ref{eq:r_neumann_diag} must be associated with a slightly more specific prescription, one instructing to sum over all collections of paths, each collection of a fixed length.

So far we have used arrows to represent paths, in spite of links being undirected. Since $R$ is symmetric, the weight of the $ij$-path is equal to that of the $ji$-path. Therefore, when only considering the weight of each path, lines offer a more appropriate representation than arrows. Yet, calculating weights is not our only goal, we also need to enumerate all the paths upon which weights need to be summed. In this paper, we prefer to use arrows because they facilitate the enumeration. For example, when $i = j$ in Eq.~\ref{eq:r_neumann_diag}, the sum runs over all closed paths starting and ending at $i$. If one of these paths is reversed, that is, if the ordering of the nodes is inverted, a different path with equal weight may be obtained. If the directionality of arrows is disregarded, the reversed and the direct paths are represented identically. Yet, both need to be summed over.

When $\nu(R_{+}) < 1$, Eqs. \ref{eq:rho_of_R} and \ref{eq:r_neumann_diag} imply that the correlation coefficient between $X_i$ and $X_j$ is
\footnotesize
\begin{equation} \label{eq:rho_diagram}
    \rho_{ij} = \frac{
    \delta_{ij} + \displaystyle{\sum_{\substack{\mathtt{p}_{ij}}}}
    \begin{tikzpicture}[->,>=Stealth,auto,shorten >=0.5pt,node distance=1.5cm,semithick,baseline=(i.base)]
        \tikzset{
            mystate/.style={fill=black,circle,inner sep=0pt,minimum width=4pt},
            myedge/.style={text opacity=0.9,opacity=0.6}}
        \node[mystate,label={above:$i$}] (i)               {};
        \node[mystate,label={above:$j$}] (j)  [right of=i] {};
        \path (i) edge [myedge,bend left,above] node {$\scriptstyle \mathtt{p}_{ij}$} (j);
    \end{tikzpicture}
    }{ \sqrt{
    \left[
    1 + \displaystyle{\sum_{\substack{\mathtt{p}_{ii}}}}
    \begin{tikzpicture}[->,>=Stealth,auto,shorten >=0.5pt,node distance=1.5cm,semithick,baseline=(i.base)]
        \tikzset{
            mystate/.style={fill=black,circle,inner sep=0pt,minimum width=4pt},
            myloopleft/.style={out=-140,in=140,looseness=20,text opacity=0.9,opacity=0.6},
            myloopright/.style={out=40,in=-40,looseness=20,text opacity=0.9,opacity=0.6}}
        \node[mystate,label={above:$i$}] (i) {};
        \path (i) edge [myloopright]  node {$\scriptstyle \mathtt{p}_{ii}$} (i);
    \end{tikzpicture}
    \right] \left[
    1 + \displaystyle{\sum_{\substack{\mathtt{p}_{jj}}}}
    \begin{tikzpicture}[->,>=Stealth,auto,shorten >=0.5pt,node distance=1.5cm,semithick,baseline=(i.base)]
        \tikzset{
            mystate/.style={fill=black,circle,inner sep=0pt,minimum width=4pt},
            myloopleft/.style={out=-140,in=140,looseness=20,text opacity=0.9,opacity=0.6},
            myloopright/.style={out=40,in=-40,looseness=20,text opacity=0.9,opacity=0.6}}
        \node[mystate,label={above:$j$}] (j) {};
        \path (j) edge [myloopright]  node {$\scriptstyle \mathtt{p}_{jj}$} (j);
    \end{tikzpicture}
    \right]
    }},
\end{equation}
\normalsize
where the two sums in the denominator are over all closed paths starting at $i$ and at $j$, respectively. From now on, we assume that $i \neq j$, and therefore omit the Kronecker delta of Eq.~\ref{eq:rho_diagram}. To exemplify the sum over paths, we now analyse a simple network. 

\subsubsection{Example: A connected pair} \label{subsubsect:example_pair_A}

Consider two variables $X_1$ and $X_2$ with matrix
\begin{equation*}
    R = \left(
    \begin{array}{cc} 
        0 & r_{12} \\
        r_{12} & 0
    \end{array}
    \right)
\end{equation*}
and associated graph
\begin{equation*}
        \begin{tikzpicture}[>=Stealth,auto,node distance=2cm,thick,baseline=(1.base)]
            \tikzset{
                mystate/.style={fill=black,circle,inner sep=0pt,minimum width=4pt},
                myedge/.style={opacity=0.2,text opacity=0.5}}

            \node[mystate,label={above:$1$}] (1) at (0,0) {};
            \node[mystate,label={above:$2$}] (2) at (2,0) {};
        
            \path   (1) edge [myedge,sloped,anchor=center,below] node {$r_{12}$} (2);
        \end{tikzpicture}.
\end{equation*}
For any $r_{12} \in (-1, 1)$, the spectral radius $\nu(R_+) < 1$, which guarantees that rearrangements of paths in each of the series of Eq.~\ref{eq:rho_diagram} are inconsequential. There are infinitely many paths $\mathtt{p}_{ij}$ from $i$ to $j$ contributing to Eq.~\ref{eq:rho_diagram}, and they can be enumerated by the number of times the link $r_{12}$ is traversed, i.e.
\begin{equation*}
    \sum_{\mathtt{p}_{ij}}
    \begin{tikzpicture}[->,>=Stealth,auto,shorten >=0.5pt,node distance=1.5cm,semithick,baseline=(i.base)]
        \tikzset{
            mystate/.style={fill=black,circle,inner sep=0pt,minimum width=4pt},
            myedge/.style={text opacity=0.9,opacity=0.6}}
        \node[mystate,label={above:$i$}] (i)               {};
        \node[mystate,label={above:$j$}] (j)  [right of=i] {};
        \path (i) edge [myedge,bend left,above] node {$\scriptstyle \mathtt{p}_{ij}$} (j);
    \end{tikzpicture}
    = r_{12} + r_{12}^3 + r_{12}^5 + \dots = \frac{r_{12}}{1 - r_{12}^2},
\end{equation*}
where we used the formula for the geometric series. Analogously, the sums over closed paths are
\begin{equation*}
\begin{split}
    \sum_{\mathtt{p}_{ii}}
    \begin{tikzpicture}[->,>=Stealth,auto,shorten >=0.5pt,node distance=1.5cm,semithick,baseline=(i.base)]
        \tikzset{
            mystate/.style={fill=black,circle,inner sep=0pt,minimum width=4pt},
            myloopleft/.style={out=-140,in=140,looseness=20,text opacity=0.9,opacity=0.6},
            myloopright/.style={out=40,in=-40,looseness=20,text opacity=0.9,opacity=0.6}}
        \node[mystate,label={above:$i$}] (i) {};
        \path (i) edge [myloopright]  node {$\scriptstyle \mathtt{p}_{ii}$} (i);
    \end{tikzpicture}
    = \sum_{\mathtt{p}_{jj}}
    \begin{tikzpicture}[->,>=Stealth,auto,shorten >=0.5pt,node distance=1.5cm,semithick,baseline=(i.base)]
        \tikzset{
            mystate/.style={fill=black,circle,inner sep=0pt,minimum width=4pt},
            myloopleft/.style={out=-140,in=140,looseness=20,text opacity=0.9,opacity=0.6},
            myloopright/.style={out=40,in=-40,looseness=20,text opacity=0.9,opacity=0.6}}
        \node[mystate,label={above:$j$}] (i) {};
        \path (i) edge [myloopright]  node {$\scriptstyle \mathtt{p}_{jj}$} (i);
    \end{tikzpicture}
    & = r_{12}^2 + r_{12}^4 + r_{12}^6 + \dots  \\
    & = \frac{r_{12}^2}{1 - r_{12}^2}.
\end{split}
\end{equation*}
Plugging both results into Eq.~\ref{eq:rho_diagram}, the (expected) outcome $\rho_{12} = r_{12}$ is obtained. 

Therefore, even in the simplest case of a network of $d = 2$ nodes, an infinite number of paths $\mathtt{p}_{ij}$, $\mathtt{p}_{ii}$ and $\mathtt{p}_{jj}$ need to be considered. In the following section we prove that Eq.~\ref{eq:rho_diagram} can be re-written only in terms of paths $\mathtt{p}_{ij}^*$, $\mathtt{p}_{ii}^*$ and $\mathtt{p}_{jj}^*$, drastically reducing the number of paths to be enumerated. To show the simplification, in Sect.~\ref{subsect:example_pair} we revisit the example of a connected pair of nodes.

\subsection{Cancellation of closed sub-paths containing both \textit{i} and \textit{j}}
\label{sec:cancellation}

Every $ij$-path contains at least one $ij^*$-path, and might also contain $ji^*$-subpaths. If the $ij$-path $\mathtt{p}_{ij}$ does not contain any $ji^*$-subpath, it can be written as the concatenation of paths from at most three different families: $\mathcal{P}^{*}_{ij}$ (family of all $ij^{*}$-paths), $\mathcal{P}^{i}_{jj}$ (family of all $jj$-paths that do not visit $i$) and $\mathcal{P}^{j}_{ii}$ (family of all $ii$-paths that do not visit $j$). Namely, $\mathtt{p}_{ij}=\mathtt{p}^j_{ii} \, \mathtt{p}^{*}_{ij} \, \mathtt{p}^i_{jj}$ or $\mathtt{p}_{ij}=\mathtt{p}^{*}_{ij} \, \mathtt{p}^i_{jj}$ or $\mathtt{p}_{ij}=\mathtt{p}^j_{ii} \, \mathtt{p}^{*}_{ij}$ or $\mathtt{p}_{ij}=\mathtt{p}^{*}_{ij}$, with $\mathtt{p}^{*}_{ij} \in \mathcal{P}^{*}_{ij}$.

On the other hand, if the $ij$-path $\mathtt{q}_{ij}$ contains at least one $ji^*$-path, it can be written as the concatenation of two paths: $\mathtt{q}_{ij}=\tilde{\mathtt{p}}_{iji} \, \mathtt{p}_{ij}$ where $\mathtt{p}_{ij}$ is an $ij$-path that does not contain any $ji^*$-subpath (i.e. of the class described in the preceding paragraph) and $\tilde{\mathtt{p}}_{iji} \in \tilde{\mathcal{P}}_{iji}$ (family of $ii$-paths that contain a $ji^*$-subpath at the end of the path). Using the concatenations described in the preceding paragraph, $\mathtt{q}_{ij}=\tilde{\mathtt{p}}_{iji} \, \mathtt{p}^j_{ii} \, \mathtt{p}^{*}_{ij} \, \mathtt{p}^i_{jj}$ or $\mathtt{q}_{ij}=\tilde{\mathtt{p}}_{iji} \, \mathtt{p}^{*}_{ij} \, \mathtt{p}^i_{jj}$ or $\mathtt{q}_{ij}=\tilde{\mathtt{p}}_{iji} \, \mathtt{p}^j_{ii} \, \mathtt{p}^{*}_{ij}$ or $\mathtt{q}_{ij}=\tilde{\mathtt{p}}_{iji} \, \mathtt{p}^{*}_{ij}$. The sum over all paths can then be written as
\begin{widetext}
\begin{align}
    \sum_{\mathtt{p}_{ij}} 
    \begin{tikzpicture}[->,>=Stealth,auto,shorten >=0.5pt,node distance=1.5cm,semithick,baseline=(i.base)]
        \tikzset{
            mystate/.style={fill=black,circle,inner sep=0pt,minimum width=4pt},
            myedge/.style={text opacity=0.9,opacity=0.6}}
        \node[mystate,label={above:$i$}] (i)               {};
        \node[mystate,label={above:$j$}] (j)  [right of=i] {};
        \path (i) edge [myedge,bend left,above] node {$\scriptstyle \mathtt{p}_{ij}$} (j);
    \end{tikzpicture}
    & = \left[1+ \sum_{\mathtt{p}_{ii}^j}
    \begin{tikzpicture}[->,>=Stealth,auto,shorten >=0.5pt,node distance=1.5cm,semithick,baseline=(i.base)]
        \tikzset{
            mystate/.style={fill=black,circle,inner sep=0pt,minimum width=4pt},
            myloopleft/.style={out=-140,in=140,looseness=20,text opacity=0.9,opacity=0.6},
            myloopright/.style={out=40,in=-40,looseness=20,text opacity=0.9,opacity=0.6}}
        \node[mystate,label={above:$i$}] (i) {};
        \path (i) edge [myloopright]  node {$\scriptstyle \mathtt{p}_{ii}^j$} (i);
    \end{tikzpicture}
    \right] \left[ \sum_{\mathtt{p}_{ij}^*}
    \begin{tikzpicture}[->,>=Stealth,auto,shorten >=0.5pt,node distance=1.5cm,semithick,baseline=(i.base)]
        \tikzset{
            mystate/.style={fill=black,circle,inner sep=0pt,minimum width=4pt},
            myedge/.style={text opacity=0.9,opacity=0.6}}
        \node[mystate,label={above:$i$}] (i)               {};
        \node[mystate,label={above:$j$}] (j)  [right of=i] {};
        \path (i) edge [myedge,bend left,above] node {$\scriptstyle \mathtt{p}_{ij}^*$} (j);
    \end{tikzpicture}
    \right]
    \left[1+ \sum_{\mathtt{p}_{jj}^i}
    \begin{tikzpicture}[->,>=Stealth,auto,shorten >=0.5pt,node distance=1.5cm,semithick,baseline=(i.base)]
        \tikzset{
            mystate/.style={fill=black,circle,inner sep=0pt,minimum width=4pt},
            myloopleft/.style={out=-140,in=140,looseness=20,text opacity=0.9,opacity=0.6},
            myloopright/.style={out=40,in=-40,looseness=20,text opacity=0.9,opacity=0.6}}
        \node[mystate,label={above:$j$}] (i) {};
        \path (i) edge [myloopright]  node {$\scriptstyle \mathtt{p}_{jj}^i$} (i);
    \end{tikzpicture}
    \right] + \nonumber \\
    & \qquad + \left[ \sum_{\tilde{\mathtt{p}}_{iji}}
    \begin{tikzpicture}[->,>=Stealth,auto,shorten >=0.5pt,node distance=1.5cm,semithick,baseline=(i.base)]
        \tikzset{
            mystate/.style={fill=black,circle,inner sep=0pt,minimum width=4pt},
            myedge/.style={text opacity=0.9,opacity=0.6}}
        \node[mystate,label={above:$i$}] (i)               {};
        \node[mystate,label={above:$j$}] (j)  [right of=i] {};
        \path (i) edge [myedge,bend left,above] node {$\scriptstyle \tilde{\mathtt{p}}_{iji}$} (j);
        \path (j) edge [myedge,bend left,below] (i);
    \end{tikzpicture}
    \right]
   \left[1+ \sum_{\mathtt{p}_{ii}^j}
    \begin{tikzpicture}[->,>=Stealth,auto,shorten >=0.5pt,node distance=1.5cm,semithick,baseline=(i.base)]
        \tikzset{
            mystate/.style={fill=black,circle,inner sep=0pt,minimum width=4pt},
            myloopleft/.style={out=-140,in=140,looseness=20,text opacity=0.9,opacity=0.6},
            myloopright/.style={out=40,in=-40,looseness=20,text opacity=0.9,opacity=0.6}}
        \node[mystate,label={above:$i$}] (i) {};
        \path (i) edge [myloopright]  node {$\scriptstyle \mathtt{p}_{ii}^j$} (i);
    \end{tikzpicture}
    \right] \left[ \sum_{\mathtt{p}_{ij}^*}
    \begin{tikzpicture}[->,>=Stealth,auto,shorten >=0.5pt,node distance=1.5cm,semithick,baseline=(i.base)]
        \tikzset{
            mystate/.style={fill=black,circle,inner sep=0pt,minimum width=4pt},
            myedge/.style={text opacity=0.9,opacity=0.6}}
        \node[mystate,label={above:$i$}] (i)               {};
        \node[mystate,label={above:$j$}] (j)  [right of=i] {};
        \path (i) edge [myedge,bend left,above] node {$\scriptstyle \mathtt{p}_{ij}^*$} (j);
    \end{tikzpicture}
    \right]
    \left[1+ \sum_{\mathtt{p}_{jj}^i}
    \begin{tikzpicture}[->,>=Stealth,auto,shorten >=0.5pt,node distance=1.5cm,semithick,baseline=(i.base)]
        \tikzset{
            mystate/.style={fill=black,circle,inner sep=0pt,minimum width=4pt},
            myloopleft/.style={out=-140,in=140,looseness=20,text opacity=0.9,opacity=0.6},
            myloopright/.style={out=40,in=-40,looseness=20,text opacity=0.9,opacity=0.6}}
        \node[mystate,label={above:$j$}] (i) {};
        \path (i) edge [myloopright]  node {$\scriptstyle \mathtt{p}_{jj}^i$} (i);
    \end{tikzpicture}
    \right]  \nonumber \\
    & = \left[ 1+\sum_{\tilde{\mathtt{p}}_{iji}}
    \begin{tikzpicture}[->,>=Stealth,auto,shorten >=0.5pt,node distance=1.5cm,semithick,baseline=(i.base)]
        \tikzset{
            mystate/.style={fill=black,circle,inner sep=0pt,minimum width=4pt},
            myedge/.style={text opacity=0.9,opacity=0.6}}
        \node[mystate,label={above:$i$}] (i)               {};
        \node[mystate,label={above:$j$}] (j)  [right of=i] {};
        \path (i) edge [myedge,bend left,above] node {$\scriptstyle \tilde{\mathtt{p}}_{iji}$} (j);
        \path (j) edge [myedge,bend left,below] (i);
    \end{tikzpicture}
    \right]
    \left[1+ \sum_{\mathtt{p}_{ii}^j}
    \begin{tikzpicture}[->,>=Stealth,auto,shorten >=0.5pt,node distance=1.5cm,semithick,baseline=(i.base)]
        \tikzset{
            mystate/.style={fill=black,circle,inner sep=0pt,minimum width=4pt},
            myloopleft/.style={out=-140,in=140,looseness=20,text opacity=0.9,opacity=0.6},
            myloopright/.style={out=40,in=-40,looseness=20,text opacity=0.9,opacity=0.6}}
        \node[mystate,label={above:$i$}] (i) {};
        \path (i) edge [myloopright]  node {$\scriptstyle \mathtt{p}_{ii}^j$} (i);
    \end{tikzpicture}
    \right] \left[ \sum_{\mathtt{p}_{ij}^*}
    \begin{tikzpicture}[->,>=Stealth,auto,shorten >=0.5pt,node distance=1.5cm,semithick,baseline=(i.base)]
        \tikzset{
            mystate/.style={fill=black,circle,inner sep=0pt,minimum width=4pt},
            myedge/.style={text opacity=0.9,opacity=0.6}}
        \node[mystate,label={above:$i$}] (i)               {};
        \node[mystate,label={above:$j$}] (j)  [right of=i] {};
        \path (i) edge [myedge,bend left,above] node {$\scriptstyle \mathtt{p}_{ij}^*$} (j);
    \end{tikzpicture}
    \right]
    \left[1+ \sum_{\mathtt{p}_{jj}^i}
    \begin{tikzpicture}[->,>=Stealth,auto,shorten >=0.5pt,node distance=1.5cm,semithick,baseline=(i.base)]
        \tikzset{
            mystate/.style={fill=black,circle,inner sep=0pt,minimum width=4pt},
            myloopleft/.style={out=-140,in=140,looseness=20,text opacity=0.9,opacity=0.6},
            myloopright/.style={out=40,in=-40,looseness=20,text opacity=0.9,opacity=0.6}}
        \node[mystate,label={above:$j$}] (i) {};
        \path (i) edge [myloopright]  node {$\scriptstyle \mathtt{p}_{jj}^i$} (i);
    \end{tikzpicture}
    \right]. 
    \label{eq:pij}
\end{align}
\end{widetext}

The separation of each path into a concatenation of two sub-paths can also be applied to the denominator of Eq.~\ref{eq:rho_diagram}. If an $ii$-path does not contain a $ji^*$-subpath it must belong to $\mathcal{P}^{j}_{ii}$. On the other hand, if an $ii$-path does contain a $ji^*$-subpath, it can be written as a concatenation of a path in $\tilde{\mathcal{P}}_{iji}$ and a path in $\mathcal{P}^{j}_{ii}$. Therefore,
\begin{equation}
\begin{split}
    \sum_{\mathtt{p}_{ii}}
    \begin{tikzpicture}[->,>=Stealth,auto,shorten >=0.5pt,node distance=1.5cm,semithick,baseline=(i.base)]
        \tikzset{
            mystate/.style={fill=black,circle,inner sep=0pt,minimum width=4pt},
            myloopleft/.style={out=-140,in=140,looseness=20,text opacity=0.9,opacity=0.6},
            myloopright/.style={out=40,in=-40,looseness=20,text opacity=0.9,opacity=0.6}}
        \node[mystate,label={above:$i$}] (i) {};
        \path (i) edge [myloopright]  node {$\scriptstyle \mathtt{p}_{ii}$} (i);
    \end{tikzpicture}
    = & \left[ 1+\sum_{\tilde{\mathtt{p}}_{iji}}
    \begin{tikzpicture}[->,>=Stealth,auto,shorten >=0.5pt,node distance=1.5cm,semithick,baseline=(i.base)]
        \tikzset{
            mystate/.style={fill=black,circle,inner sep=0pt,minimum width=4pt},
            myedge/.style={text opacity=0.9,opacity=0.6}}
        \node[mystate,label={above:$i$}] (i)               {};
        \node[mystate,label={above:$j$}] (j)  [right of=i] {};
        \path (i) edge [myedge,bend left,above] node {$\scriptstyle \tilde{\mathtt{p}}_{iji}$} (j);
        \path (j) edge [myedge,bend left,below] (i);
    \end{tikzpicture}
    \right] \times \\
    & \times \left[1+ \sum_{\mathtt{p}_{ii}^j}
    \begin{tikzpicture}[->,>=Stealth,auto,shorten >=0.5pt,node distance=1.5cm,semithick,baseline=(i.base)]
        \tikzset{
            mystate/.style={fill=black,circle,inner sep=0pt,minimum width=4pt},
            myloopleft/.style={out=-140,in=140,looseness=20,text opacity=0.9,opacity=0.6},
            myloopright/.style={out=40,in=-40,looseness=20,text opacity=0.9,opacity=0.6}}
        \node[mystate,label={above:$i$}] (i) {};
        \path (i) edge [myloopright]  node {$\scriptstyle \mathtt{p}_{ii}^j$} (i);
    \end{tikzpicture}
    \right].
\end{split}
    \label{eq:pii}
\end{equation}
Interchanging $i$ and $j$,
\begin{equation}
\begin{split}
    \sum_{\mathtt{p}_{jj}}
    \begin{tikzpicture}[->,>=Stealth,auto,shorten >=0.5pt,node distance=1.5cm,semithick,baseline=(i.base)]
        \tikzset{
            mystate/.style={fill=black,circle,inner sep=0pt,minimum width=4pt},
            myloopleft/.style={out=-140,in=140,looseness=20,text opacity=0.9,opacity=0.6},
            myloopright/.style={out=40,in=-40,looseness=20,text opacity=0.9,opacity=0.6}}
        \node[mystate,label={above:$j$}] (j) {};
        \path (j) edge [myloopright]  node {$\scriptstyle \mathtt{p}_{jj}$} (j);
    \end{tikzpicture}
    = & \left[ 1+ \sum_{\tilde{\mathtt{p}}_{jij}}
    \begin{tikzpicture}[->,>=Stealth,auto,shorten >=0.5pt,node distance=1.5cm,semithick,baseline=(i.base)]
        \tikzset{
            mystate/.style={fill=black,circle,inner sep=0pt,minimum width=4pt},
            myedge/.style={text opacity=0.9,opacity=0.6}}
        \node[mystate,label={above:$j$}] (j)               {};
        \node[mystate,label={above:$i$}] (i)  [right of=j] {};
        \path (j) edge [myedge,bend left,above] node {$\scriptstyle \tilde{\mathtt{p}}_{jij}$} (i);
        \path (i) edge [myedge,bend left,below] (j);
    \end{tikzpicture}
    \right] \times \\
    & \times \left[1+ \sum_{\mathtt{p}_{jj}^i}
    \begin{tikzpicture}[->,>=Stealth,auto,shorten >=0.5pt,node distance=1.5cm,semithick,baseline=(i.base)]
        \tikzset{
            mystate/.style={fill=black,circle,inner sep=0pt,minimum width=4pt},
            myloopleft/.style={out=-140,in=140,looseness=20,text opacity=0.9,opacity=0.6},
            myloopright/.style={out=40,in=-40,looseness=20,text opacity=0.9,opacity=0.6}}
        \node[mystate,label={above:$j$}] (j) {};
        \path (j) edge [myloopright]  node {$\scriptstyle \mathtt{p}_{jj}^i$} (j);
    \end{tikzpicture}
    \right].
\end{split}
    \label{eq:pjj}
\end{equation}
\noindent where $\tilde{\mathtt{p}}_{jij} \in \tilde{\mathcal{P}}_{jij}$, the set of $jj$-paths that terminate in a $ij^*$-path. 

The symmetry of the problem defines a one-to-one correspondence between the elements of $\tilde{\mathcal{P}}_{iji}$ and of $\tilde{\mathcal{P}}_{jij}$. By definition, each $\tilde{\mathtt{p}}_{iji}$ path is closed, and ends in a $ji^*$-subpath. Its partner in $\tilde{\mathcal{P}}_{jij}$ is the path that traverses $\tilde{\mathtt{p}}_{iji}$ in the opposite direction, but beginning in the first step of the $ji^*$-subpath. This is a $jj$-path that ends in a $ij^*$-subpath, and thus belongs to $\tilde{\mathcal{P}}_{jij}$. A similar reasoning, interchanging $i$ and $j$, relates paths in $\tilde{\mathcal{P}}_{jij}$ to paths in $\tilde{\mathcal{P}}_{iji}$, implying that the correspondence is one-to-one. Furthermore, since the two related paths contain the same edges (just in different order), they have the same weight, so
\begin{align}
    & 
   \sum_{\tilde{\mathtt{p}}_{iji}}
    \begin{tikzpicture}[->,>=Stealth,auto,shorten >=0.5pt,node distance=1.5cm,semithick,baseline=(i.base)]
        \tikzset{
            mystate/.style={fill=black,circle,inner sep=0pt,minimum width=4pt},
            myedge/.style={text opacity=0.9,opacity=0.6}}
        \node[mystate,label={above:$i$}] (i)               {};
        \node[mystate,label={above:$j$}] (j)  [right of=i] {};
        \path (i) edge [myedge,bend left,above] node {$\scriptstyle \tilde{\mathtt{p}}_{iji}$} (j);
        \path (j) edge [myedge,bend left,below] (i);
    \end{tikzpicture}
    = \sum_{\tilde{\mathtt{p}}_{jij}}
    \begin{tikzpicture}[->,>=Stealth,auto,shorten >=0.5pt,node distance=1.5cm,semithick,baseline=(i.base)]
        \tikzset{
            mystate/.style={fill=black,circle,inner sep=0pt,minimum width=4pt},
            myedge/.style={text opacity=0.9,opacity=0.6}}
        \node[mystate,label={above:$j$}] (j)               {};
        \node[mystate,label={above:$i$}] (i)  [right of=j] {};
        \path (j) edge [myedge,bend left,above] node {$\scriptstyle \tilde{\mathtt{p}}_{jij}$} (i);
        \path (i) edge [myedge,bend left,below] (j);
    \end{tikzpicture}
    \label{eq:equality}
\end{align}

Replacing Eqs.~\ref{eq:pij},~\ref{eq:pii}, and~\ref{eq:pjj} in Eq.~\ref{eq:rho_diagram}, and using Eq.~\ref{eq:equality}, the contributions from paths in $\tilde{\mathcal{P}}_{iji}$ cancel out, and
\small
\begin{equation} \label{eq:rho_diagram_3}
\begin{split}
    \rho_{ij} =  &  \left[1+ \sum_{\mathtt{p}_{ii}^j}
    \begin{tikzpicture}[->,>=Stealth,auto,shorten >=0.5pt,node distance=1.5cm,semithick,baseline=(i.base)]
        \tikzset{
            mystate/.style={fill=black,circle,inner sep=0pt,minimum width=4pt},
            myloopleft/.style={out=-140,in=140,looseness=20,text opacity=0.9,opacity=0.6},
            myloopright/.style={out=40,in=-40,looseness=20,text opacity=0.9,opacity=0.6}}
        \node[mystate,label={above:$i$}] (i) {};
        \path (i) edge [myloopright]  node {$\scriptstyle \mathtt{p}_{ii}^j$} (i);
    \end{tikzpicture}
    \right]^{1/2}   \left[ \sum_{\mathtt{p}_{ij}^*}
    \begin{tikzpicture}[->,>=Stealth,auto,shorten >=0.5pt,node distance=1.5cm,semithick,baseline=(i.base)]
        \tikzset{
            mystate/.style={fill=black,circle,inner sep=0pt,minimum width=4pt},
            myedge/.style={text opacity=0.9,opacity=0.6}}
        \node[mystate,label={above:$i$}] (i)               {};
        \node[mystate,label={above:$j$}] (j)  [right of=i] {};
        \path (i) edge [myedge,bend left,above] node {$\scriptstyle \mathtt{p}_{ij}^*$} (j);
    \end{tikzpicture}
    \right] \times \\
    & \times \left[1+\sum_{\mathtt{p}_{jj}^i}
    \begin{tikzpicture}[->,>=Stealth,auto,shorten >=0.5pt,node distance=1.5cm,semithick,baseline=(i.base)]
        \tikzset{
            mystate/.style={fill=black,circle,inner sep=0pt,minimum width=4pt},
            myloopleft/.style={out=-140,in=140,looseness=20,text opacity=0.9,opacity=0.6},
            myloopright/.style={out=40,in=-40,looseness=20,text opacity=0.9,opacity=0.6}}
        \node[mystate,label={above:$j$}] (j) {};
        \path (j) edge [myloopright]  node {$\scriptstyle \mathtt{p}_{jj}^i$} (j);
    \end{tikzpicture}
    \right]^{1/2}
\end{split}
\end{equation}
\normalsize
This equation suggests an interpretation of marginal correlations as the capacity to transmit information along the edges of the partial correlation graph. The outer factors represent communications between either $i$ or $j$ and the rest of the network, and the middle factor describes information sent by $i$ and received by $j$ (or equivalently, sent by $j$ and received by $i$), sometimes implicating additional nodes. 

The factorisation of Eq.~\ref{eq:rho_diagram_3} is easy to interpret, but not so easy to compute. Therefore, we now develop another expression, involving paths where $i$ and $j$ only appear as endpoints. To derive it, we use $\mathtt{p}_{aa}^{b*}$ to denote $aa^*$-paths that do not visit $b$. Expanding the sum in the first factor of Eq.~\ref{eq:rho_diagram_3} in  the number of times $i$ is visited
\begin{align}
    1 + \sum_{\mathtt{p}_{ii}^{j}}
    \begin{tikzpicture}[->,>=Stealth,auto,shorten >=0.5pt,node distance=1.5cm,semithick,baseline=(i.base)]
        \tikzset{
            mystate/.style={fill=black,circle,inner sep=0pt,minimum width=4pt},
            myloopleft/.style={out=-140,in=140,looseness=20,text opacity=0.9,opacity=0.6},
            myloopright/.style={out=40,in=-40,looseness=20,text opacity=0.9,opacity=0.6}}
        \node[mystate,label={above:$i$}] (i) {};
        \path (i) edge [myloopright]  node {$\scriptstyle \mathtt{p}_{ii}^j$} (i);
    \end{tikzpicture}
    &= 1  + \sum_{\mathtt{p}_{ii}^{j*}}
    \begin{tikzpicture}[->,>=Stealth,auto,shorten >=0.5pt,node distance=1.5cm,semithick,baseline=(i.base)]
        \tikzset{
            mystate/.style={fill=black,circle,inner sep=0pt,minimum width=4pt},
            myloopleft/.style={out=-140,in=140,looseness=20,text opacity=0.9,opacity=0.6},
            myloopright/.style={out=40,in=-40,looseness=20,text opacity=0.9,opacity=0.6}}
        \node[mystate,label={above:$i$}] (i) {};
        \path (i) edge [myloopright]  node {$\scriptstyle \mathtt{p}_{ii}^{j*}$} (i);
    \end{tikzpicture}
    + \nonumber \\
    & \quad + \sum_{\mathtt{p}_{ii}^{j*}} \sum_{\mathtt{p'}_{ii}^{j*}}
    \begin{tikzpicture}[->,>=Stealth,auto,shorten >=0.5pt,node distance=1.5cm,semithick,baseline=(i.base)]
        \tikzset{
            mystate/.style={fill=black,circle,inner sep=0pt,minimum width=4pt},
            myloopleft/.style={out=-140,in=140,looseness=20,text opacity=0.9,opacity=0.6},
            myloopright/.style={out=40,in=-40,looseness=20,text opacity=0.9,opacity=0.6}}
        \node[mystate,label={above:$i$}] (i) {};
        \path (i) edge [myloopright]  node {$\scriptstyle \mathtt{p}_{ii}^{j*}$} (i);
        \path (i) edge [myloopleft]  node {$\scriptstyle \mathtt{p'}_{ii}^{j*}$} (i);
    \end{tikzpicture}
    + \dots \nonumber \\
        &= \sum_{\ell = 0}^{\infty} \left[ \sum_{\mathtt{p}_{ii}^{j*}}
    \begin{tikzpicture}[->,>=Stealth,auto,shorten >=0.5pt,node distance=1.5cm,semithick,baseline=(i.base)]
        \tikzset{
            mystate/.style={fill=black,circle,inner sep=0pt,minimum width=4pt},
            myloopleft/.style={out=-140,in=140,looseness=20,text opacity=0.9,opacity=0.6},
            myloopright/.style={out=40,in=-40,looseness=20,text opacity=0.9,opacity=0.6}}
        \node[mystate,label={above:$i$}] (i) {};
        \path (i) edge [myloopright]  node {$\scriptstyle \mathtt{p}_{ii}^{j*}$} (i);
    \end{tikzpicture}
    \right]^\ell \nonumber \\ 
    & = \frac{1}{1 - \displaystyle{\sum_{\substack{\mathtt{p}_{ii}^{j*}}}}
    \begin{tikzpicture}[->,>=Stealth,auto,shorten >=0.5pt,node distance=1.5cm,semithick,baseline=(i.base)]
        \tikzset{
            mystate/.style={fill=black,circle,inner sep=0pt,minimum width=4pt},
            myloopleft/.style={out=-140,in=140,looseness=20,text opacity=0.9,opacity=0.6},
            myloopright/.style={out=40,in=-40,looseness=20,text opacity=0.9,opacity=0.6}}
        \node[mystate,label={above:$i$}] (i) {};
        \path (i) edge [myloopright]  node {$\scriptstyle \mathtt{p}_{ii}^{j*}$} (i);
    \end{tikzpicture}
    }, \label{eq:loop_generator}
\end{align}
where the convergence of the geometric series is guaranteed (see Appendix~\ref{appendixA}) by the fact that
\begin{equation} \label{eq:menorqueuno}
    0 \le \sum_{\mathtt{p}_{ii}^{j*}}
    \begin{tikzpicture}[->,>=Stealth,auto,shorten >=0.5pt,node distance=1.5cm,semithick,baseline=(i.base)]
        \tikzset{
            mystate/.style={fill=black,circle,inner sep=0pt,minimum width=4pt},
            myloopleft/.style={out=-140,in=140,looseness=20,text opacity=0.9,opacity=0.6},
            myloopright/.style={out=40,in=-40,looseness=20,text opacity=0.9,opacity=0.6}}
        \node[mystate,label={above:$i$}] (i) {};
        \path (i) edge [myloopright]  node {$\scriptstyle \mathtt{p}_{ii}^{j*}$} (i);
    \end{tikzpicture}
    < 1.
\end{equation}
Since Eq.~\ref{eq:loop_generator} also applies when interchanging $i$ and $j$, Eq.~\ref{eq:rho_diagram_3} can be written as
\small
\begin{equation} \label{eq:rho_diagram_4}
    \rho_{ij} = \frac{\displaystyle{\sum_{\substack{\mathtt{p}_{ij}^{*}}}}
    \begin{tikzpicture}[->,>=Stealth,auto,shorten >=0.5pt,node distance=1.5cm,semithick,baseline=(i.base)]
        \tikzset{
            mystate/.style={fill=black,circle,inner sep=0pt,minimum width=4pt},
            myedge/.style={text opacity=0.9,opacity=0.6}}
        \node[mystate,label={above:$i$}] (i)               {};
        \node[mystate,label={above:$j$}] (j)  [right of=i] {};
        \path (i) edge [myedge,bend left,above] node {$\scriptstyle \mathtt{p}_{ij}^{*}$} (j);
    \end{tikzpicture}
    }{ \sqrt{
    \left[
    1 - \displaystyle{\sum_{\substack{{\mathtt{p}_{ii}^{j*}}}}}
    \begin{tikzpicture}[->,>=Stealth,auto,shorten >=0.5pt,node distance=1.5cm,semithick,baseline=(i.base)]
        \tikzset{
            mystate/.style={fill=black,circle,inner sep=0pt,minimum width=4pt},
            myloopleft/.style={out=-140,in=140,looseness=20,text opacity=0.9,opacity=0.6},
            myloopright/.style={out=40,in=-40,looseness=20,text opacity=0.9,opacity=0.6}}
        \node[mystate,label={above:$i$}] (i) {};
        \path (i) edge [myloopright]  node {$\scriptstyle \mathtt{p}_{ii}^{j*}$} (i);
    \end{tikzpicture}
    \right] \left[
    1 - \displaystyle{\sum_{\substack{{\mathtt{p}_{jj}^{i*}}}}}
    \begin{tikzpicture}[->,>=Stealth,auto,shorten >=0.5pt,node distance=1.5cm,semithick,baseline=(i.base)]
        \tikzset{
            mystate/.style={fill=black,circle,inner sep=0pt,minimum width=4pt},
            myloopleft/.style={out=-140,in=140,looseness=20,text opacity=0.9,opacity=0.6},
            myloopright/.style={out=40,in=-40,looseness=20,text opacity=0.9,opacity=0.6}}
        \node[mystate,label={above:$j$}] (j) {};
        \path (j) edge [myloopright]  node {$\scriptstyle \mathtt{p}_{jj}^{i*}$} (j);
    \end{tikzpicture}
    \right]
    }}\end{equation}
\normalsize
This expression only includes paths in which $i$ and $j$ appear exclusively at the endpoints.  

In principle, Eq.~\ref{eq:rho_diagram_4} is only guaranteed to hold when $\nu(R_+) < 1$ because this was needed to show that some paths cancel out. Even though this condition is stronger than $\nu(R) < 1$, it can be shown (Appendix~\ref{appendixB}) that Eq.~\ref{eq:rho_diagram_4} also holds when $\nu(R) < 1$, as long as the series involved are calculated ordering the paths by length. 

The path expansions described above cannot be used if $\nu(R) \geq 1$. However, it is possible to obtain a general path expansion, valid even when $\nu(R) \geq 1$, that depends on a free parameter $q$. At variance with the case $\nu(R) < 1$, the general expansion requires the introduction of self-loops, which makes explicit summation more difficult. Since the examples considered below satisfy $\nu(R) < 1$, the derivation of the general expansion is left to Appendix~\ref{app:gen_expansion}. 

\section{Examples}
\label{sect:ejemplos}

In this section we discuss four examples. The simplicity of the first two enables marginal and partial correlations to be explicitly related by the traditional method of matrix inversion and rescaling. We still discuss them here to gain intuition about the series expansion in terms of paths, and to envision paths as routes to transmit information. The third and fourth examples, instead, show that the expansion provides a computational tool that facilitates calculations of correlations when matrix inversion does not yield a closed expression.

\subsection{A pair} 
\label{subsect:example_pair}

Here we revisit the example of Sect.~\ref{subsubsect:example_pair_A} of a pair of variables $X_1$ and $X_2$ with partial correlation $r_{12}$. We recompute the marginal correlation $\rho_{12}$ using Eq.~\ref{eq:rho_diagram_4} instead of Eq.~\ref{eq:rho_diagram}. Since there are no closed paths from node $1$ that do not visit node $2$ and vice versa, the sums in the denominator of Eq.~\ref{eq:rho_diagram_4} vanish. The only path from $1$ to $2$ with no intermediate stop at either $1$ or $2$ is the direct path $1 \to 2$. Hence, the sum in the numerator is the single weight $r_{12}$, and $\rho_{12} = r_{12}$, as before. Yet, the new calculation is significantly simpler, since it involves a single path.

\subsection{A three-variable chain} 
\label{ex:3_chain}

In this example, the variables $X_1, X_2, X_3$ form a chain, in the sense that the partial correlation graph is
\begin{equation*}
    \begin{tikzpicture}[>=Stealth,auto,node distance=2cm,thick,baseline=(1.base)]
            \tikzset{
                mystate/.style={fill=black,circle,inner sep=0pt,minimum width=4pt},
                myedge/.style={opacity=0.2,text opacity=0.5}}

            \node[mystate,label={above:$1$}] (1) at (0,0) {};
            \node[mystate,label={above:$2$}] (2) at (2,0) {};
            \node[mystate,label={above:$3$}] (3) at (4,0) {};
        
            \path   (1) edge [myedge,sloped,anchor=center,below] node {$r_{12}$} (2)
                    (2) edge [myedge,sloped,anchor=center,below] node {$r_{23}$} (3);
    \end{tikzpicture}
\end{equation*}
The partial correlation matrix $R$ has $r_{13} = r_{31} = 0$. For $\mathbb{1}$ $- R$ to be positive definite, the inequality $r_{12}^2 + r_{23}^2 < 1$ must be satisfied, also guaranteeing that $\nu(R) < 1$. 

In this simple case, marginal correlations can be calculated by matrix inversion and rescaling, yielding
\begin{align} \label{eq:chain3}
    \rho_{12} = &\frac{r_{12}}{\sqrt{1 - r_{23}^2}}, \qquad 
    \rho_{23} = \frac{r_{23}}{\sqrt{1 - r_{12}^2}}, \nonumber \\
    \rho_{13} = & \frac{r_{12}r_{23}}{\sqrt{(1 - r_{12}^2)(1 - r_{23}^2)}}.
\end{align}
Interestingly, $\rho_{13} = \rho_{12} \ \rho_{23}$, since all paths connecting nodes $1$ and $3$ pass through $2$. The result generalises to networks of arbitrary structure, provided that the node here labelled as $2$ be a separating node for those labelled $1$ and $3$, i.e., all paths connecting $1$ and $3$ pass through $2$ (Sect.~\ref{sec:marginalisation}).

The matrix element $\rho_{12}$ contains a factor $(1 - r_{23}^2)^{-\sfrac{1}{2}}$ that is greater than unity. Therefore, the presence of the third node in the chain enhances the marginal correlation between the first two. In order to understand this amplification factor, we derive the value of $\rho_{12}$ as resulting from the series expansion of Eq.~\ref{eq:rho_diagram_4}. Just as in the previous example, there is a single path from node $1$ to node $2$: the direct path $1 \to 2$ with weight $r_{12}$. There are no closed paths from $1$ that do not visit $2$, so the first sum in the denominator of Eq.~\ref{eq:rho_diagram_4} is zero. However, there is one closed path starting at $2$ without instances of $1$ or $2$ in the middle: the path $2 \to 3 \to 2$ with weight $r_{23}^2$. Plugging the three sums over paths in Eq.~\ref{eq:rho_diagram_4} yields the correct $\rho_{12}$ value shown in Eq.~\ref{eq:chain3}. The factor $(1 - r_{23}^2)^{-1}$ is the sum over all the closed paths starting at $2$ that do not visit $1$, but are allowed to visit $2$ in between. The information transmitted between nodes $1$ and $2$ reverberates in the connection $2 \leftrightarrow 3$ with a geometric damping factor, effectively amplifying the marginal interaction between $1$ and $2$.

We now derive the value of $\rho_{13}$ from the series expansion of Eq.~\ref{eq:rho_diagram_4}. Again, there is a single path that contributes to the numerator: the path $1 \to 2 \to 3$ with weight $r_{12} \ r_{23}$. The sums in the denominator also have a single contribution each, from the paths $1 \to 2 \to 1$ and $3 \to 2 \to 3$ with weights $r_{12}^2$ and $r_{23}^2$, respectively. Putting all these terms together, $\rho_{13} =  r_{12} r_{23} [(1 - r_{12}^2)(1 - r_{23}^2)]^{-\sfrac{1}{2}}$. This correlation involves two steps of information transmission, each of which reverberates by looping with the remaining node, thereby enhancing the marginal interaction.

\subsection{Two non-interacting variables connected by many confounding factors} \label{ex:one_many_one}

In this example the partial correlations of the variables $X_1, \dots, X_d$ are defined by the graphical model

\begin{equation*}
    \begin{tikzpicture}[>=Stealth,auto,node distance=2cm,thick,baseline=(1.base)]
        \tikzset{
            mystate/.style={fill=black,circle,inner sep=0pt,minimum width=4pt},
            myedge/.style={opacity=0.2,text opacity=0.5}}

        \node[mystate,label={left:$1$}] (1) at (0,0) {};
        \node[mystate,label={above:$2$}] (2) at (2.5,2) {};
        \node[mystate,label={above:$3$}] (3) at (2.5,1) {};
        \node[label={[label distance=-0.42cm]90:$\vdots$}] (dots) at (2.5,0) {};
        \node[mystate,label={below:$d-2$}] (5) at (2.5,-1) {};
        \node[mystate,label={below:$d-1$}] (6) at (2.5,-2) {};
        \node[mystate,label={right:$d$}] (7) at (5,0) {};
        
        \path   (1) edge [myedge,bend left,anchor=center,above] node {$r$} (2)
                (2) edge [myedge,bend left,anchor=center,above] node {$r$} (7);
        \path   (1) edge [myedge,bend left,looseness=0.6,in=160,anchor=center,above] node {$r$} (3)
                (3) edge [myedge,bend left,looseness=0.6,out=20,anchor=center,above] node {$r$} (7);
        \path   (1) edge [myedge,opacity=0.2] node {} (dots)
                (dots) edge [myedge,opacity=0.2] node {} (7);
        \path   (1) edge [myedge,bend right,looseness=0.6,in=200,anchor=center,below] node {$r$} (5)
                (5) edge [myedge,bend right,looseness=0.6,out=-20,anchor=center,below] node {$r$} (7);
        \path   (1) edge [myedge,bend right,anchor=center,below] node {$r$} (6)
                (6) edge [myedge,bend right,anchor=center,below] node {$r$} (7);
    \end{tikzpicture},
\end{equation*}
where, for simplicity, all the non-vanishing partial correlations are equal to $r$. The requirement that the precision matrix be positive definite implies that $(d-2)\,r^2 < \sfrac{1}{2}$. The marginal correlation between $X_1$ and $X_d$ is
\begin{equation} \label{eq:rho12paralelo}
\rho_{1d} = \frac{(d-2) \, r^2}{1 - (d-2) \, r^2}.
\end{equation}
Proving this result by inverting the precision matrix requires cumbersome algebraic manipulations (Appendix~\ref{app:net_corr_boring}). A simpler option, however, is to obtain this result from the series expansion of Eq.~\ref{eq:rho_diagram_4}. The relevant terms are
\begin{itemize}
    \item[-] $\displaystyle{\sum_{\substack{\mathtt{p}_{1d}^*}}}
        \begin{tikzpicture}[->,>=Stealth,auto,shorten >=0.5pt,node distance=1.5cm,semithick,baseline=(i.base)]
            \tikzset{
                mystate/.style={fill=black,circle,inner sep=0pt,minimum width=4pt},
                myedge/.style={text opacity=0.9,opacity=0.6}}
            \node[mystate,label={above:$1$}] (i)               {};
            \node[mystate,label={above:$d$}] (j)  [right of=i] {};
            \path (i) edge [myedge,bend left,above] node {$\scriptstyle \mathtt{p}_{1d}^*$} (j);
        \end{tikzpicture}
        = \displaystyle{\sum_{\substack{k=2}}^{d-1}} \, r_{1k} r_{kd} = (d-2) \, r^2$,
    \item[-] $\displaystyle{\sum_{\substack{\mathtt{p}_{11}^{d*}}}}
        \begin{tikzpicture}[->,>=Stealth,auto,shorten >=0.5pt,node distance=1.5cm,semithick,baseline=(i.base)]
            \tikzset{
                mystate/.style={fill=black,circle,inner sep=0pt,minimum width=4pt},
                myloopleft/.style={out=-140,in=140,looseness=20,text opacity=0.9,opacity=0.6},
                myloopright/.style={out=40,in=-40,looseness=20,text opacity=0.9,opacity=0.6}}
            \node[mystate,label={above:$1$}] (i) {};
            \path (i) edge [myloopright]  node {$\scriptstyle \mathtt{p}_{11}^{d*}$} (i);
        \end{tikzpicture}
        = \displaystyle{\sum_{\substack{k=2}}^{d-1}} \, r_{1k}^2 = (d-2) \, r^2$,
    \item[-] $\displaystyle{\sum_{\substack{\mathtt{p}_{dd}^{1*}}}}
        \begin{tikzpicture}[->,>=Stealth,auto,shorten >=0.5pt,node distance=1.5cm,semithick,baseline=(i.base)]
            \tikzset{
                mystate/.style={fill=black,circle,inner sep=0pt,minimum width=4pt},
                myloopleft/.style={out=-140,in=140,looseness=20,text opacity=0.9,opacity=0.6},
                myloopright/.style={out=40,in=-40,looseness=20,text opacity=0.9,opacity=0.6}}
            \node[mystate,label={above:$d$}] (i) {};
            \path (i) edge [myloopright]  node {$\scriptstyle \mathtt{p}_{dd}^{1*}$} (i);
        \end{tikzpicture}
        = \sum_{k=2}^{d-1} r_{kd}^2 = (d-2) \, r^2$.
\end{itemize}
The first term describes the $d-2$ paths that go from node $1$ to node $d$, each with a single intermediate node. The second and third terms describe the $d-2$ closed paths of the form $1 \to k \to 1$ or $d \to k \to d$, with $k = 2, 3, \dots, d-1$. 

Generalising the calculation of $\rho_{1d}$ using the series expansion to cases in which the partial correlations vary from link to link is straightforward. Yet, the inversion of the resulting precision matrix is far from straightforward  --- we are not aware of any closed expression that yields $\Omega^{-1}$ for arbitrary $d$. Therefore, the expansion proposed here should not only be regarded as a theoretical way to conceive the transmission of information along paths, but also as a practical tool for computational purposes.

\subsection{Chain of arbitrary length} \label{ex:markov_chain}

In this example the direct interactions of the variables $X_1, X_2, \dots, X_d$ define a linear chain
\begin{equation*}
        \begin{tikzpicture}[>=Stealth,auto,node distance=2cm,thick,baseline=(1.base)]
            \tikzset{
                mystate/.style={fill=black,circle,inner sep=0pt,minimum width=4pt},
                myedge/.style={opacity=0.2,text opacity=0.5}}

            \node[mystate,label={above:$1$}] (1) at (0,0) {};
            \node[mystate,label={above:$2$}] (2) at (1.5,0) {};
            \node (dotsL) at (3,0) {};
            \node[label={[label distance=-0.33cm]90:$\dots$}] (dots) at (3.3,0) {};
            \node (dotsR) at (3.55,0) {};
            \node[mystate,label={above:$d-1$}] (n-1) at (5.05,0) {};
            \node[mystate,label={above:$d$}] (n) at (6.55,0) {};
        
            \path   (1) edge [myedge,sloped,anchor=center,below] node {$r$} (2)
                    (2) edge [myedge,sloped,anchor=center,below] node {$r$} (dotsL)
                    (dotsR) edge [myedge,sloped,anchor=center,below] node {$r$} (n-1)
                    (n-1) edge [myedge,sloped,anchor=center,below] node {$r$} (n);
        \end{tikzpicture}
\end{equation*}
This graph underlies various systems of interest, such as first-order autoregressive processes and martingales, as discussed in Sect.~\ref{sec:separatingnode}. 

The partial correlation $r$ must satisfy the condition $|r| \le \sfrac{1}{2}$ to ensure that the precision matrix be positive definite irrespective of $d$, which also guarantees that $\nu(R) < 1$  \citep{artner2022shape}.  As far as we know, no simple expression can be obtained for $\rho_{ij}$ by explicit matrix inversion, yet a recurrent solution can be derived from our path expansion. We first derive a recurrence relation for the marginal correlation $\rho_{1d}$ between the endpoints of the chain, and then use this result to compute a relation for the marginal correlation between two arbitrary points.

\subsubsection{Marginal correlation between the endpoints of the chain}
\label{sec:chain}

To simplify the notation, in this example we call the sums in Eq.~\ref{eq:rho_diagram_4}
\begin{equation*}
    c_d \equiv \sum_{\mathtt{p}_{1d}^{*}}
    \begin{tikzpicture}[->,>=Stealth,auto,shorten >=0.5pt,node distance=1.5cm,semithick,baseline=(i.base)]
        \tikzset{
            mystate/.style={fill=black,circle,inner sep=0pt,minimum width=4pt},
            myedge/.style={text opacity=0.9,opacity=0.6}}
        \node[mystate,label={above:$1$}] (i)               {};
        \node[mystate,label={above:$d$}] (j)  [right of=i] {};
        \path (i) edge [myedge,bend left,above] node {$\scriptstyle \mathtt{p}_{1d}^{*}$} (j);
    \end{tikzpicture}
\end{equation*}
and 
\begin{equation*}
    \ell_d \equiv
    \sum_{\mathtt{p}_{11}^{d*}}
    \begin{tikzpicture}[->,>=Stealth,auto,shorten >=0.5pt,node distance=1.5cm,semithick,baseline=(i.base)]
        \tikzset{
            mystate/.style={fill=black,circle,inner sep=0pt,minimum width=4pt},
            myloopleft/.style={out=-140,in=140,looseness=20,text opacity=0.9,opacity=0.6},
            myloopright/.style={out=40,in=-40,looseness=20,text opacity=0.9,opacity=0.6}}
        \node[mystate,label={above:$1$}] (i) {};
        \path (i) edge [myloopright]  node {$\scriptstyle \mathtt{p}_{11}^{d*}$} (i);
    \end{tikzpicture}
    = 
    \sum_{\mathtt{p}_{dd}^{1*}}
    \begin{tikzpicture}[->,>=Stealth,auto,shorten >=0.5pt,node distance=1.5cm,semithick,baseline=(i.base)]
        \tikzset{
            mystate/.style={fill=black,circle,inner sep=0pt,minimum width=4pt},
            myloopleft/.style={out=-140,in=140,looseness=20,text opacity=0.9,opacity=0.6},
            myloopright/.style={out=40,in=-40,looseness=20,text opacity=0.9,opacity=0.6}}
        \node[mystate,label={above:$d$}] (i) {};
        \path (i) edge [myloopright]  node {$\scriptstyle \mathtt{p}_{dd}^{1*}$} (i);
    \end{tikzpicture}.
\end{equation*}
The two sums over closed paths in Eq.~\ref{eq:rho_diagram_4} coincide, since the chain is invariant to the inversion of the order of the nodes. The marginal correlation is
\begin{equation*}
    \rho_d \equiv \rho_{1d} = \frac{c_d}{1 - \ell_d}.
\end{equation*}

The sums $c_d$ and $\ell_d$ are hard to compute in a closed form, but can be written in terms of the sums $c_{d-1}$ and $\ell_{d-1}$ associated to a chain with one fewer node. Indeed, every path $\mathtt{p}_{1d}^*$ can be separated into three sub-paths: A first part up to the first arrival to the node $d-1$, a second (optional) part that includes back-and-forth journeys from node $d-1$ to the left side of chain until the last return to node $d-1$, and a final part from $d-1$ to $d$. Therefore,
\begin{equation} \label{eq:chain_paths_rec}
    c_d = \frac{r \, c_{d-1}}{1 - \ell_{d-1}}.
\end{equation}
At the same time, every closed path $\mathtt{p}_{11}^{d*}$ either visits node $d-1$ or does not. If it does not, then it is included in the sum $\ell_{d-1}$ of a shorter chain. If it does, then it can be separated into three sub-paths: A first part up to the first arrival to node $d-1$, an (optional) second part that includes back-and-forth oscillations from $d-1$ to the left side of the chain until node $d-1$ is visited for the last time, and finally, the return from node $d-1$ to node $1$. Therefore,
\begin{equation} \label{eq:chain_loops_rec}
    \ell_{d} = \ell_{d-1} + \frac{c_{d-1}^2}{1 - \ell_{d-1}}.
\end{equation}
Eqs.~\ref{eq:chain_paths_rec} and ~\ref{eq:chain_loops_rec} define a recurrence relation for the sums $c_d$ and $\ell_d$. Since a chain with $d=2$ nodes has a single $\mathtt{p}_{1d}^*$ path (the direct path) and no $\mathtt{p}_{11}^{*d}$ paths, the initial condition of the recurrence is $c_2 = r$ and $\ell_2 = 0$.
Equations~\ref{eq:chain_paths_rec} and ~\ref{eq:chain_loops_rec} yield the following recurrence relation for the marginal correlation 
\begin{equation*} \label{eq:chain_rho_rec}
    \rho_{d} = \frac{\rho_{d-1}^2}{\rho_{d-2} \left(1 - \rho_{d-1}^2 \right)}
\end{equation*}
in terms of the marginal correlation between the endpoints of shorter chains, where the initial conditions are $\rho_1 = 1$ and $\rho_2 = r$.

\subsubsection{Correlation between two arbitrary nodes} \label{ex:markov_chain_b}

Without loss of generality, we assume that $i < j$. Since the paths $\mathtt{p}_{ij}^*$ only contain nodes $i$ and $j$ as endpoints, the sum in the numerator of Eq.~\ref{eq:rho_diagram_4} only contains the sector of the chain that stretches from $i$ to $j$. Since this sector is itself a chain of $j-i+1$ nodes,
\begin{equation*}
    \sum_{\mathtt{p}_{ij}^*}
    \begin{tikzpicture}[->,>=Stealth,auto,shorten >=0.5pt,node distance=1.5cm,semithick,baseline=(i.base)]
        \tikzset{
            mystate/.style={fill=black,circle,inner sep=0pt,minimum width=4pt},
            myedge/.style={text opacity=0.9,opacity=0.6}}
        \node[mystate,label={above:$i$}] (i)               {};
        \node[mystate,label={above:$j$}] (j)  [right of=i] {};
        \path (i) edge [myedge,bend left,above] node {$\scriptstyle \mathtt{p}_{ij}^{*}$} (j);
    \end{tikzpicture}
    = c_{j-i+1},
\end{equation*}
where $c_k$ follows the recurrence relation given by Eq.~\ref{eq:chain_paths_rec}. At the same time, any closed path $\mathtt{p}_{ii}^{j*}$ must either remain within the $1$-to-$i$ sector, or within the $i$-to-$j$. Therefore,
\begin{equation*}
    \sum_{\mathtt{p}_{ii}^{j*}}
    \begin{tikzpicture}[->,>=Stealth,auto,shorten >=0.5pt,node distance=1.5cm,semithick,baseline=(i.base)]
        \tikzset{
            mystate/.style={fill=black,circle,inner sep=0pt,minimum width=4pt},
            myloopleft/.style={out=-140,in=140,looseness=20,text opacity=0.9,opacity=0.6},
            myloopright/.style={out=40,in=-40,looseness=20,text opacity=0.9,opacity=0.6}}
        \node[mystate,label={above:$i$}] (i) {};
        \path (i) edge [myloopright]  node {$\scriptstyle \mathtt{p}_{ii}^{j*}$} (i);
    \end{tikzpicture}
    = \ell_{i+1} + \ell_{j-i+1},
\end{equation*}
where $\ell_k$ follows the recurrence relation of Eq.~\ref{eq:chain_loops_rec}, and the subscript of $\ell_{i+1}$ stems from the fact that closed paths in the $1$-to-$i$ sector can reach node $1$. Applying the same argument to the $\mathtt{p}_{jj}^{i*}$ paths yields
\begin{equation*}
    \sum_{\mathtt{p}_{jj}^{i*}}
    \begin{tikzpicture}[->,>=Stealth,auto,shorten >=0.5pt,node distance=1.5cm,semithick,baseline=(i.base)]
        \tikzset{
            mystate/.style={fill=black,circle,inner sep=0pt,minimum width=4pt},
            myloopleft/.style={out=-140,in=140,looseness=20,text opacity=0.9,opacity=0.6},
            myloopright/.style={out=40,in=-40,looseness=20,text opacity=0.9,opacity=0.6}}
        \node[mystate,label={above:$j$}] (i) {};
        \path (i) edge [myloopright]  node {$\scriptstyle \mathtt{p}_{jj}^{i*}$} (i);
    \end{tikzpicture}
    = \ell_{j-i+1} + \ell_{d-j+2}.
\end{equation*}
Hence, the marginal correlation  between nodes $i$ and $j$ is 
\small
\begin{equation} \label{eq:chain_corr_ij}
    \rho_{ij} = \frac{c_{j-i+1}}{\sqrt{\left( 1 - \ell_{j-i+1} - \ell_{i+1} \right) \left( 1 - \ell_{j-i+1} - \ell_{d-j+2} \right)}}.
\end{equation}
\normalsize

\subsubsection{Correlation length} \label{ex:markov_chain_corr_length}

The asymptotic behaviour of correlations can be derived from Eq.~\ref{eq:chain_corr_ij}. The decay of marginal interactions is typically characterised by the \emph{correlation length} $\xi$ (see for example \citep[Chapter~3.5]{goldenfeld2018lectures}), defined as
\small
\begin{equation} \label{eq:chain_corr_length_def}
    \xi \equiv \lim_{|i-j| \to \infty} - \frac{|i-j|}{\ln |\rho_{ij}|}.
\end{equation}
\normalsize
The motivation behind this definition is to capture the exponential decay of correlations between variables that are far away in the graph, i.e. $|\rho_{ij}| \sim \exp(-|i-j|/\xi)$ when $|i-j| \gg 1$. A divergent correlation length implies that $\rho_{ij}$ decays slower than exponentially, and that fluctuations are correlated across the entire system -- a typical signature of a phase transition.

The limit in Eq.~\ref{eq:chain_corr_length_def} is calculated for an infinite chain in Appendix \ref{app:corr_length}, yielding
\small
\begin{equation} \label{eq:chain_corr_length}
    \xi = \left[\ln \left( \frac{1 + \sqrt{1 - 4\,r^2}}{2 \, |r|} \right)\right]^{-1}.
\end{equation}
\normalsize
The correlation length diverges for $|r| \to \sfrac{1}{2}$. This divergence resembles the one exhibited in one dimensional statistical systems with first neighbours interactions (i.e. a Markov chain), in the zero temperature limit \citep[Chapter~4]{zinn2007phase}. Since the divergence occurs exactly at the boundary of the parameter space $|r| \le \sfrac{1}{2}$, linear chains only exhibit a single phase, i.e., no transition to a different phase is possible.  

The computation of $\xi$ for a chain with periodic boundary conditions (i.e. a ring) is identical to the derivation of Eq.~\ref{eq:chain_corr_length}. Therefore, the present analysis of the large-distance behaviour of correlations is also valid for graphs with ring-like structure.

\section{The role of separating nodes}
\label{sec:separatingnode}

The topology of the partial correlation network determines the structure of the marginal correlations. In particular, imposing zero weights on certain edges constrains the possible values of the marginal correlations in a non-trivial manner. A network topology that induces particularly tractable correlations is one containing a separating node, that is, a node whose removal increases the number of connected components in the graph. Consider a connected partial correlation graph $G$ composed of a set of nodes $\mathcal{U}$. The node $k \in \mathcal{U}$ is a separating node if, when pruning all the edges that connect $k$ to other nodes, $G$ is no longer connected. The original set $\mathcal{U}$, hence, can be partitioned as $\mathcal{U} = \mathcal{I} \cup \{ k \} \cup \mathcal{J}$, so that the severance of node $k$ leaves the subsets $\mathcal{I}$ and $\mathcal{J}$ in different components of $G$. The definition of a separating node, therefore, pertains to the structure of the partial correlation network. In this section, we state the consequences on the marginal correlations. 

The main result is a bidirectional implication: The node $k \in \mathcal{U}$ is a separating node for $G$ if and only if the set of nodes $\mathcal{U}$ can be partitioned into three disjoint subsets $\mathcal{U} = \mathcal{I} \cup \{k\} \cup \mathcal{J}$ such that $\rho_{ij} = \rho_{ik} \, \rho_{kj}$ for every $i \in \mathcal{I}$ and $j \in \mathcal{J}$. The forward implication states that the factorisation of marginal correlations is a necessary condition for the existence of a separating node. This statement is proved in Sect.~\ref{sec:sep_node_necessity}. The converse statement asserts the sufficiency of the factorisation of marginal correlations for the existence of a separating node, and is demonstrated in Appendix~\ref{app:sep_node_sufficiency}. 

The one-to-one relation between the factorisation of the marginal correlations and the existence of a separating node is particularly useful for martingales, that is, random processes that satisfy ${\rm E}(X_t \, | X_{t-1}, X_{t-1}, \dots) = \alpha \, X_{t-1}$ for any $t$ and some constant $\alpha$. This condition suffices to demonstrate that the marginal correlation $\rho_{t_1 t_3}$ factorises as $\rho_{t_1 t_3} = \rho_{t_1 t_2}\,\rho_{t_2 t_3}$ for any indices such that $t_1 < t_2 < t_3$ (see Appendix~\ref{app:martingale_fact_prop}). Therefore, all intermediate nodes are separating nodes, which means that the partial correlation graph associated to a martingale is always a linear chain. Moreover, any system with a linear chain graph satisfies the factorisation property of marginal correlations.

\section{Changing the structure of the network}
\label{sec:alterations}

The example of Sect.~\ref{ex:3_chain} can be conceived of as appending an extra node to a pair of connected variables, thereby constructing a trio with the shape of a 3-node chain. As a result, the marginal correlation between the original pair is amplified. In this section, we study the changes in marginal and partial correlations under more general alterations of the network. Alterations may consist of additions (new nodes are added to the network), deletions (nodes are eliminated), or a combination of both (some nodes are added, other deleted). Alterations are important because, in most applications, the network under study is only an effective description of the system of interest. Choosing the network implies deciding which variables are represented. Moreover, it also implies deciding whether the non-represented nodes are cut out of the network, so that they can exert no influence on the remaining nodes, or whether they are still present and exerting their influence, but not included in the description. The study of alterations in the structure of the network relates these different alternative descriptions. If the set of variables of one description is included in that of another, we here call the first description {\it reduced}, and the second, {\it detailed}.

There is no need to discuss additions and deletions separately, since an addition of nodes can be seen as a reversed deletion, and \textit{vice versa}. This section relates the correlations (marginal and partial) of a detailed description with those of a reduced description. The derived relations hold irrespective of which network was the original, and which the altered. If the original contained fewer nodes, we speak of an addition, and if it contained more, of a deletion.

To describe deletions, we start with a graph $G$ defined on a set of nodes $\mathcal{U}$ and partial correlations $r_{ij}$, and eliminate a subset of nodes $\mathcal{S} \subset \mathcal{U}$, leaving only the nodes in the complementary set $\mathcal{T} = \mathcal{U} - \mathcal{S}$ with a new graph $G'$ with links $r'_{ij}$. There are different ways in which nodes in $\mathcal{S}$ can be eliminated. Here we focus on two alternatives. The first consists of pruning all the connections that tie the nodes in $\mathcal{S}$ to the rest of the network without altering the partial correlations between the remaining nodes, for which $r'_{ij} = r_{ij}$. This strategy is here called {\sl node severance} (Sect.~\ref{sec:conditioning}), and as a result, the number of paths connecting any two remaining nodes is reduced. Therefore, node severance modifies the marginal correlations ($\rho'_{ij} \ne \rho_{ij}$). The second alternative is to marginalise out a subset of nodes, removing them from the description  (Sect.~\ref{sec:marginalisation}).  While these nodes are no longer described, they continue to influence and be influenced by other nodes in the network. In this alternative, the marginal correlations of the remaining nodes remain unchanged, but the partial correlations need to be modified to encompass the effect of the nodes that were marginalised out.

\subsection{Node severance}
\label{sec:conditioning}

In this section, the pruning of the connections that link a subset $\mathcal{S}$ of nodes to the rest of the network is done in such a way that the partial correlations $r_{ij}$ between the remaining nodes do not change. This intervention represents a local action on the network, effectively disconnecting the severed nodes. For those distributions in which conditional and partial correlations coincide (Sect.~\ref{sec:notation}), this type of node severance is equivalent to conditioning on the values of nodes in $\mathcal{S}$.

When nodes are thus severed, the graph $G'$ is the sub-graph of $G$ that comprises the remaining nodes, i.e. nodes in $\mathcal{T}$, so that $r'_{ij} = r_{ij}$, for all $i, j \in \mathcal{T}$. The marginal correlations $\rho'_{ij}$ can still be computed using Eq.~\ref{eq:rho_diagram_4}, but now, the sums only include paths in $G'$, that is, paths that do not visit nodes in $\mathcal{S}$.

Reducing the number of paths may either increase or decrease the marginal correlation $\rho'_{ij}$ between two nodes. The denominator of Eq.~\ref{eq:rho_diagram_4} always increases when the closed paths that visit nodes in $\mathcal{S}$ are excluded from the sum (Appendix \ref{appendixA}). The numerator of Eq.~\ref{eq:rho_diagram_4}, however, may vary in either direction, since the excluded terms may contribute with the same or the opposite sign as the remaining terms. In the following subsections some special cases in which the numerator remains unchanged are discussed. In these cases, node severance always reduces the marginal correlations between the remaining nodes. 

The utility of this section is not restricted to the case of severing the nodes in $\mathcal{S}$ from a graph $G$ defined on the set $\mathcal{U}$. The results also hold when adding new nodes $\mathcal{S}$ to a pre-defined network of nodes $\mathcal{T}$ and graph $G'$, thereby producing a larger graph $G \supset G'$ defined on $\mathcal{U} = \mathcal{T} \cup \mathcal{S}$. Each new node has to be provided with the value of all its partial correlations to all other nodes. This kind of addition can be viewed as the inverse operation of node severance. 

\subsubsection{Severing the nodes of a sub-networks appended to a separating node}
\label{sec:appended}

\begin{figure*}[htbp]
        \centering
        \begin{tikzpicture}[>=Stealth,auto,node distance=1cm,thick]
            \tikzset{
                mystate/.style={fill=black,circle,inner sep=0pt,minimum width=4pt},
                myedge/.style={opacity=0.2,text opacity=0.5},
                myedge2/.style={text opacity=0.9,opacity=0.6}}

            \node[mystate,label={above:$i$}] (ia) at (0,0) {};
            \node[mystate] (2a) at ($(ia) + (0,1.2)$) {};
            \node[mystate,label={above:$j$}] (3a) at ($(ia) + (1,.6)$) {};
            \node[mystate] (4a) at ($(ia) + (-1.25,0)$) {};
            \node[mystate] (5a) at ($(4a) + (0,1.2)$) {};

            \node[mystate] (6a) at ($(ia) + (0,-1)$) {};
            \node[mystate] (7a) at ($(ia) + (-.6,-1)$) {};
            \node[mystate] (8a) at ($(ia) + (.6,-1)$) {};

            \path   (ia) edge [myedge] node {} (3a)
                    (3a) edge [myedge] node {} (2a)
                    (2a) edge [myedge] node {} (5a)
                    (5a) edge [myedge] node {} (4a)
                    (4a) edge [myedge] node {} (ia)    
                    (4a) edge [myedge] node {} (2a);
            \path   (ia) edge [myedge] node {} (7a)
                    (7a) edge [myedge] node {} (6a)
                    (6a) edge [myedge] node {} (8a)
                    (8a) edge [myedge] node {} (ia)
                    (ia) edge [myedge] node {} (6a);

            \draw[thick,rounded corners,dotted,opacity=0.5] ($(5a) + (-.4,.25)$) rectangle ($(ia) + (1.4,-.25)$);
            \node[label={$\mathcal{T}$}] (Ta) at ($(3a) + (.75,-0.3)$) {};

            \draw[thick,rounded corners,dotted,opacity=0.5] ($(ia) + (-.95,-0.5)$) rectangle ($(ia) + (.95,-1.75)$);
            \node[label={$\mathcal{S}$}] (Sa) at ($(ia) + (1.25,-1.6)$) {};
            \draw[myedge2,postaction={decorate},decoration={markings,mark=at position 0.61 with {\arrow{>}}}] plot [smooth,tension=0.5] coordinates {($(ia) + (-0.17,-0.08)$) ($(7a) + (-0.15,0.05)$) ($(7a) + (0,-0.18)$) ($(6a) + (0,-0.2)$) ($(8a) + (0,-0.18)$) ($(8a) + (0.15,0.05)$) ($(ia) + (0.15,-0.05)$)};
            \node[label={$\scriptstyle \mathtt{p}_{i\mathcal{S}i}^*$}] (pii) at ($(6a) + (0.1,-.9)$) {};

            \node[label={\large (a)}] (a) at ($(5a) + (-.65,.4)$) {};
            \node[label={\large (b)}] (b) at ($(a) + (5.2,0)$) {};

            \node[mystate,label={left:$i$}] (1b) at (3.8,-.55) {};
            \node[mystate] (2b) at ($(1b) + (1,.5)$) {};
            \node[mystate] (3b) at ($(1b) + (1,-.5)$) {};
            \node[mystate] (4b) at ($(1b) + (2,0)$) {};
            \node[mystate,label={right:$j$}] (5b) at ($(1b) + (3,0)$) {};
            
            \node[mystate,label={left:$k$}] (ab) at ($(1b) + (0,1.25)$) {};
            \node[mystate,label={above:$\ell$}] (bb) at ($(1b) + (1,1.25)$) {};
            \node[mystate] (cb) at ($(1b) + (2,1.25)$) {};
            \node[mystate] (db) at ($(1b) + (3,1.25)$) {};
            \node[mystate] (eb) at ($(1b) + (2,2)$) {};
        
            \path   (1b) edge [myedge] node {} (2b)
                    (2b) edge [myedge] node {} (4b);
            \path   (1b) edge [myedge] node {} (3b)
                    (3b) edge [myedge] node {} (4b);
            \path   (2b) edge [myedge] node {} (3b);
            \path   (4b) edge [myedge] node {} (5b);
            \path   (1b) edge [myedge] node {} (ab)
                    (ab) edge [myedge] node {} (bb)
                    (bb) edge [myedge] node {} (cb)
                    (cb) edge [myedge] node {} (db)
                    (db) edge [myedge] node {} (5b)
                    (5b) edge [myedge] node {} (cb);
            \path   (eb) edge [myedge] node {} (bb)
                    (eb) edge [myedge] node {} (cb)
                    (eb) edge [myedge] node {} (db);

            \draw[thick,rounded corners,dotted,opacity=0.5] ($(1b) + (0.5,-0.75)$) rectangle ($(1b) + (2.25,0.75)$);
            \node[label={$\mathcal{S}$}] (Sb) at ($(1b) + (1.375,-1.4)$) {};

            \draw[myedge2,postaction={decorate},decoration={markings,mark=at position 0.95 with {\arrow{>}}}] plot [smooth,tension=0.4] coordinates {($(1b) + (0.1,0.15)$) ($(2b) + (-0.1,0.1)$) ($(2b) + (0.15,0.05)$) ($(3b) + (0.15,0.3)$) ($(3b) + (0.22,0)$) ($(4b) + (0,-0.16)$) ($(5b) + (-0.1,-0.12)$)};
            \node[label={$\scriptstyle \mathtt{p}_{i\mathcal{S}j}^*$}] (piSj) at ($(5b) + (-.2,-.85)$) {};

            \node[label={\large $G$}] (Gb) at ($(1b) + (.4,1.8)$) {};

            \draw[->,thick] ($(db) + (.6,-0.625)$) -- ($(db) + (3.7,-0.625)$);
            \node[text width=3cm,align=center] (arrow_text) at ($(db) + (2,-0.6)$) {\footnotesize marginalisation over nodes in $\mathcal{S}$};

            \node[mystate,label={left:$i$}] (1c) at ($(5b) + (4.4,0)$) {};
            \node[mystate,label={right:$j$}] (5c) at ($(1c) + (3,0)$) {};
            
            \node[mystate,label={left:$k$}] (ac) at ($(1c) + (0,1.25)$) {};
            \node[mystate,label={above:$\ell$}] (bc) at ($(1c) + (1,1.25)$) {};
            \node[mystate] (cc) at ($(1c) + (2,1.25)$) {};
            \node[mystate] (dc) at ($(1c) + (3,1.25)$) {};
            \node[mystate] (ec) at ($(1c) + (2,2)$) {};

            \path   (1c) edge [red!90!black,opacity=0.5,thick,right,text opacity=0.8] node {} (ac)
                    (ac) edge [myedge] node {} (bc)
                    (bc) edge [myedge] node {} (cc)
                    (cc) edge [myedge] node {} (dc)
                    (dc) edge [red!90!black,opacity=0.5,thick] node {} (5c)
                    (5c) edge [red!90!black,opacity=0.5,thick] node {} (cc);
            \path   (ec) edge [myedge] node {} (bc)
                    (ec) edge [myedge] node {} (cc)
                    (ec) edge [myedge] node {} (dc);
            \path   (1c) edge [red!80!black,bend right,looseness=0.5,opacity=0.8,thick,dashed,sloped,anchor=center,below] node {$r_{ij}'$} (5c);

            \node[label={\large $G'$}] (G'c) at ($(1c) + (.4,1.8)$) {};
        \end{tikzpicture}
    \caption{Two examples of eliminating a subset $\mathcal{S}$ of nodes from a network, and the resulting effect on the correlations of the remaining nodes $\mathcal{T}$. (a): When $\mathcal{S}$ is appended to $\mathcal{T}$ through a single node $i \in \mathcal{T}$, severing the nodes in $\mathcal{S}$ decreases the marginal correlations $\rho_{ij}$ (Sect.~\ref{sec:appended}), and marginalising over $\mathcal{S}$ increases the partial correlations $r_{ij}$ (Sect.~\ref{sec:appended2}). The closed path $\mathtt{p}_{i\mathcal{S}i}^*$ starts at $i$ and only traverses nodes in $\mathcal{S}$. (b): When $\mathcal{S}$ is connected to $\mathcal{T}$ through two nodes $i, j \in \mathcal{T}$, marginalisation over $\mathcal{S}$ modifies the partial correlations $r_{ik}$ and $r_{jk}$ (red edges), introduces a new connection $r_{ij}$ (red dashed edge), and leaves the remaining links (grey) unchanged. The path $\mathtt{p}_{i\mathcal{S}j}^*$ connects $i$ and $j$ traversing entirely through $\mathcal{S}$, and as such, contributes to the weight $r_{ij}'$ of the new edge (Eq.~\ref{eq:nesting_transf}).}
    \label{fig:net_struct}
\end{figure*}
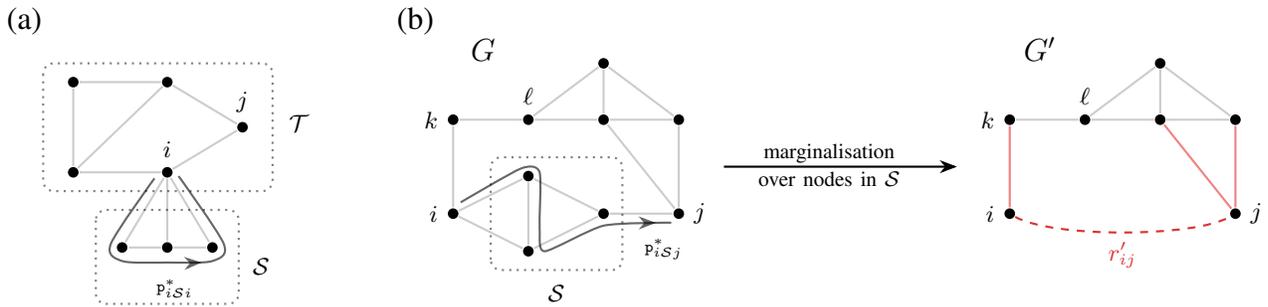

Consider a connected graph $G$ defined on a set $\mathcal{U}$ that contains a separating node $i$, i.e. a node that belongs to all the paths connecting any node in a subset $\mathcal{S}$ with any node of the complementary set $\mathcal{T}$ (Fig.~\ref{fig:net_struct}a). In this situation, the following two statements are true:
\begin{enumerate}
\item Severing the nodes of $\mathcal{S}$ always results in decreasing the absolute value of the marginal correlation $\rho_{ij}$ between the separating node $i$ and any other node in $\mathcal{T}$. That is, in the reduced graph $G'$ obtained after severing, $|\rho'_{ij}| \le |\rho_{ij}|$. 

\item  After severing, the contribution that node $i$ makes to the marginal correlation $\rho'_{jk}$ between any two nodes $j, k \ne i$ within $\mathcal{T}$ is smaller than before.
\end{enumerate}

To prove the first statement, we observe that when calculating $\rho_{ij}$ with Eq.~\ref{eq:rho_diagram_4}, no nodes in $\mathcal{S}$ contribute to the sum in the numerator, since the separating node $i$ can only be the first or the last of the path. Therefore, the numerator required for calculating $\rho_{ij}$ coincides with that for $\rho'_{ij}$. The denominator for $\rho'_{ij}$, however, is larger, since the sum over closed paths $\mathtt{p}_{ii}^{j*}$ no longer contains those passing through $\mathcal{S}$, which make a positive contribution to the sum (Eq.~\ref{eq:menorqueuno}). Figure~\ref{fig:net_struct}a shows an example of an excluded path $\mathtt{p}_{i\mathcal{S}i}^{*}$, i.e. an $ii^*$-path in the detailed graph $G$ that only visits nodes in $\mathcal{S}$. The remaining factor in the denominator, the one with paths $\mathtt{p}_{jj}^{i*}$, has the same value before and after severing, since it contains no paths travelling through $\mathcal{S}$.  In summary, the numerators of $\rho_{ij}$ and $\rho'_{ij}$ are equal, but the denominator for $\rho'_{ij}$ is larger, so $|\rho'_{ij}| < |\rho_{ij}|$. Therefore, sub-graphs that are connected to the rest of the network through a single node always increase the marginal correlations between the separating node and other elements of $\mathcal{T}$. The set $\mathcal{S}$ can be viewed as acting as a resonant cavity that amplifies the influence of node $i$ on the others.

The second statement stems from the fact that, when calculating the marginal correlation $\rho_{jk}$ between nodes $j, k \in \mathcal{T}$ (different from $i$), every $jk^*$-path visiting node $i$ can be enlarged to include an arbitrary number of paths of type $\mathtt{p}_{i\mathcal{S}i}^*$ (i.e. $ii^*$-paths that only visit nodes in $\mathcal{S}$). The sum over all such paths is a geometric series that produces a factor greater than $1$, which increases the (absolute value of the) weight of every path that visits $i$, compared to the corresponding weight when $\mathcal{S}$ is absent from the graph. The severance of the set $\mathcal{S}$ eliminates all detours inside $\mathcal{S}$, thereby reducing the absolute value of the weight of every path involving  $i$.

\subsubsection{Application: The amplifying effect of prolonging a linear network}
\label{sec:revlin}

\begin{figure}[h]
    \centering
    \includegraphics[width=0.4\textwidth]{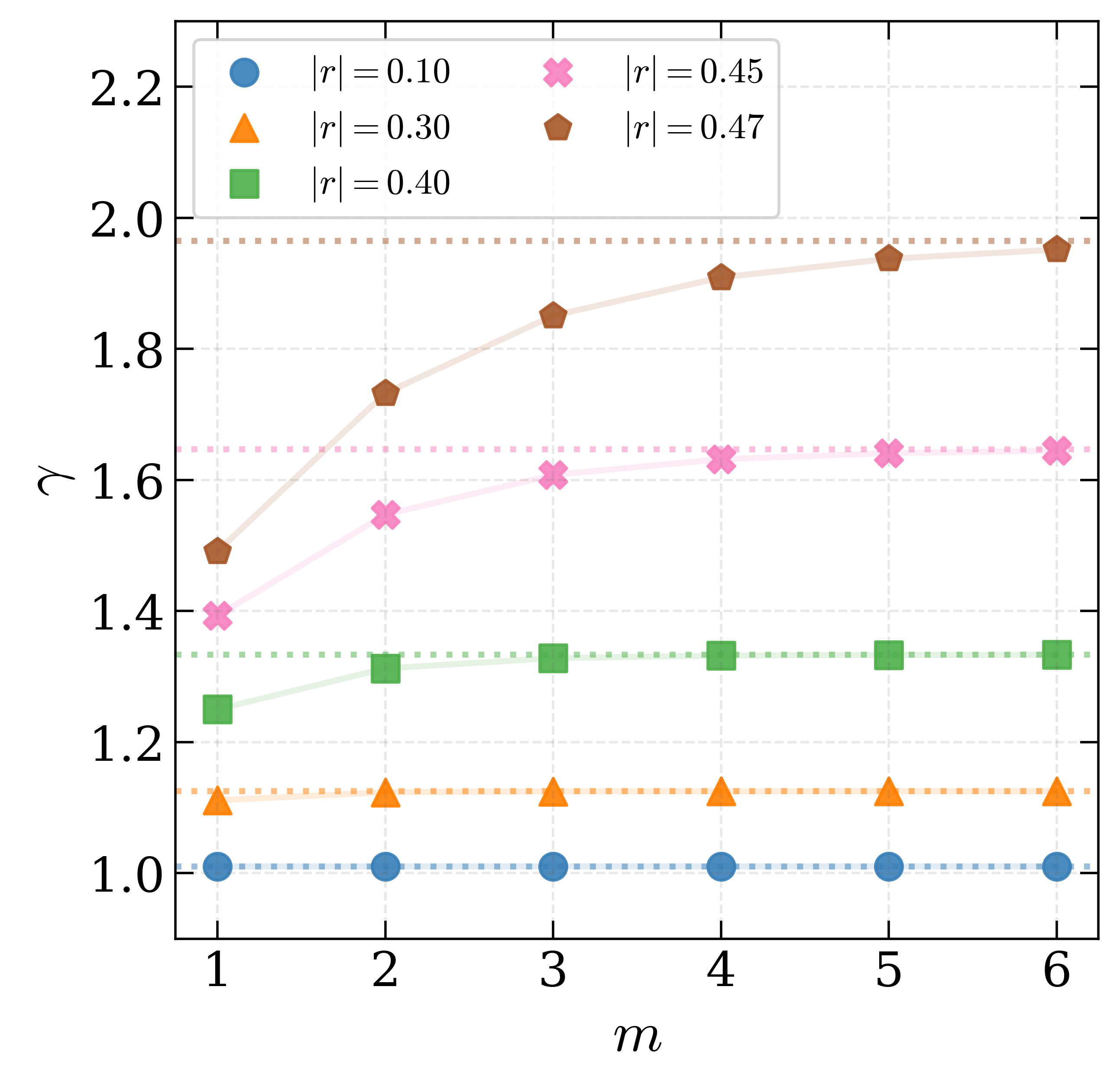}
    \caption{Amplification factor $\gamma$, defined as the ratio  $\rho_{ij}(m) /$ $\rho_{ij}(0)$, displayed as a function of the number of nodes  $m$ appended to each end of the chain stretching from $i$ to $j$, for different absolute values of the partial correlation $r$, and $k = 10$ intermediate nodes. Dotted lines indicate the value of $\gamma$ in the limit $m \to \infty$, computed using Eq.~\ref{eq:chain_l_inf}. For $|r| < 0.4$, only  the first appended node produces a substantial amplification, the subsequent ones having only a marginal effect. Partial correlations with larger absolute value yield amplification factors that more evidently display a gradual increase with $m$.}
    \label{fig:gamma_factor}
\end{figure}

As mentioned earlier, adding nodes to a graph without modifying its existing connections is the inverse operation of severing nodes. Therefore, the first statement of the previous section also asserts that appending a sub-network to either $i$ or $j$ amplifies the value of $\rho_{ij}$. This section discusses the effect of prolonging a chain graphical model, where nodes $i$ and $j$ are connected by a chain of $k$ nodes, and then, an additional $m$-node chain is appended to each side:
\begin{equation*}
        \begin{tikzpicture}[>=Stealth,auto,node distance=2cm,thick,baseline=(dots1.base)]
            \tikzset{
                mystate/.style={fill=black,circle,inner sep=0pt,minimum width=4pt},
                myedge/.style={opacity=0.2,text opacity=0.5}}

            \node[label={[label distance=-0.68cm]0:$m$}] (dots1) at (0,0) {};
            \draw[thick,rounded corners,dotted] (-0.67,-0.4) rectangle (0.13,0.4);
            
            \node[mystate,label={above:$i$}] (2) at (1.5,0) {};
            
            \node (dots2L) at (3,0) {};
            \node[label={[label distance=-0.35cm]0:$k$}] (dots2) at (3.27,0) {};
            \node (dots2R) at (3.55,0) {};
            \draw[thick,rounded corners,dotted] (2.87,-0.4) rectangle (3.67,0.4);
            
            \node[mystate,label={above:$j$}] (n-1) at (5.05,0) {};
            
            \node[label={[label distance=-0.69cm]180:$m$}] (dots3) at (6.55,0) {};
            \draw[thick,rounded corners,dotted] (6.42,-0.4) rectangle (7.22,0.4);
        
            \path   (dots1) edge [myedge,sloped,anchor=center,below] node {$r$} (2)
                    (2) edge [myedge,sloped,anchor=center,below] node {$r$} (dots2L)
                    (dots2R) edge [myedge,sloped,anchor=center,below] node {$r$} (n-1)
                    (n-1) edge [myedge,sloped,anchor=center,below] node {$r$} (dots3);
        \end{tikzpicture}
\end{equation*}
This graphical model corresponds to a chain of $d = 2 + k + 2 \, m$ variables, the partial correlations of which are all equal to $r$. As in Sect.~\ref{ex:markov_chain}, the condition $|r| \le \sfrac{1}{2}$ guarantees that the precision matrix is positive definite. The amplification factor $\gamma = \rho_{ij}(m)/\rho_{ij}(0)$ is the ratio between the marginal correlation $\rho_{ij}(m)$ between nodes $i$ and $j$ in the longer chain, and the marginal correlation $\rho_{ij}(0)$ in the original one ($m = 0$). The marginal correlation between $i$ and $j$ is given by Eq.~\ref{eq:chain_corr_ij}, so
\begin{equation*}
    \gamma = \frac{1 - \ell_{k + 2}}{1 - \ell_{k + 2} - \ell_{m+2}},
\end{equation*}
where $\ell_k$ follows the recurrence relation of Eq.~\ref{eq:chain_loops_rec}. Quite remarkably, the value of $\gamma$ does not depend on the sign of $r$, because neither does $\ell_k$. The amplification factor $\gamma$ is shown in Fig.~\ref{fig:gamma_factor} as a function of $m$, for different values of $|r|$ and a fixed number of $k = 10$ intermediate nodes. When $|r|$ is not too close to its maximal value $0.5$, the propagation of information along the appended chain is marginally effective, so only the first added node amplifies the marginal correlation significantly. Additional nodes make no more than a modest contribution, causing $\gamma$ to be essentially independent of $m$. Yet, in this regime the amplification can still be noticeable. For example, for $|r| = 0.1$, adding a single node at each side increases $\rho_{ij}$ by $11\%$. As $|r|$ becomes closer to $0.5$, the dependence on $m$ becomes stronger, for example, if $|r| = 0.47$, adding a single node increases $\rho_{ij}$ by a $49\%$, reaching up to $95\%$ for $m=6$.

An amplification factor that is always greater than unity implies that appending nodes to the linear chain invariably enhances the marginal correlations between elements of the original chain, in accordance to the discussion of Sect.~\ref{sec:appended}. Therefore, $\rho_{ij}$ cannot be entirely understood in terms of the ability to transmit information directly from $i$ to $j$. The addition of peripheral detours allows for reverberating activity at either end of the chain, amplifying the influence that nodes $i$ and $j$ exert on each other.

\subsection{Marginalising a subset of nodes}
\label{sec:marginalisation}

When marginalising a subset $\mathcal{S}$ of nodes, the partial correlations between the remaining variables change, whereas the marginal correlations are unaffected. This property sometimes allows us to derive relations between the marginal correlations of pairs of nodes in complex networks by analysing simpler networks. By selectively marginalising certain nodes, the structure of a network may be stripped of many of its details, leaving only a backbone of the original graph.

\subsubsection{Partial correlations of the reduced model in terms of those of the detailed model} 

To express the partial correlations of the reduced model as a function of those of the detailed model, in Appendix~\ref{app:marg_part_corr} we use the transformation properties of the covariance and the precision matrices under marginalisation, and derive their effect on the partial correlations using Eq.~\ref{eq:rij_omega}. The result is that the partial correlation $r_{ij}'$ of the reduced model can be expressed entirely in terms of paths in $G$ that only visit nodes in $\mathcal{S}$, the marginalised subset. Using $p_{i\mathcal{S}j}^*$ to denote an $ij^*$-path in $G$ traversing only nodes in $\mathcal{S}$ we get (Appendix~\ref{app:marg_part_corr})
\small
\begin{equation} \label{eq:nesting_transf}
    r'_{ij} =
    \frac{r_{ij} + \displaystyle{\sum_{\substack{\mathtt{p}_{i\mathcal{S}j}^{*}}}}
    \begin{tikzpicture}[->,>=Stealth,auto,shorten >=0.5pt,node distance=1.5cm,semithick,baseline=(i.base)]
        \tikzset{
            mystate/.style={fill=black,circle,inner sep=0pt,minimum width=4pt},
            myedge/.style={text opacity=0.9,opacity=0.6}}
        \node[mystate,label={above:$i$}] (i)               {};
        \node[mystate,label={above:$j$}] (j)  [right of=i] {};
        \path (i) edge [myedge,bend left,above] node {$\scriptstyle \mathtt{p}_{i\mathcal{S}j}^{*}$} (j);
    \end{tikzpicture}
    }{ \sqrt{
    \left[
    1 - \displaystyle{\sum_{\substack{\mathtt{p}_{i\mathcal{S}i}^*}}}
    \begin{tikzpicture}[->,>=Stealth,auto,shorten >=0.5pt,node distance=1.5cm,semithick,baseline=(i.base)]
        \tikzset{
            mystate/.style={fill=black,circle,inner sep=0pt,minimum width=4pt},
            myloopleft/.style={out=-140,in=140,looseness=20,text opacity=0.9,opacity=0.6},
            myloopright/.style={out=40,in=-40,looseness=20,text opacity=0.9,opacity=0.6}}
        \node[mystate,label={above:$i$}] (i) {};
        \path (i) edge [myloopright]  node {$\scriptstyle \mathtt{p}_{i\mathcal{S}i}^*$} (i);
    \end{tikzpicture}
    \right] \left[
    1 - \displaystyle{\sum_{\substack{\mathtt{p}_{j\mathcal{S}j}^*}}}
    \begin{tikzpicture}[->,>=Stealth,auto,shorten >=0.5pt,node distance=1.5cm,semithick,baseline=(i.base)]
        \tikzset{
            mystate/.style={fill=black,circle,inner sep=0pt,minimum width=4pt},
            myloopleft/.style={out=-140,in=140,looseness=20,text opacity=0.9,opacity=0.6},
            myloopright/.style={out=40,in=-40,looseness=20,text opacity=0.9,opacity=0.6}}
        \node[mystate,label={above:$j$}] (j) {};
        \path (j) edge [myloopright]  node {$\scriptstyle \mathtt{p}_{j\mathcal{S}j}^*$} (j);
    \end{tikzpicture}
    \right]
    }}. 
\end{equation}
\normalsize
Comparing Eqs.~\ref{eq:rho_diagram_4} and \ref{eq:nesting_transf}, we see that marginalisation induces a certain duality between marginal and partial correlations. The partial correlation $r'_{ij}$ between two nodes $i, j \in \mathcal{T}$ that remain after marginalising over the nodes in $\mathcal{S}$ is also the marginal correlation between those same nodes, when severing the nodes in $\mathcal{T} - \{i, j\}$. In other words, the partial correlation $r'_{ij}$ in the marginalised graph is also the marginal correlation between $i$ and $j$ of the sub-graph of $G$ generated by nodes in $\{i, j\} \cup \mathcal{S}$.

This property is trivially observed when the set $\mathcal{S}$ includes all the nodes in the network except $i$ and $j$, so that the graph obtained after marginalisation is just a $2$-node network. In this case, $r'_{ij}$ coincides with $\rho_{ij}$, and Eq.~\ref{eq:nesting_transf} reduces to Eq.~\ref{eq:rho_diagram_4}. 

An important property of marginalisation is that, if the nodes $i$ and $j$ are not connected to any  nodes in $\mathcal{S}$, their partial correlation remains unchanged after marginalisation, as the numerator of Eq.~~\ref{eq:nesting_transf} only contains the direct link $r_{ij}$, and the sums in the denominator vanish. 

\subsubsection{Necessary condition for the existence of a separating node}
\label{sec:sep_node_necessity}

As an example of how marginalisation can be used to derive relations between the different $\rho_{ij}$, consider a graph in which there is a separating node $k$ for nodes $i$ and $j$, that is, all the paths connecting $i$ and $j$ include the node $k$. If we marginalise all the nodes that are different from $i$, $j$ and $k$, the reduced network has the structure of a $3$-node chain $i-k-j$, implying that $r_{ij}' = 0$, as in Sect.~\ref{ex:3_chain}. Therefore, the matrix of marginal correlations can be written exactly as in Eq.~\ref{eq:chain3}, but replacing all the $r$ by $r'$. From there, we conclude that $\rho_{ij} = \rho_{ik} \, \rho_{kj}$. This factorisation was originally derived for a $3$-node chain, but the fact that marginalisation leaves marginal correlations unaffected, extends its validity to any network in which $k$ is a separating node. In particular, it holds for linear chains, where all nodes except those at the extremes are separating nodes.

\subsubsection{Marginalising out the nodes of a sub-network connected to one of the nodes composing a pair}
\label{sec:appended2}

In Sect.~\ref{sec:appended} and Fig.~\ref{fig:net_struct}a, we concluded that conditioning on the nodes of a sub-network appended to a separating node $i$ diminishes the marginal correlations $\rho'_{ij}$ between $i$ and any of the nodes remaining in the reduced network $G'$. If the elimination is instead performed through marginalisation, the partial correlation $r'_{ij}$ between $i$ and any of the remaining nodes increases. To prove this statement, note that in Eq.~\ref{eq:nesting_transf}, the numerator is simply $r_{ij}$, whereas in the denominator, the sum over closed paths $\mathtt{p}_{i\mathcal{S}i}^*$ lies between $0$ and $1$ (Eq.~\ref{eq:menorqueuno}), and the sum over paths $\mathtt{p}_{j\mathcal{S}j}^*$ vanishes, since no such paths exist. Therefore, the denominator is less than $1$, so $|r'_{ij}| > |r_{ij}|$. This argument holds also for more general networks, as long as $\mathcal{S}$ is not connected to $j$. In other words, when marginalising a sub-network $\mathcal{S}$, every partial correlation $r_{ij}'$ between two remaining nodes increases, provided that one (and only one) of them is connected to $\mathcal{S}$.

\subsubsection{Marginalising a subset of variables connected to the two nodes composing a pair}

Figure~\ref{fig:net_struct}b shows an example of a graph $G$ defined on a set $\mathcal{U}$, in which the subset $\mathcal{S} \subset \mathcal{U}$ is connected to its complement $\mathcal{T} = \mathcal{U} - \mathcal{S}$ through two nodes $i$ and $j$. As is always the case, eliminating the subset $\mathcal{S}$ by marginalisation modifies the value of the partial correlations $r'_{ik}$ and $r'_{jk}$ between either  $i$ or $j$ and the rest of the reduced network. The argument of the previous section proves that the absolute value of the partial correlation increases: $|r_{ik}'| > |r_{ik}|$. In addition, the elimination of $\mathcal{S}$ changes the value of the direct link $r'_{ij}$. In fact, if in the original graph $G$ the nodes $i$ and $j$ are disconnected (as in the example of Fig.~\ref{fig:net_struct}b), the new link $r'_{ij}$ can be thought of as an effective partial correlation, summarising the effect of $\mathcal{S}$. If $i$ and $j$ are initially connected, the value of the connection typically changes, so that $r'_{ij} \ne r_{ij}$. The change can either increase or decrease the absolute value of the connection, depending on whether the original link and the new contribution have equal or opposite signs.

\subsubsection{The minimal equivalent model}

The elimination of a set of nodes $\mathcal{S}$ by marginalisation produces a new link $r'_{ij}$ between the nodes $i$ and $j$ that effectively replaces the action of the nodes in $\mathcal{S}$ (Fig~\ref{fig:net_struct}b). In this section, rather than generating new effective links, we develop a strategy to produce new effective nodes that supplant the marginalised variables. The new nodes, here called \textit{latent}, are often fewer than the marginalised ones, making this approach useful for constructing compact equivalent models, in terms of latent variables. 

\begin{figure*}
    \centering
    \begin{tikzpicture}[>=Stealth,auto,node distance=1cm,thick]
            \tikzset{
                mystate/.style={fill=black,circle,inner sep=0pt,minimum width=4pt},
                myedge/.style={opacity=0.2,text opacity=0.5},
                myedge2/.style={text opacity=0.9,opacity=0.6}}

            \node[mystate] (1) at (0,0) {};
            \node[mystate] (2) at ($(1) + (-.4,-.5)$) {};
            \node[mystate] (3) at ($(1) + (.4,-.5)$) {};

            \path   (1) edge [myedge] node {} (2)
                    (2) edge [myedge] node {} (3)
                    (3) edge [myedge] node {} (1);

            \draw[thick,rounded corners,dotted,opacity=0.3] ($(1) + (-.8,.4)$) rectangle ($(3) + (.4,-.4)$);
            \node[label={$\mathcal{S}$}] (S) at ($(1) + (1.1,-.5)$) {};

            \node[mystate] (4) at ($(2) + (0,-2)$) {};
            \node[mystate] (5) at ($(4) + (.8,0)$) {};
            \node[mystate] (6) at ($(4) + (-.4,-.5)$) {};
            \node[mystate] (7) at ($(4) + (.4,-.5)$) {};
            \node[mystate] (8) at ($(4) + (1.2,-.5)$) {};

            \path   (4) edge [myedge] node {} (5)
                    (5) edge [myedge] node {} (8)
                    (8) edge [myedge] node {} (7)
                    (7) edge [myedge] node {} (5)
                    (5) edge [myedge] node {} (6)
                    (6) edge [myedge] node {} (4)
                    (6) edge [myedge] node {} (7);
            \draw[myedge] (4) to[bend left,out=60,in=90,looseness=1.5] node {} (8);

            \draw[thick,rounded corners,dotted,opacity=0.3] ($(4) + (-.8,.55)$) rectangle ($(8) + (.4,-.4)$);
            \node[label={$\mathcal{T}$}] (T) at ($(5) + (1.1,-.7)$) {};

            \node[label={\large $G$}] (G) at ($(1) + (-1.5,-0.8)$) {};

            \path   (2) edge [myedge] node {} (4)
                    (2) edge [myedge] node {} (6);
            \draw[myedge] plot [smooth,tension=1] coordinates {($(1)$) ($(4) + (-.5,1.7)$) ($(6)$)};

            \node[mystate] (1b) at ($(1) + (5.5,0)$) {};
            \node[mystate] (2b) at ($(1b) + (-.4,-.5)$) {};
            \node[mystate] (3b) at ($(1b) + (.4,-.5)$) {};

            \path   (1b) edge [myedge] node {} (2b)
                    (2b) edge [myedge] node {} (3b)
                    (3b) edge [myedge] node {} (1b);

            \draw[thick,rounded corners,dotted,opacity=0.3] ($(1b) + (-.8,.4)$) rectangle ($(3b) + (.4,-.4)$);
            \node[label={$\mathcal{S}$}] (Sb) at ($(1b) + (1.1,-.5)$) {};

            \node[mystate] (4b) at ($(2b) + (0,-2)$) {};
            \node[mystate] (5b) at ($(4b) + (.8,0)$) {};
            \node[mystate] (6b) at ($(4b) + (-.4,-.5)$) {};
            \node[mystate] (7b) at ($(4b) + (.4,-.5)$) {};
            \node[mystate] (8b) at ($(4b) + (1.2,-.5)$) {};

            \path   (4b) edge [myedge] node {} (5b)
                    (5b) edge [myedge] node {} (8b)
                    (8b) edge [myedge] node {} (7b)
                    (7b) edge [myedge] node {} (5b)
                    (5b) edge [myedge] node {} (6b)
                    (6b) edge [myedge] node {} (4b)
                    (6b) edge [myedge] node {} (7b);
            \draw[myedge] (4b) to[bend left,out=60,in=90,looseness=1.5] node {} (8b);

            \draw[thick,rounded corners,dotted,opacity=0.3] ($(4b) + (-.8,.55)$) rectangle ($(8b) + (.4,-.4)$);
            \node[label={$\mathcal{T}$}] (Tb) at ($(5b) + (1.1,-.7)$) {};

            \node[label={\large $\tilde{G}$}] (G_mono) at ($(G) - (1) + (1b)$) {};

            \node[circle,fill=red!90!black,draw=black,semithick,inner sep=0pt,minimum width=5pt,label={left:\textcolor{red!80!black}{$Y_1$}}] (Y1b) at ($(2b) + (-.2,-.8)$) {};
            \node[circle,fill=blue!90!black,draw=black,semithick,inner sep=0pt,minimum width=5pt,label={right:\textcolor{blue!80!black}{$Y_2$}}] (Y2b) at ($(3b) + (.2,-.8)$) {};

            \draw[red!90!black,opacity=0.3] (Y1b) to[bend left,out=80,in=120,looseness=1.5] node {} (1b);
            \draw[red!90!black,opacity=0.3] (Y1b) to[bend left,out=30,in=140,looseness=1] node {} (2b);
            \draw[red!90!black,opacity=0.3] (Y1b) to[bend left,out=-20,in=-150,looseness=1] node {} (4b);
            \draw[red!90!black,opacity=0.3] (Y1b) to[bend left,out=-40,in=-130,looseness=1] node {} (6b);

            \draw[blue!90!black,opacity=0.3] (Y2b) to[bend right,out=320,in=240,looseness=1.5] node {} (1b);
            \draw[blue!90!black,opacity=0.3] (Y2b) to[bend left,out=20,in=160,looseness=1] node {} (2b);
            \draw[blue!90!black,opacity=0.3] (Y2b) to[bend right,out=-20,in=-150,looseness=1] node {} (4b);
            \draw[blue!90!black,opacity=0.3] (Y2b) to[bend right,out=-45,in=220,looseness=1] node {} (6b);

            \node[mystate] (4c) at ($(4b) + (5.5,0)$) {};
            \node[mystate] (5c) at ($(4c) + (.8,0)$) {};
            \node[mystate] (6c) at ($(4c) + (-.4,-.5)$) {};
            \node[mystate] (7c) at ($(4c) + (.4,-.5)$) {};
            \node[mystate] (8c) at ($(4c) + (1.2,-.5)$) {};

            \path   (4c) edge [myedge] node {} (5c)
                    (5c) edge [myedge] node {} (8c)
                    (8c) edge [myedge] node {} (7c)
                    (7c) edge [myedge] node {} (5c)
                    (5c) edge [myedge] node {} (6c)
                    (6c) edge [myedge] node {} (4c)
                    (6c) edge [myedge] node {} (7c);
            \draw[myedge] (4c) to[bend left,out=60,in=90,looseness=1.5] node {} (8c);

            \draw[thick,rounded corners,dotted,opacity=0.3] ($(4c) + (-.8,.55)$) rectangle ($(8c) + (.4,-.4)$);
            \node[label={$\mathcal{T}$}] (Tc) at ($(5c) + (1.1,-.7)$) {};

            \node[label={\large $G'$}] (G_prime) at ($(G) - (4) + (4c)$) {};
            
            \node[circle,fill=red!90!black,draw=black,semithick,inner sep=0pt,minimum width=5pt,label={left:\textcolor{red!80!black}{$Y_1$}}] (Y1c) at ($(Y1b) + (4c) - (4b)$) {};
            \node[circle,fill=blue!90!black,draw=black,semithick,inner sep=0pt,minimum width=5pt,label={right:\textcolor{blue!80!black}{$Y_2$}}] (Y2c) at ($(Y2b) + (4c) - (4b)$) {};

            \draw[red!90!black,opacity=0.4,densely dashed] (Y1c) to[bend left,out=-20,in=-150,looseness=1] node {} (4c);FigureFigure
            \draw[red!90!black,opacity=0.4,densely dashed] (Y1c) to[bend left,out=-40,in=-130,looseness=1] node {} (6c);

            \draw[blue!90!black,opacity=0.4,densely dashed] (Y2c) to[bend right,out=-20,in=-150,looseness=1] node {} (4c);
            \draw[blue!90!black,opacity=0.4,densely dashed] (Y2c) to[bend right,out=-45,in=220,looseness=1] node {} (6c);
            
            \draw[magenta!80!black,opacity=0.5,densely dashed] (Y1c) to[bend left,looseness=1] node {} (Y2c);

            \node[label={\large (a)}] (a) at ($(1) + (-1.5,.35)$) {};
            \node[label={\large (b)}] (b) at ($(a) + (1b) - (1)$) {};
            \node[label={\large (c)}] (c) at ($(b) + (4c) - (4b)$) {};
    \end{tikzpicture}
    \caption{Logical scheme used to derive a minimal equivalent model. (a): Graph $G$, composed of a set of nodes $\mathcal{U}$, that can be partitioned into a subset $\mathcal{S}$ of nodes to be marginalised out, and its complement $\mathcal{T}$, with those that remain. (b): Graph $\tilde{G}$, including the nodes in $G$, and two additional latent variables $Y_1$ and $Y_2$, mediating all the connections between $\mathcal{S}$ and $\mathcal{T}$. The connections to and from the latent variables must be chosen so that, when the latent variables are marginalised, the graph $G$ in (a) is obtained. (c): Graph $G'$, obtained when $\mathcal{S}$ is marginalised from $\tilde{G}$. In $G'$, the marginal and partial correlations within $\mathcal{T}$ coincide with those of $G$.}
    \label{fig:minimal}
\end{figure*}
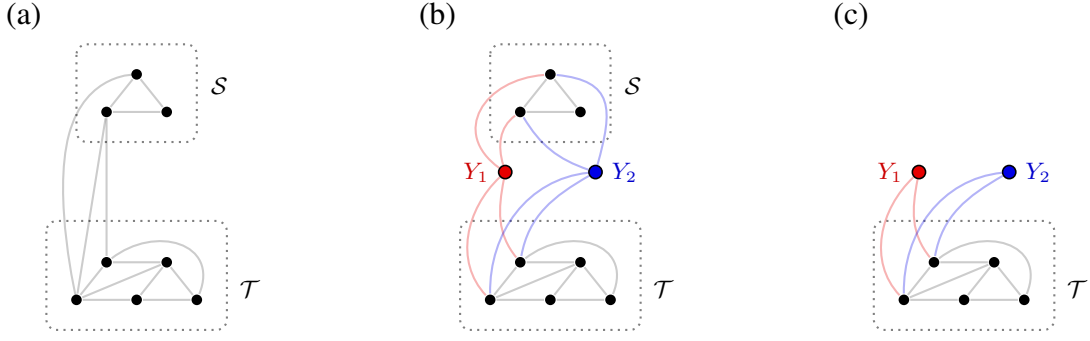

Let $G$ be the graph associated to the set $\mathcal{U}$ with partial correlations $r_{ij}$, as illustrated in Fig.~\ref{fig:minimal}a. Equation~\ref{eq:nesting_transf} states that if the variables in $\mathcal{S}$ are marginalised out, the partial correlations between the variables in $\mathcal{T}$ change from $r_{ij}$ to $r'_{ij}$. Our goal here, however, is to preserve not only the original marginal correlations $\rho_{ij}$ (as always happens with marginalisation), but also the partial correlations $r_{ij}$. Therefore, the new model cannot consist solely of the variables in $\mathcal{T}$. Additional latent variables $Y_1, Y_2, \dots, Y_\ell$ must be included to replace the  marginalised ones. The addition of latent variables can be thought of as the inverse operation of removing variables through marginalisation. To devise the smallest equivalent model, we require that the number of latent variables be minimised. 

To construct the effective model, imagine that the full graph $G$ associated to the detailed description $\mathcal{U}$ results from marginalising the latent variables $Y_1, \dots, Y_\ell$ of an even larger graph $\tilde{G}$ defined over the set $\mathcal{U} \cup \{Y_1, \dots, Y_\ell\}$ (Fig.~\ref{fig:minimal}b). In $\tilde{G}$, there are no direct connections between elements of $\mathcal{S}$ and those of $\mathcal{T}$. The connections between two nodes within $\mathcal{S}$ or two nodes within $\mathcal{T}$ are the same as those in the graph $G$. Each of the new latent variables $Y_1, \dots, Y_\ell$ can in principle be connected to all the nodes in $\mathcal{S}$ and also to all the nodes in $\mathcal{T}$. Let $d_\mathcal{S}$ be the number of nodes in $\mathcal{S}$, and $\tilde{\boldsymbol{a}}^u$ be a $d_\mathcal{S}$-dimensional vector containing the partial correlation between the latent variable $Y_u$ and each node in $\mathcal{S}$.  Similarly, if $d_\mathcal{T}$ is the number of nodes in $\mathcal{T}$, let the $d_\mathcal{T}$-dimensional vector $\tilde{\boldsymbol{b}}^u$ contain the partial correlations between the latent variable $Y_u$ and each node in $\mathcal{T}$. The question is whether such vectors can be defined so as to get the graph $G$ when marginalising over the latent variables. Equation~\ref{eq:nesting_transf} implies that, if such vectors exist, then
\begin{equation}  \label{eq:monios}
    r_{ij} =
    \frac{\displaystyle{\sum_{\substack{u = 1}}^\ell
    \tilde{a}^u_{i}} \ \tilde{b}^u_{j}
    }{ \sqrt{
    \left[
    1 - \displaystyle{\sum_{\substack{v = 1}}^\ell} \big( \tilde{a}^v_{i} \big)^2
    \right] \left[
    1 - \displaystyle{\sum_{\substack{w = 1}}^\ell} \big(\tilde{b}^w_{j}\big)^2
    \right]
    }},
\end{equation}
for $i \ \in \ \mathcal{S}$ and $j \ \in \ \mathcal{T}$.
Fortunately, this problem has an explicit solution. Let the $d_\mathcal{S} \times d_\mathcal{T}$ matrix $Q$ of entries $q_{ij}$, where $i \in \mathcal{S}$ and $j \in \mathcal{T}$, contain the elements of $R$ that represent the partial correlations $r_{ij}$ of the original graph $G$. If $\mu$ is the rank of $Q$, its singular value decomposition implies that a set of $\mu \le {\rm min}\{d_\mathcal{S}, d_\mathcal{T}\}$ orthonormal $d_\mathcal{S}$-dimensional vectors $\boldsymbol{a}^1, \dots, \boldsymbol{a}^\mu$ exists, as well as another collection of $\mu$ orthonormal, $d_\mathcal{T}$-dimensional vectors $\boldsymbol{b}^1, \dots, \boldsymbol{b}^\mu$, such that
\begin{equation*}
    q_{ij} = \sum_{u = 1}^\mu \sigma^2_u \ a^u_i \ b^u_ j,
\end{equation*}
for some non-zero $\sigma_1, \dots, \sigma_\mu \in \mathbb{R}$. Therefore, by taking the number $\ell$ of latent variables as equal to $\mu$, and defining the partial correlations to the latent variables $\tilde{\boldsymbol{a}}^u$ and $\tilde{\boldsymbol{b}}^u$ as 
\begin{equation} \label{eq:newconnections}
\begin{split}
    \tilde{a}^u_i &= \frac{\sigma_u \, a^u_i}{\sqrt{1 + \displaystyle{\sum_{\substack{v = 1}}^\mu} \sigma_v^2 \left( a_i^v \right)^2}}   \\  \tilde{b}^u_j &= \frac{\sigma_u \, b^u_j}{\sqrt{1 + \displaystyle{\sum_{\substack{v = 1}}^\mu} \sigma_v^2 \left( b_j^v \right)^2}},
\end{split}
\end{equation}
Eq.~\ref{eq:monios} holds, i.e. $r_{ij} = q_{ij}$ for any $i \in \mathcal{S}$ and $j \in \mathcal{T}$.

The singular value decomposition of a matrix is not unique. However, the number of non-zero principal values is indeed unique. Therefore, the minimal number of latent variables is equal to the rank $\mu$ of $Q$, and any attempt to reduce it further makes the fulfilment of Eq.~\ref{eq:monios} impossible. The partial correlations between the latent variables and the original variables are not uniquely determined, as there may be multiple choices for the vectors $\boldsymbol{a}^u$ and $\boldsymbol{b}^u$, as well as for the sign of $\sigma_u$. Nonetheless, any of them provides a valid solution to the problem.

Now that we have an enlarged network $\tilde{G}$ that, from the point of view of the nodes in $\mathcal{T}$, is completely equivalent to the original network $G$, the variables describing the nodes in $\mathcal{S}$ can be safely marginalised out (Fig.~\ref{fig:minimal}c), so that the equivalent model is obtained. This operation modifies the partial correlations $\tilde{\boldsymbol{b}}^j$, but not the ones inside $\mathcal{T}$.

In order to get a useful reduction, the number of latent variables $\mu$ must be smaller than the number of marginalised variables $d_\mathcal{S}$. Since $\mu$ is the rank of $Q$, it cannot exceed $d_\mathcal{S}$. Yet, to achieve a proper reduction, $\mu$ must be strictly smaller than $d_\mathcal{S}$, and the smaller, the more efficient the reduction. If in the original graph $G$, some nodes of $\mathcal{S}$ are disconnected from all the nodes in $\mathcal{T}$, some of the rows of $Q$ vanish. Similarly, if some of the nodes in $\mathcal{T}$ are disconnected from all the nodes in $\mathcal{S}$, some columns of $Q$ vanish. Both these scenarios favour a small rank. This suggests that useful reductions can be expected when the connections between $\mathcal{S}$ and $\mathcal{T}$ are sparse. In terms of the topology of $G$, the number of latent variables $\mu$ is bounded by
\begin{equation*}
    \mu \le \min 
    \Bigg\{
    \begin{aligned}
        &\text{\# nodes in $\mathcal{S}$ connected to $\mathcal{T}$} \\
        &\text{\# nodes in $\mathcal{T}$ connected to $\mathcal{S}$}
    \end{aligned}
    \Bigg\}.
\end{equation*}

In summary,  to construct the minimal equivalent model, the following steps need to be taken:

\begin{enumerate}
    \item Partition the initial set of nodes $\mathcal{U}$ into two subsets: $\mathcal{S}$ with the $d_{\mathcal{S}}$ nodes to be marginalised, and the complementary subset $\mathcal{T}$ with the $d_{\mathcal{T}}$ nodes to be retained.
    \item Extract from the partial correlation matrix $R$ the submatrix $Q \in \mathbb{R}^{d_{\mathcal{S}}\times d_{\mathcal{T}}}$ containing the partial correlations between  elements of $\mathcal{S}$ and elements of $\mathcal{T}$.
    \item Perform a singular value decomposition on the matrix $Q$, from which the number $\mu$ of non-zero singular values can be determined, along with the singular values $\sigma_1^2, \dots, \sigma_\mu^2$ themselves, and the associated singular vectors $\bm{a}^1, \dots, \bm{a}^\mu \in \mathbb{R}^{d_{\mathcal{S}}}$ and $\bm{b}^1, \dots, \bm{b}^\mu \in \mathbb{R}^{d_{\mathcal{T}}}$.  
    \item Use Eqs.~\ref{eq:newconnections} to define the partial correlations $\tilde{\bm{a}}^i$ and $\tilde{\bm{b}}^j$ between the new latent variables $Y_1, \dots, Y_\mu$ and the nodes in $\mathcal{S}$ and $\mathcal{T}$, respectively. At this stage, take the latent variables $Y_j$ to be disconnected from each other, that is, assume that the partial correlations between different $Y$-variables vanish.
    \item Marginalise out the nodes in $\mathcal{S}$. This operation modifies the value of the partial correlations between the latent variables $Y^j$ and the nodes in $\mathcal{T}$, as well as the connections between different latent variables. Use Eq.~\ref{eq:nesting_transf} to recalculate these connections by summing over paths that travel within $\mathcal{S}$.
\end{enumerate}

\section{Relation between paths and mutual information}
\label{sec:info}

In this section, the variables $\{X_1, \dots,$ $ X_d\}$ are assumed to be governed by a Gaussian density. We will demonstrate that the sums over closed paths in the denominator of Eq.~\ref{eq:rho_diagram_4} can be written in terms of conditional mutual informations. Let $\{X_{\kappa}\} \subset \{X_1, \dots, X_d\}$ be the set of variables whose index $\kappa$ is neither $i$ nor $j$. The conditional mutual information $I(X_i; \{X_{\kappa}\} | X_j)$ can be written in terms of conditional differential entropies as \citep{Cover1991}
\begin{equation*}
    I(X_i; \{X_{\kappa}\} | X_j) = h(X_i | X_j) - h(X_i | X_j, \{ X_\kappa \}), 
\end{equation*}
with conditional differential entropies (in nats) 
\begin{equation*}
\begin{split}
    h(X_i | X_j) &= \frac{1}{2} \ln\left[ 2 \pi e {\rm Var}(X_i | X_j) \right]
    \\
    h(X_i | X_j, \{X_\kappa\}) &= \frac{1}{2} \ln\left[ 2 \pi e {\rm Var}(X_i | X_j, \{X_\kappa\}) \right].
\end{split}
\end{equation*}
Therefore, using nats,
\begin{equation*}
    I(X_i; \{X_{\kappa}\} | X_j) = \frac{1}{2} \ln \left[ \frac{{\rm Var}(X_i | X_j)}{{\rm Var}(X_i | X_j, \{X_\kappa\})} \right].
\end{equation*}
Comparing this result with Eq.~\ref{eq:loops_var_quotient}, the sum over closed paths can be written as
\begin{equation} \label{eq:loops_and_mut_info}
    \sum_{\mathtt{p}_{ii}^{j*}}
    \begin{tikzpicture}[->,>=Stealth,auto,shorten >=0.5pt,node distance=1.5cm,semithick,baseline=(i.base)]
        \tikzset{
            mystate/.style={fill=black,circle,inner sep=0pt,minimum width=4pt},
            myloopleft/.style={out=-140,in=140,looseness=20,text opacity=0.9,opacity=0.6},
            myloopright/.style={out=40,in=-40,looseness=20,text opacity=0.9,opacity=0.6}}
        \node[mystate,label={above:$i$}] (i) {};
        \path (i) edge [myloopright]  node {$\scriptstyle \mathtt{p}_{ii}^{j*}$} (i);
    \end{tikzpicture}
    = 1- e^{- 2 \, I(X_i; \{X_{\kappa}\} | X_j)}.
\end{equation}
Equation~\ref{eq:loops_and_mut_info} shows that  the sum over closed paths is a monotonically increasing function of the conditional mutual information $I(X_i; \{X_{\kappa}\} | X_j)$. The exponential of the mutual information between two variables measures the maximum number of distinct values one variable can assume, while ensuring that the corresponding values of the other variable remain in separate, non-overlapping groups.
Therefore, $\exp[I(X_i; \{X_{\kappa}\} | X_j)]$ quantifies the number of distinguishable regions in the space of $X_i$-values that can be encoded by the remaining nodes in the network, or equivalently, the ability of node $i$ to leave its footprint on the state of the rest of the nodes (excluding $j$). This ability, by recurrently feeding back to node $i$ along the closed paths, reflects the capacity of $i$ to use the rest of the network as a resonant cavity. In light of Eq.~\ref{eq:loops_and_mut_info}, Eq.~\ref{eq:rho_diagram_4} becomes
\begin{equation*}
    \rho_{ij} = \mathrm{e}^{I(X_i; \{X_\kappa\}|X_j)} \left[ \displaystyle{\sum_{\substack{\mathtt{p}_{ij}^{*}}}}
    \begin{tikzpicture}[->,>=Stealth,auto,shorten >=0.5pt,node distance=1.5cm,semithick,baseline=(i.base)]
        \tikzset{
            mystate/.style={fill=black,circle,inner sep=0pt,minimum width=4pt},
            myedge/.style={text opacity=0.9,opacity=0.6}}
        \node[mystate,label={above:$i$}] (i)               {};
        \node[mystate,label={above:$j$}] (j)  [right of=i] {};
        \path (i) edge [myedge,bend left,above] node {$\scriptstyle \mathtt{p}_{ij}^{*}$} (j);
    \end{tikzpicture} \right] \mathrm{e}^{I(X_j; \{X_\kappa\}|X_i)},
\end{equation*}
implying that the three relevant ingredients determining the value of $\rho_{ij}$ are the  opportunities to transmit information offered by the network (the term in square brackets), and the capacity of the network to encode the activity of $X_i$ and $X_j$ (the two exponentials). 

The relation between sums over closed paths and conditional mutual information is more general. If the set $\{X_1, \dots, X_d\}$ is partitioned into three disjoint sets $\mathcal{A}$, $\mathcal{B}$ and $\mathcal{Z}$, the conditional mutual information $I(\mathcal{A}; \mathcal{B} | \mathcal{Z})$ between variables in $\mathcal{A}$ and variables in $\mathcal{B}$ conditioned on variables in $\mathcal{Z}$ can be written in terms of closed paths that remain in $\mathcal{A} \cup \mathcal{B}$, i.e. closed paths that do not visit nodes in $\mathcal{Z}$. In Appendix~\ref{app:info}, the explicit relation is derived.

\section{Convergence of truncated path expansions}
\label{sect:convergence}

\begin{figure}[h]
    \centering
    \includegraphics[width=0.45\textwidth]{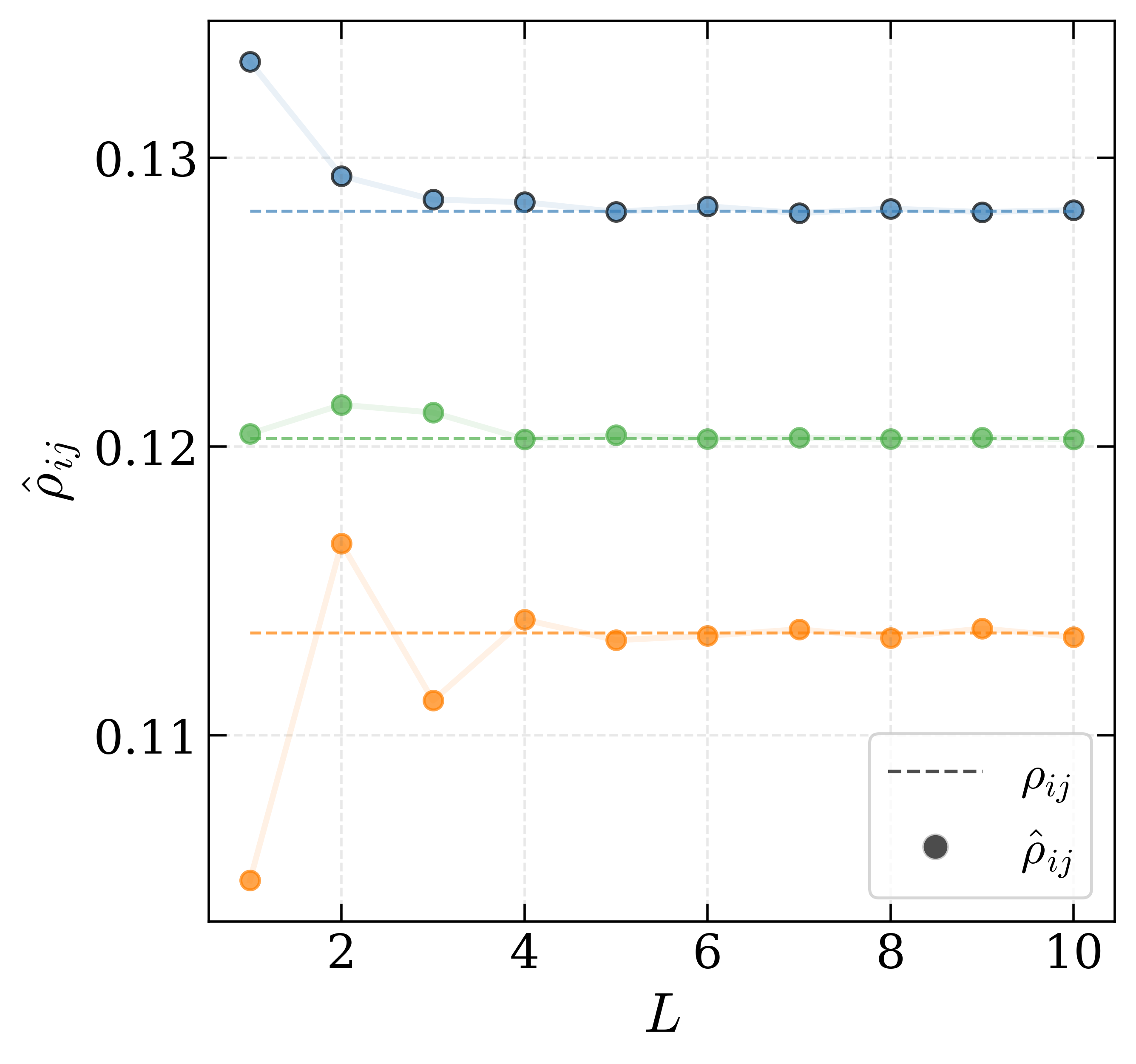}
    \caption{Analysis of the convergence properties of the series in Eq.~\ref{eq:rho_diagram_4}, for three randomly generated $R$ matrices with spectral radius $\nu(R)<1$, as a function of the length of the paths included in the sum.}
    \label{fig:conv}
\end{figure}

The cases in which the infinite sum can be explicitly computed are rare, so in most applications the expansion must be truncated for actual calculations. It is therefore important to assess the rate of convergence of the expansion, as slow convergence can result in a truncated series that approximates the final result poorly. To this end, we conducted a numerical analysis to characterise the approximation accuracy of truncated series. Sparse $R$ matrices may be suspected to converge rapidly, as they contain fewer paths connecting any two nodes. To assess convergence in a worst-case scenario, we chose to examine the effect of truncation in random matrices $R$ with no null entries, that is, non-sparse matrices. 

To generate the matrices, we began by sampling $n = 1000$ data points $\{\boldsymbol{x}_1, \dots, \boldsymbol{x}_n \}$ from a $d = 100$ dimensional Gaussian distribution (i.e.~each $\boldsymbol{x}_i \in \mathbb{R}^d$), with zero mean and covariance equal to the identity matrix. From the generated data, we computed the sampled covariance matrix
\begin{equation*}
    S = \frac{1}{n} \sum_{i=1}^n \left( \boldsymbol{x}_i - \bar{\boldsymbol{x}} \right) \left( \boldsymbol{x}_i - \bar{\boldsymbol{x}} \right)^T,
\end{equation*}
where $\bar{\boldsymbol{x}} = \sum_{i=1}^n \boldsymbol{x}_i / n$ is the sample mean. The random partial correlation matrix $R$ was obtained from Eq.~\ref{eq:omega_lambda_r}, by defining $\Omega \equiv S^{-1}$. A small number of matrices thus generated had a spectral radius $\nu(R) \ge 1$, and were discarded. The convergence properties of matrices with $\nu(R) \ge 1$ are discussed in Appendix~\ref{app:gen_expansion}. This algorithm was used to sample partial correlation matrices on which to test the convergence properties of the expansion. The results of the test were robust to changes in the sampled matrix, and within each matrix, to the choice of the matrix element to be analysed.

To characterise the way in which paths of varying length contributed to the total sum, we defined the approximate marginal correlation $\hat{\rho}_{ij}(L)$ as the one obtained when the three sums in Eq.~\ref{eq:rho_diagram_4} were truncated, so as to only include paths of length at most equal to $L$. Therefore, the true marginal correlation $\rho$ is the limit of $\hat{\rho}(L)$ when $L \to +\infty$. Figure~\ref{fig:conv} displays $\hat{\rho}_{ij}$ as a function of $L$ for three different pairs of indices $i$ and $j$, each from a different sampled partial correlation matrix. Even though Eq.~\ref{eq:rho_diagram_4} may contain arbitrarily long paths when applied to the examples of Fig.~\ref{fig:conv}, computing $\hat{\rho}(L)$ including only paths of length up to $L = 5$ yields an excellent approximation to the true correlation $\rho$. 

\section{Discussion}

Marginal correlations and partial correlations between a pair of variables differ in the way in which they relate to the remaining variables in the system. The former describe net statistical dependencies, whereas the latter describe direct dependencies. Partial correlations are easy to conceive in a mechanistic conception of the system, when it is possible to imagine the process by which two variables interact, while discounting the influence of the remaining ones. Marginal correlations, instead, describe the concerted effect of the whole collection of variables on the statistical dependencies between a specific pair. They are difficult to conceive from a mechanistic understanding, but have the virtue of providing a summarised view of what happens when all pairs of variables interact simultaneously. This paper derives a mathematical relation between the two, structured in terms of the paths through which information can be transmitted through the network, each path weighted by a specific measure of its communication ability. 

\subsubsection*{Relation with other expansions}

Other series expansions with diagrammatic interpretations have been employed in the past. For example, in field theories, Green functions can be computed with path integrals, the expansion of which involves terms that can be represented as Feynman diagrams \cite{ZinnJustin02}. Another example is given by high-temperature expansions \cite{Georges1991, Maillard2019}, in which the system’s free energy is written as a sum over powers of the inverse temperature, also invoking diagrammatic interpretations (see, for example, \citep[Part~III, Chapter~1]{amit2005field}). Both frameworks expand quantities from which several variables of interest can be computed, including marginal covariances and marginal correlations. Our goal is less ambitious, since we focus on expanding marginal correlations alone. The advantage, however, is that the convergence of our series can be proved for arbitrary probability distributions.

Our expansion differs from the high-temperature expansion in two key aspects: the range of applicability, and the variables in powers of which the series is formulated. In high-temperature expansions, the zeroth-order term typically corresponds to the non-interacting case, akin to Eq.~\ref{eq:r_neumann}. These expansions are often applied to specific spin-glass models with particular Hamiltonians, each yielding a distinct collection of diagrams. The hypothesis of thermal equilibrium implies the use of the Gibbs distribution, so focusing on a specific Hamiltonian is tantamount to selecting a specific joint probability distribution. Our series expansion and its diagramatics, instead, are not restricted to specific probability distributions. The distribution is relevant, but only in determining the values of the partial correlations, which in turn, set the weight of each path. Moreover, our expansion is in terms of partial correlations, rather than the inverse temperature, in line with our objective of interpreting partial correlations as building blocks of marginal correlations.

The expansion proposed in this paper also bears some apparent similarity with the expansion of the moment-generating or cumulant-generating function in terms of Feynman diagrams. However, again, there are two main differences. First, the zeroth-order term in our expansion is different from that of quantum field theories. While a few non-Gaussian formulations exist \citep{kuhn2018expansion}, Feynman diagrams are mostly used to expand around quadratic actions, in which the non-interacting case is defined by the saddle-point equations being linear in the fields. Instead, in our expansion, the zeroth-order or ``non-interacting'' case typically refers to independent variables, where the joint distribution factorises into a product of marginals, none of which need to be Gaussian  (see for example Eqs.~\ref{eq:factor_mod} and \ref{eq:interpretarij}). Second, as we already mentioned, our expansion involves powers of partial correlations. Feynman diagrams are ultimately written in terms of propagators, which represent the covariances of the free theory. In other words, expansions via Feynman diagrams around a quadratic action are expansions in powers of covariances, rather than partial correlations.

\subsubsection*{The expansion as an operational tool}

The expansion is also useful to perform actual calculations. In Sect.~\ref{sect:ejemplos} we presented examples in which we were unable  to invert the precision matrix, at least for arbitrary dimensionality, and yet, the exact calculation of the marginal correlations with the series expansion was possible. In more general situations, the expansion is particularly useful when the underlying graph has structural symmetries that can be exploited to calculate the summation over paths. Graphs with sparse partial correlation matrices, or with large difference in the magnitude of the matrix components, are also convenient, since only a few terms in the expansion become relevant. 

\subsubsection*{Convergence properties}

Even for non-sparse matrices, the rapid convergence observed in Sect.~\ref{sect:convergence} provides a promising perspective regarding the possibility of using the expansion to approximate the marginal correlation matrix in the cases in which the precision matrix is known, by truncating the series, thereby considering only paths that contain a bounded number of nodes. The rapid convergence is an important property, since the number of paths grows exponentially with their length. Yet, the weight of each path decreases exponentially. As long as the partial correlations are small enough (and they need to be small to ensure the matrices be positive definite), these two effects compensate, so as to yield series that converge, and often do so rapidly.

The expansion is proven to be always convergent when the spectral radius is smaller than unity (Appendix \ref{appendixB}). When the spectral radius is larger, the expansion can be modified in order to yield a convergent series (Appendix \ref{app:gen_expansion}). Therefore, the theoretical perspective of marginal correlations as a sum of contributions from different paths holds irrespective of the number of variables and of the value of the partial correlations. For practical calculations, however, the expansion for the case $\nu(R) \ge 1$ includes self-loops, making the calculation significantly more cumbersome, and limiting the operational applicability.

\subsubsection*{The expansion as a way to gain intuition} 

In our view, the most significant value of the derived expansion lies in its capacity to provide intuition about the role of each node, each link, and each path in the collection of pairwise interactions between variables. For instance, the amplification effect by which sub-networks appended to a single node enhance the marginal correlations involving that node reveals that external variables may modify interactions even when they are not mediators. Similarly,  separating nodes in the graph have been shown to impose strong constraints in the values that marginal correlations can take. The intuition gained by the expansion in terms of paths becomes particularly valuable when exploring counterfactual scenarios, such as predicting how the network’s behaviour would change if its structure were altered by specific interventions, such as pruning certain links or appending another network to a given node or set of nodes. Importantly, the expansion in terms of paths also provides a systematic method for constructing compact effective models, when searching for simplified descriptions.


\begin{thebibliography}{26}
\providecommand{\natexlab}[1]{#1}
\providecommand{\url}[1]{\texttt{#1}}
\expandafter\ifx\csname urlstyle\endcsname\relax
  \providecommand{\doi}[1]{doi: #1}\else
  \providecommand{\doi}{doi: \begingroup \urlstyle{rm}\Url}\fi

\bibitem[Matthews(2000)]{Matthews2000}
Robert~Andrew Matthews.
\newblock Storks deliver babies (p= 0.008).
\newblock \emph{Teaching Statistics}, 22\penalty0 (6):\penalty0 36--38, 2000.
\newblock \doi{10.1111/1467-9639.00013}.

\bibitem[Lauritzen(1996)]{lauritzen1996}
Steffen~L. Lauritzen.
\newblock \emph{Graphical Models}.
\newblock Oxford University Press, 1996.

\bibitem[Baba et~al.(2004)Baba, Shibata, and Sibuya]{baba2004partial}
Kunihiro Baba, Ritei Shibata, and Masaaki Sibuya.
\newblock Partial correlation and conditional correlation as measures of conditional independence.
\newblock \emph{Australian \& New Zealand Journal of Statistics}, 46\penalty0 (4):\penalty0 657--664, 2004.

\bibitem[Cramér(1946)]{cramer1946mathematical}
Harald Cramér.
\newblock \emph{Mathematical methods of statistics}, volume~26.
\newblock Princeton university press, 1946.

\bibitem[Alves et~al.(2019)Alves, Foulon, Karolis, Bzdok, Margulies, Volle, and Thiebaut~de Schotten]{alves2019improved}
Pedro~Nascimento Alves, Chris Foulon, Vyacheslav Karolis, Danilo Bzdok, Daniel~S Margulies, Emmanuelle Volle, and Michel Thiebaut~de Schotten.
\newblock An improved neuroanatomical model of the default-mode network reconciles previous neuroimaging and neuropathological findings.
\newblock \emph{Communications biology}, 2\penalty0 (1):\penalty0 370, 2019.

\bibitem[Millington and Niranjan(2020)]{millington2020partial}
Tristan Millington and Mahesan Niranjan.
\newblock Partial correlation financial networks.
\newblock \emph{Applied Network Science}, 5:\penalty0 1--19, 2020.

\bibitem[von Klipstein et~al.(2021)von Klipstein, Borsboom, and Arntz]{von2021exploratory}
Lino von Klipstein, Denny Borsboom, and Arnoud Arntz.
\newblock The exploratory value of cross-sectional partial correlation networks: Predicting relationships between change trajectories in borderline personality disorder.
\newblock \emph{PloS one}, 16\penalty0 (7):\penalty0 e0254496, 2021.

\bibitem[Zhang et~al.(2021)Zhang, Zhan, Wu, and Zhang]{zhang2021partial}
Jinting Zhang, F~Benjamin Zhan, Xiu Wu, and Daojun Zhang.
\newblock Partial correlation analysis of association between subjective well-being and ecological footprint.
\newblock \emph{Sustainability}, 13\penalty0 (3):\penalty0 1033, 2021.

\bibitem[Werme et~al.(2022)Werme, van~der Sluis, Posthuma, and de~Leeuw]{werme2022integrated}
Josefin Werme, Sophie van~der Sluis, Danielle Posthuma, and Christiaan~A de~Leeuw.
\newblock An integrated framework for local genetic correlation analysis.
\newblock \emph{Nature genetics}, 54\penalty0 (3):\penalty0 274--282, 2022.

\bibitem[Williams and Rast(2020)]{williams2020back}
Donald~R Williams and Philippe Rast.
\newblock Back to the basics: Rethinking partial correlation network methodology.
\newblock \emph{British Journal of Mathematical and Statistical Psychology}, 73\penalty0 (2):\penalty0 187--212, 2020.

\bibitem[Baba and Sibuya(2005)]{baba2005}
Kunihiro Baba and Masaaki Sibuya.
\newblock Equivalence of partial and conditional correlation coefficients.
\newblock \emph{Journal of the Japanese Statistical Society}, 35\penalty0 (1):\penalty0 1--19, 2005.

\bibitem[Meyer(2010)]{Meyer2010}
Carl~D. Meyer.
\newblock \emph{Matrix analysis and applied linear algebra}.
\newblock Society for Industrial and Applied Mathematics, 2010.

\bibitem[Artner et~al.(2022)Artner, Wellingerhof, Lafit, Loossens, Vanpaemel, and Tuerlinckx]{artner2022shape}
Richard Artner, Paul~P Wellingerhof, Ginette Lafit, Tim Loossens, Wolf Vanpaemel, and Francis Tuerlinckx.
\newblock The shape of partial correlation matrices.
\newblock \emph{Communications in Statistics-Theory and Methods}, 51\penalty0 (12):\penalty0 4133--4150, 2022.

\bibitem[Goldenfeld(2018)]{goldenfeld2018lectures}
Nigel Goldenfeld.
\newblock \emph{Lectures on phase transitions and the renormalization group}.
\newblock CRC Press, 2018.

\bibitem[Zinn-Justin(2007)]{zinn2007phase}
Jean Zinn-Justin.
\newblock \emph{Phase transitions and renormalization group}.
\newblock Oxford University Press, 2007.

\bibitem[Cover and Thomas(1991)]{Cover1991}
Thomas~M. Cover and Joy~A. Thomas.
\newblock \emph{Elements of Information Theory}.
\newblock Wiley-Interscience, New York, 1991.

\bibitem[Zinn-Justin(2002)]{ZinnJustin02}
Jean Zinn-Justin.
\newblock \emph{{Quantum Field Theory and Critical Phenomena}}.
\newblock Oxford University Press, 06 2002.
\newblock \doi{10.1093/acprof:oso/9780198509233.001.0001}.

\bibitem[Georges and Yedidia(1991)]{Georges1991}
A.~Georges and J.~S. Yedidia.
\newblock How to expand around mean-field theory using high-temperature expansions.
\newblock \emph{Journal of Physics A: Mathematical and General}, 24\penalty0 (9):\penalty0 2173, 1991.
\newblock \doi{10.1088/0305-4470/24/9/024}.

\bibitem[Maillard et~al.(2019)Maillard, Foini, Lage~Castellanos, Krzakala, M{\'e}zard, and Zdeborov{\'a}]{Maillard2019}
Antoine Maillard, Laura Foini, Alejandro Lage~Castellanos, Florent Krzakala, Marc M{\'e}zard, and Lenka Zdeborov{\'a}.
\newblock High-temperature expansions and message passing algorithms.
\newblock \emph{Journal of Statistical Mechanics: Theory and Experiment}, 2019, 2019.
\newblock \doi{10.1088/1742-5468/ab4bbb}.

\bibitem[Amit and Martin-Mayor(2005)]{amit2005field}
Daniel~J Amit and Victor Martin-Mayor.
\newblock \emph{Field theory, the renormalization group, and critical phenomena: graphs to computers}.
\newblock World Scientific Publishing Company, 2005.

\bibitem[K{\"u}hn and Helias(2018)]{kuhn2018expansion}
Tobias K{\"u}hn and Moritz Helias.
\newblock Expansion of the effective action around non-gaussian theories.
\newblock \emph{Journal of Physics A: Mathematical and Theoretical}, 51\penalty0 (37):\penalty0 375004, 2018.

\bibitem[Zhang(2006)]{zhang2006schur}
Fuzhen Zhang.
\newblock \emph{The Schur complement and its applications}, volume~4.
\newblock Springer Science \& Business Media, 2006.

\bibitem[Horn and Johnson(1985)]{horn2012matrix}
Roger~A Horn and Charles~R Johnson.
\newblock \emph{Matrix analysis}.
\newblock Cambridge university press, 1985.

\bibitem[Krantz and Parks(2002)]{krantz2002primer}
Steven~G Krantz and Harold~R Parks.
\newblock \emph{A primer of real analytic functions}.
\newblock Springer Science \& Business Media, 2002.

\bibitem[Stanley(2015)]{stanley2015catalan}
Richard~P. Stanley.
\newblock \emph{Catalan Numbers}.
\newblock Cambridge University Press, 2015.

\bibitem[Petersen et~al.(2008)Petersen, Pedersen, et~al.]{petersen2008matrix}
Kaare~Brandt Petersen, Michael~Syskind Pedersen, et~al.
\newblock The matrix cookbook.
\newblock \emph{Technical University of Denmark}, 7\penalty0 (15):\penalty0 510, 2008.

\end{thebibliography}



\appendix

\section{Convergence of the *-path expansions}
\label{appendixA}
Here we show that if $\nu(R) < 1$, the sum over paths $\mathtt{p}_{ii}^{j*}$ lies between $0$ and $1$, and therefore the geometric series in Eq.~\ref{eq:loop_generator} converges. For any set of variables, 
\footnotesize
\begin{align}
    \sum_{\mathtt{p}_{ii}^{j*}} &
    \begin{tikzpicture}[->,>=Stealth,auto,shorten >=0.5pt,node distance=1.5cm,semithick,baseline=(i.base)]
        \tikzset{
            mystate/.style={fill=black,circle,inner sep=0pt,minimum width=4pt},
            myloopleft/.style={out=-140,in=140,looseness=20,text opacity=0.9,opacity=0.6},
            myloopright/.style={out=40,in=-40,looseness=20,text opacity=0.9,opacity=0.6}}
        \node[mystate,label={above:$i$}] (i) {};
        \path (i) edge [myloopright]  node {$\scriptstyle \mathtt{p}_{ii}^{j*}$} (i);
    \end{tikzpicture} = \nonumber \\
    &= \sum_{\ell = 0}^{\infty} \, \sum_{\kappa_0 \notin \{i,j\}} \dots \sum_{\kappa_{\ell} \notin \{i,j\}} r_{i \kappa_0} r_{\kappa_0 \kappa_1} \dots r_{\kappa_{\ell-1} \kappa_\ell} r_{\kappa_\ell i} \nonumber \\
    &= \sum_{\ell = 0}^{\infty} \, \sum_{\kappa_0, \kappa_{\ell} \notin \{i,j\}} r_{i \kappa_0} \left[\left(R_\mathcal{K}\right)^{\ell}\right]_{\kappa_0 \kappa_\ell} r_{\kappa_\ell i} \nonumber \\
    &= \sum_{\alpha, \beta \notin \{i,j\}} r_{i \alpha} \left[\left(\mathbb{1} - R_\mathcal{K}\right)^{-1}\right]_{\alpha \beta} r_{\beta i} \nonumber \\
    &= \frac{\sum_{\alpha, \beta \notin \{i,j\}} \omega_{i \alpha} \left[\left(\Omega_\mathcal{K}\right)^{-1}\right]_{\alpha \beta} \omega_{\beta i}}{\omega_{ii}} \nonumber \\
    &= \frac{\omega_{ii} - \left[\Omega / \Omega_\mathcal{K}\right]_{ii}}{\omega_{ii}} \nonumber \\
    &= 1 - \frac{{\rm Var}(Y_i | \{ Y_k : \, k \ne i \})}{{\rm Var}(Y_i | Y_j)}, \label{eq:loops_var_quotient}
\end{align}
\normalsize
where $R_\mathcal{K}$ and $\Omega_\mathcal{K}$ are the sub-matrices of $R$ and $\Omega$ obtained by deleting columns and rows $i$ and $j$, and $\Omega / \Omega_\mathcal{K}$ is the Schur complement of the block $\Omega_\mathcal{K}$ of $\Omega$ \citep{zhang2006schur}. To go from the second to the third line we used the Neumann series, which is guaranteed to hold, since $R$ is symmetric and thus $\nu(R_\mathcal{K}) \le \nu(R) < 1$ \cite[Chapter~4]{horn2012matrix}. Without loss of generality, in the last line we interpreted the inverse of the diagonal elements of $\Omega$ as conditional variances of Gaussian random variables $Y_1, \dots, Y_d$ with precision matrix $\Omega$. These variables may differ from the $X_i$ defining our original problem, since those were not required to be Gaussian. Still, the $Y_i$ can be invoked for the sake of demonstrating the inequality below. The variance of a variable never increases when conditioning on additional variables, i.e. $0 < {\rm Var}(Y_i | \{ Y_k : \, k \ne i \}) \le {\rm Var}(Y_i | Y_j)$. Therefore
\begin{equation*}
    0 \le \sum_{\mathtt{p}_{ii}^{j*}}
    \begin{tikzpicture}[->,>=Stealth,auto,shorten >=0.5pt,node distance=1.5cm,semithick,baseline=(i.base)]
        \tikzset{
            mystate/.style={fill=black,circle,inner sep=0pt,minimum width=4pt},
            myloopleft/.style={out=-140,in=140,looseness=20,text opacity=0.9,opacity=0.6},
            myloopright/.style={out=40,in=-40,looseness=20,text opacity=0.9,opacity=0.6}}
        \node[mystate,label={above:$i$}] (i) {};
        \path (i) edge [myloopright]  node {$\scriptstyle \mathtt{p}_{ii}^{j*}$} (i);
    \end{tikzpicture}
    < 1.
\end{equation*}
If $\nu(R) \ge 1$, the partial correlation matrix $R$ needs to be replaced for a re-scaled version $R(q)$ (Appendix~\ref{app:gen_expansion}). When weighing the edges of the graph $G(q)$ with $R(q)$, a similar calculation shows that $q$ can always be chosen such that the sum over paths $\mathtt{p}_{ii}^{j*}$ lies between $0$ and $1$.

\section{Convergence when \protect{$\nu(R) < 1$}}
\label{appendixB}
Here we show that Eq.~\ref{eq:rho_diagram_4} only requires $\nu(R) < 1$, instead of the stronger condition $\nu(R_+) < 1$. In the derivation of Eq.~\ref{eq:rho_diagram_4}, the convergence of the Neumann series required $\nu(R) < 1$, yielding Eq.~\ref{eq:rho_diagram}. The stronger condition $\nu(R_+) < 1$ was then needed to show that some paths cancel out. Nevertheless, each of the series appearing in Eq.~\ref{eq:rho_diagram_4} can be shown to converge with just the weaker condition $\nu(R) < 1$. For example, the numerator of Eq.~\ref{eq:rho_diagram_4} is
\footnotesize
\begin{align*} 
    \sum_{\mathtt{p}_{ij}^{*}} &
    \begin{tikzpicture}[->,>=Stealth,auto,shorten >=0.5pt,node distance=1.5cm,semithick,baseline=(i.base)]
        \tikzset{
            mystate/.style={fill=black,circle,inner sep=0pt,minimum width=4pt},
            myedge/.style={text opacity=0.9,opacity=0.6}}
        \node[mystate,label={above:$i$}] (i)               {};
        \node[mystate,label={above:$j$}] (j)  [right of=i] {};
        \path (i) edge [myedge,bend left,above] node {$\scriptstyle \mathtt{p}_{ij}^{*}$} (j);
    \end{tikzpicture} = \nonumber \\
    &= r_{ij} + \sum_{\ell = 0}^{\infty} \, \sum_{\kappa_0 \notin \{i,j\}} \dots \sum_{\kappa_{\ell} \notin \{i,j\}} r_{i \kappa_0} r_{\kappa_0 \kappa_1} \dots r_{\kappa_{\ell-1} \kappa_\ell} r_{\kappa_\ell j} \nonumber \\
    &= r_{ij} + \sum_{\ell = 0}^{\infty} \, \sum_{\kappa_0, \kappa_{\ell} \notin \{i,j\}} r_{i \kappa_0} \left[\left(R_\mathcal{K}\right)^{\ell}\right]_{\kappa_0 \kappa_\ell} r_{\kappa_\ell j} \nonumber \\
    &= r_{ij} + \sum_{\alpha, \beta \notin \{i,j\}} r_{i \alpha} \left[ \left( \mathbb{1} - R_\mathcal{K}\right)^{-1} \right]_{\alpha \beta} r_{\beta j},
\end{align*}
\normalsize
where the convergence of the Neumann series only requires $\nu(R_\mathcal{K}) < 1$, which also guarantees that the infinite and finite sums can be interchanged. Applying similar manipulations to the denominator,
\begin{widetext}
    \footnotesize
    \begin{equation} \label{eq:p*_paths_conv}
        \rho_{ij} = \frac{r_{ij} + \displaystyle{\sum_{\substack{\alpha, \beta \notin \{i,j\}}}} r_{i \alpha} \left[ \left( \mathbb{1} - R_\mathcal{K}\right)^{-1} \right]_{\alpha \beta} r_{\beta j}}{\sqrt{\left[ 1 - \displaystyle{\sum_{\substack{\alpha, \beta \notin \{i,j\}}}} r_{i \alpha} \left[ \left( \mathbb{1} - R_\mathcal{K}\right)^{-1} \right]_{\alpha \beta} r_{\beta i} \right] \left[1 -  \displaystyle{\sum_{\substack{\alpha, \beta \notin \{i,j\}}}} r_{j \alpha} \left[\left( \mathbb{1} - R_\mathcal{K}\right)^{-1} \right]_{\alpha \beta} r_{\beta j} \right]}}.
    \end{equation}
    \normalsize
\end{widetext}
The right-hand side of Eq.~\ref{eq:p*_paths_conv} is a function of the elements of $R$ that are above the diagonal, and this function is analytic in the domain $\mathcal{D} = \left\{R \in \mathbb{R}^{d(d-1)/2} : \nu(R) < 1 \right\}$. Moreover, the right-hand side of Eq.~\ref{eq:rho_of_R} is also analytic in $\mathcal{D}$, and, as shown above, it is equal to the right-hand side of Eq.~\ref{eq:p*_paths_conv} in the sub-domain $\mathcal{D}_+ = \left\{R \in \right. $ $ \left. \mathbb{R}^{d(d-1)/2} : \nu(R_+) < 1 \right\} \subset \mathcal{D}$. The identity theorem \cite{krantz2002primer} then guarantees that both functions are equal in $\mathcal{D}$, i.e. the region where $\nu(R) < 1$. This means that Eq.~\ref{eq:p*_paths_conv}, and therefore Eq.~\ref{eq:rho_diagram_4}, only require the weaker condition $\nu(R) < 1$.

\section{Path expansion for general \textit{R}}
\label{app:gen_expansion}

If $R$ does not satisfy the condition $\nu(R)<1$, Eq.~\ref{eq:r_neumann} cannot be used to obtain a series expansion of $(\mathbb{1} - R)^{-1}$. However, an expansion can still be obtained by defining the auxiliary matrix
\begin{equation} \label{eq:Rq}
    R(q) \coloneqq (1-q)\mathbb{1}+q R,
\end{equation}
\noindent where $q$ is a free real parameter. Choosing $q$ so that $\nu(R(q))<1$ enables the use of the Neumann to invert $\mathbb{1}-R(q)$. However, Eq.~\ref{eq:Rq} implies that
\begin{equation} \label{eq:erredecu}
(\mathbb{1}-R)^{-1}=q \, (\mathbb{1}-R(q))^{-1}.
\end{equation}
Except for the multiplicative constant $q$, the series for $(\mathbb{1}-R)^{-1}$ can also be considered as a sum over paths weights, just as in Eq.~\ref{eq:rho_diagram_4}, but replacing $r_{ij}$ by $r_{ij}(q)=(1-q) \, \delta_{ij}+q \, r_{ij}$. In other words, the paths now belong to a new graph $G(q)$ where all edge weights are multiplied by $q$, and there is a self-loop in every vertex, with weight $1-q$. The original $G$ had no such loops. Equations~\ref{eq:rho_diagram_3} and ~\ref{eq:rho_diagram_4} are still valid, but now, when counting paths, the self-loops must be taken into account.

As explained before, if paths are ordered by length, the condition $\nu(R(q)) < 1$ guarantees the validity of Eqs.~\ref{eq:rho_diagram_3} and ~\ref{eq:rho_diagram_4}.  Since $R(q)=(1-q)\mathbb{1}+q R$, this condition is satisfied when
\begin{equation}
    0<q<\frac{2}{1-\lambda_{min}(R)}=\frac{2}{1+\nu(R)}.
    \label{eq:q}
\end{equation}
The left-hand side of Eq.~\ref{eq:erredecu} does not depend on $q$, so the right-hand side cannot depend on $q$ either. Therefore, if the series expansion of $[1 - R(q)]^{-1}$ converges, when multiplied by the constant $q$, the result must also be independent of $q$. This may seem to imply that all $q$ values that make $\nu(R(q)) < 1$ are equally useful. Yet, small $q$-values (i.e. close to the lower bound) may  slow down the convergence of the series significantly. Therefore, when using the truncated series expansion as an approximate computational tool, in order to reduce the number of terms needed to achieve a desired accuracy, $q$ should be chosen close to the upper bound of Eq.~\ref{eq:q}.    

\begin{figure}
    \centering
    \includegraphics[width=0.45\textwidth]{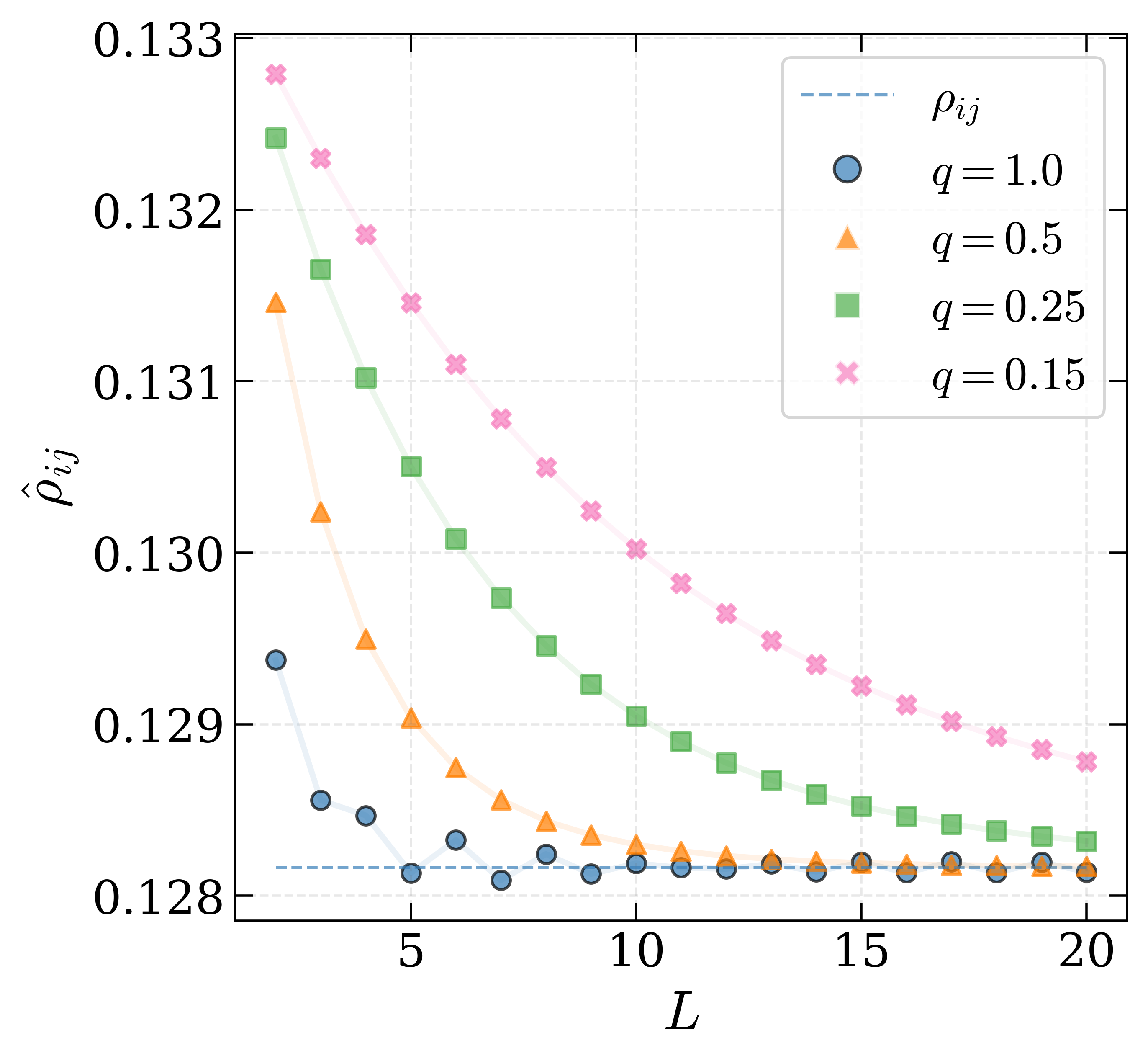}
    \caption{Analysis of the convergence properties of the series in Eq.~\ref{eq:rho_diagram_4}, as a function of the length of the paths included in the sum, for different $q$-values.}
\label{fig:conv_b}
\end{figure}

In Fig.~\ref{fig:conv_b}, the effect of $q$ is illustrated by re-scaling the $R$ matrix of the example of the upper curve of Fig.~\ref{fig:conv}. Admittedly, in this example re-scaling is not strictly needed, since $\nu(R) < 1$. We still choose to discuss this case because the full range of possible $q$ values can be explored, from $q = 0$ to $q = 1$. As $q$ decreases, longer paths need to be included to obtain a satisfactory approximation of the asymptotic correlation value, i.e. the convergence becomes slower. The same effect was observed in our numerical explorations of $R$ matrices with spectral radius $\nu(R) > 1$.

\section{Example of Sect. \ref{ex:one_many_one}. Computing marginal correlations by matrix inversion}

\label{app:net_corr_boring}

Here we arrive to Eq.~\ref{eq:rho12paralelo} by inverting the precision matrix $\Omega = \mathbb{1} - R$ associated to the graphical model introduced in example \ref{ex:one_many_one}. The marginal correlation is
\begin{align}
    \rho_{1d}
    &= \frac{\left(\Omega^{-1}\right)_{1d}}{\sqrt{\left(\Omega^{-1}\right)_{11} \left(\Omega^{-1}\right)_{dd}}} \nonumber \\
    & = \frac{{\rm adj}(\Omega)_{1d}}{\sqrt{{\rm adj}(\Omega)_{11} \, {\rm adj}(\Omega)_{dd}}}, \label{eq:app_one_many_one_rho}
\end{align}
where ${\rm adj}(\Omega)$ is the adjugate of $\Omega$ and we used the fact that $\Omega^{-1} = \det(\Omega)^{-1} {\rm adj}(\Omega)$.  The adjugate ${\rm adj}(\Omega)$ is the transpose of the cofactor matrix of $\Omega$, i.e.
\begin{equation*}
    {\rm adj}(\Omega)_{ij} = (-1)^{i+j} M_{ji},
\end{equation*}
where $M_{ji}$ is the determinant of the $(d-1) \times (d-1)$ matrix that results from deleting row $i$ and column $j$ of $\Omega$. Since
\footnotesize
\begin{equation*}
    \Omega = 
    \left(
    \begin{array}{cccccc}
        1 & r & r & \dots & r & 0 \\
        r & 1 & 0 & \dots & 0 & r \\
        \vdots & 0 & 1 & 0 & \dots & \vdots \\
        r & 0 & 0 & \ddots & 0 & r \\
        r & 0 & \dots & 0 & 1 & r \\
        0 & r & r & \dots & r & 1
    \end{array}
    \right),
\end{equation*}
\normalsize
then
\footnotesize
\begin{equation*}
    M_{d1} = \det
    \left(
    \begin{array}{ccccc}
        r & r & \dots & r & 0 \\
        1 & 0 & \dots & 0 & r \\
        0 & \ddots & 0 & 0 & \vdots \\
        \vdots & 0 & 1 & \dots & r \\
        0 & 0 & \dots & 1 & r
    \end{array}
    \right).
\end{equation*}
\normalsize
Calling $a_d = M_{d1}$, and expanding the determinant along the first column,
\begin{equation*}
    a_d = (-1)^{d-1} \, r^2 - a_{d-1},
\end{equation*}
where $a_{d-1}$ is the determinant of the $(d-2) \times (d-2)$ matrix that results from deleting the last row and first column of a precision matrix with the same structure as $\Omega$ but of dimension $d-1$. Noting that $a_2 = 0$, the solution to the recurrence yields
\begin{equation*}
    a_d = (d-2) \, (-1)^{d-1} \, r^2,
\end{equation*}
which in turn means that
\begin{equation} \label{eq:adj_1d}
    {\rm adj}(\Omega)_{1d} = (d-2) \, r^2.
\end{equation}
Similar algebraic manipulations yield
\begin{equation} \label{eq:adj_11}
    {\rm adj}(\Omega)_{11} = {\rm adj}(\Omega)_{dd} = 1 - (d-2) \, r^2.
\end{equation}
Inserting Eqs.~\ref{eq:adj_1d} and \ref{eq:adj_11} into Eq.~\ref{eq:app_one_many_one_rho} produces the desired result of Eq.~\ref{eq:rho12paralelo}.

\noindent
\section{Correlation length of an infinite chain}
\label{app:corr_length}

Here we compute the correlation length $\xi$, as defined in Eq.~\ref{eq:chain_corr_length_def}, of the linear chain presented in Sect.~\ref{ex:markov_chain}. We first calculate the sum $\ell_d$ over closed paths $\mathtt{p}_{11}^{d*}$ in the $d \to \infty$ limit, i.e.
\begin{equation*}
    \ell_\infty = \lim_{d \to \infty} \ell_d.
\end{equation*}
To compute the sum, we group together the closed paths with the same weight. Every closed path has weight $r^{2n}$ for some $n \ge 1$, because the number $n$ of steps to one side of the chain must equal the number of steps to the other side. The number of paths with weight $r^{2n}$ for a fixed $n$ is equal to the number of sequences of right- and left-steps that only visit the starting node in the last step. Since the chain is infinite, this number is the Catalan number $C_{n-1}$ \citep{stanley2015catalan}. Therefore,
\begin{equation} \label{eq:chain_l_inf}
    \ell_{\infty}
    = \sum_{n=1}^{\infty} C_{n-1} \, r^{2n}
    = \frac{1-\sqrt{1-4\,r^2}}{2},
\end{equation}
where the infinite sum was replaced by the generating function of the Catalan numbers \citep{stanley2015catalan}. To extend the argument to a finite chain, the paths that reach the end of the chain need to be discarded.

In an infinite chain, the inverse of the correlation length is
\begin{align}
    \xi^{-1} 
    &= \lim_{|i-j| \to \infty} - \frac{1}{|i-j|} \ln |\rho_{ij}| \nonumber \\
    &= \lim_{n \to \infty} - \frac{1}{n} \ln | c_{n+1} | + \frac{1}{n} \ln ( 1 - \ell_{n+1} - \ell_{\infty}) \nonumber \\
    &= \lim_{n \to \infty} - \frac{1}{n} \left[ n \ln |r| - \sum_{k=2}^{n} \ln(1-\ell_k) \right] \nonumber \\
    &= -\ln |r| + \ln(1-\ell_{\infty}) \nonumber \\
    &= \ln \left( \frac{1 + \sqrt{1 - 4\,r^2}}{2 \, |r|} \right). \label{eq:chain_corr_length_app}
\end{align}
In the second line, Eq.~\ref{eq:chain_corr_ij} was used in the limit where both $i, d \to \infty$. The second term was discarded, since the numerator converges to a finite value for any $|r| < \sfrac{1}{2}$. In the third line, the recursion given by Eq.~\ref{eq:chain_paths_rec} allowed us to write $c_{n+1}$ in terms of the sums $\ell_k$.

\section{Sufficient condition for the existence of a separating node}
\label{app:sep_node_sufficiency}

In Sect.~\ref{sec:separatingnode} we stated a necessary and sufficient condition for $k$ to be a separating node for the partial correlation graph, namely, that when partitioning the set of nodes $\mathcal{U}$ into $\mathcal{U} = \mathcal{I} \cup \{k\} \cup \mathcal{J}$ the equality $\rho_{ij} = \rho_{ik} \, \rho_{kj}$ holds for every $i \in \mathcal{I}$ and $j \in \mathcal{J}$. Here we show that the condition is sufficient.

To prove sufficiency, it is enough to show that the factorisation of marginal correlations implies that $r_{ij} = 0$ for every $i \in \mathcal{I}$ and $j \in \mathcal{J}$. Since the partial correlations can be written in terms of the marginal correlation matrix $P$ as
\footnotesize
\begin{equation*}
    r_{ij} = -\frac{(P^{-1})_{ij}}{\sqrt{(P^{-1})_{ii}(P^{-1})_{jj}}},
\end{equation*}
\normalsize
we need to show that the sub-matrix $(P^{-1})_\mathcal{IJ}$ vanishes. This sub-matrix contains the elements of $P^{-1}$ between nodes in $\mathcal{I}$ and in $\mathcal{J}$.

The factorisation condition implies that the marginal correlation matrix is
\footnotesize
\begin{equation*}
    \renewcommand{\arraystretch}{1.4}
    P = \left(
    \begin{NiceArray}{cc:c}[margin]
        P_\mathcal{I} & \boldsymbol{\rho}_{\mathcal{I}k} & \boldsymbol{\rho}_{\mathcal{I}k} \, \boldsymbol{\rho}_{\mathcal{J}k}^T \\
        \boldsymbol{\rho}_{\mathcal{I}k}^T & 1 & \boldsymbol{\rho}_{\mathcal{J}k}^T \\
        \dashedline
        \boldsymbol{\rho}_{\mathcal{J}k} \, \boldsymbol{\rho}_{\mathcal{I}k}^T & \boldsymbol{\rho}_{\mathcal{J}k} & P_\mathcal{J} \\
    \end{NiceArray}
    \right) = \left(
    \begin{NiceArray}{c:c}[margin]
        A & B \\
        \dashedline
        B^T & C \\
    \end{NiceArray}
    \right),
\end{equation*}
\normalsize
where the matrices $P_\mathcal{I}$ and $P_\mathcal{J}$ contain marginal correlations between pairs in the same set, and the vectors $\boldsymbol{\rho}_{\mathcal{I}k}$ and $\boldsymbol{\rho}_{\mathcal{J}k}$, marginal correlations between the nodes in one of the two sets and $k$. The matrices $A$, $B$ and $C$ were implicitly defined to match the four sub-matrices divided by the dashed lines. Using the block inversion formula \citep{petersen2008matrix}, the sub-matrix $(P^{-1})_{(\mathcal{I}k)\mathcal{J}}$ containing elements between a node in $\mathcal{I} \cup \{k\}$ and a node in $\mathcal{J}$ is
\footnotesize
\begin{align} \label{eq:inv_marg_corr_matrix_IkJ}
    (P^{-1})_{(\mathcal{I}k)\mathcal{J}} &= \left(
    \begin{array}{c}
        (P^{-1})_{\mathcal{IJ}} \\
        (P^{-1})_{k\mathcal{J}}
    \end{array}
    \right) \nonumber \\
    &= - A^{-1} B \left( C - B^T A^{-1} B \right)^{-1},
\end{align}
\normalsize
where all the inverses exist because $P$ is positive-definite. For our purposes, it suffices to compute the first product
\footnotesize
\begin{equation*}
    A^{-1} B = \left(
    \begin{array}{cc}
        P_\mathcal{I} & \boldsymbol{\rho}_{\mathcal{I}k} \\
        \boldsymbol{\rho}_{\mathcal{I}k}^T & 1
    \end{array}
    \right)^{-1} \left(
    \begin{array}{c}
        \boldsymbol{\rho}_{\mathcal{I}k}  \\
        1
    \end{array}
    \right) \boldsymbol{\rho}_{\mathcal{J}k}^T
    = \left(
    \begin{array}{c}
        \mathbb{0}  \\
        \boldsymbol{\rho}_{\mathcal{J}k}^T
    \end{array}
    \right),
\end{equation*}
\normalsize
where $\mathbb{0}$ is the null matrix. Plugging this result in Eq.~\ref{eq:inv_marg_corr_matrix_IkJ}, the elements of $(P^{-1})_{\mathcal{IJ}}$ are shown to vanish, which in turn implies that $r_{ij} = 0$ for every $i \in \mathcal{I}$ and $j \in \mathcal{J}$. That is, there are no edges between nodes in $\mathcal{I}$ and $\mathcal{J}$. Excluding the trivial case where the nodes of $\mathcal{I}$ and $\mathcal{J}$ are in different connected components, this result implies that $k$ is a separating node for $\mathcal{I}$ and $\mathcal{J}$.

\section{Factorisation of marginal correlations in martingales}
\label{app:martingale_fact_prop}

Here we show that a discrete-time martingale process $X_t$, where ${\rm E}(X_t \, | X_{t-1}, X_{t-2}, \dots) = \alpha \, X_{t-1}$, with finite second moments satisfies that $\rho_{t_1 t_3} = \rho_{t_1 t_2} \, \rho_{t_2 t_3}$ for any indices such that $t_1 < t_2 < t_3$. For simplicity, we assume that ${\rm E}(X_t) = 0$ for any $t$, but the proof can be easily generalised to martingales with arbitrary mean values. The covariance between $X_{t_1}$ and $X_{t_3}$ is
\begin{equation*}
    {\rm E}(X_{t_1} \, X_{t_3}) = {\rm E}[X_{t_1} \, {\rm E}(X_{t_3}\,|X_{t_1})]
\end{equation*}
where
\begin{align*}
    {\rm E}(X_{t_3}\,|X_{t_1})
    &= {\rm E}[ {\rm E}(X_{t_3} \,|X_{t_3-1}, \dots X_{t_1} ) \,|X_{t_1}] \\
    &= \alpha \, {\rm E}(X_{t_3-1}\,|X_{t_1}) \\
    &= \alpha^{t_3-t_1} \, X_{t_1}.
\end{align*}
In the first equality we first averaged over $X_{t_3}$ for fixed $X_{t_1}, \dots, X_{t_3-1}$ (inner expectation value) and then averaged over the intermediate variables $X_{t_1+1}, \dots, X_{t_3-1}$ for fixed $X_{t_1}$ (outer expectation value). The marginal correlation $\rho_{t_1 t_3}$ then is
\begin{align*}
    \rho_{t_1 t_3}
    &= \frac{{\rm E}(X_{t_1} \, X_{t_3})}{\sqrt{{\rm E}(X_{t_1}^2) \, {\rm E}(X_{t_3}^2)}} \\
    &= \alpha^{t_3-t_2+t_2-t_1} \, \sqrt{\frac{{\rm E}(X_{t_1}^2)}{{\rm E}(X_{t_3}^2)} \frac{{\rm E}(X_{t_2}^2)}{{\rm E}(X_{t_2}^2)}} \\
    &= \rho_{t_1 t_2} \, \rho_{t_2 t_3}.
\end{align*}

\section{Partial correlations of a marginalised model}
\label{app:marg_part_corr}

Given a graph $G$ defined on a set of nodes $\mathcal{U}$ and partial correlations $r_{k\ell}$, we here derive the partial correlations $r_{ij}'$ of the reduced graph $G'$, which results from marginalising a subset of nodes $\mathcal{S} \subset \mathcal{U}$.

Denoting as $\mathcal{T} = \mathcal{U} - \mathcal{S}$ the set of remaining nodes, the precision matrix of the marginalised model is $\Omega' = \left( C_\mathcal{T} \right)^{-1}$, where $C_\mathcal{T}$ is the sub-matrix of $C$ that carries the elements corresponding to pairs of nodes in $\mathcal{T}$. The precision matrix $\Omega_\mathcal{U}$ of the original model can be partitioned into blocks as
\begin{equation*}
    \Omega_\mathcal{U} =
    \left(\begin{array}{cc} \Omega_\mathcal{T} & \Omega_\mathcal{TS} \\ \Omega_\mathcal{ST} & \Omega_\mathcal{S}\end{array} \right),
\end{equation*}
where $\Omega_\mathcal{T}$ and $\Omega_\mathcal{S}$ carry the elements corresponding to pairs of nodes in the same set, and $\Omega_\mathcal{TS}$ and $\Omega_{ST} = \Omega_\mathcal{TS}^T$ carry the elements of pairs in different sets. We can compute the sub-matrix $C_\mathcal{T}$ in terms of the blocks of $\Omega$ using the block inversion formula \citep{petersen2008matrix}, which in turn leads to
\begin{equation*}
    \Omega' = C_\mathcal{T}^{-1} = \Omega_\mathcal{T} - \Omega_\mathcal{TS} \, \Omega_{S}^{-1} \, \Omega_{ST}.
\end{equation*}
Using Eq.~\ref{eq:rij_omega}, the partial correlation $r'_{ij}$ between nodes $i$ and $j$ in $\mathcal{T}$ then is
\begin{widetext}
\footnotesize
\begin{equation} \label{eq:part_corr_marg_model}
    r'_{ij} = \frac{r_{ij} + \sum_{k,\ell} r_{ik} \left[ \left( \mathbb{1} - R_\mathcal{S} \right)^{-1} \right]_{k\ell} r_{\ell j}}{\sqrt{\left[ 1 - \sum_{k,\ell} r_{ik} \left[ \left( \mathbb{1} - R_\mathcal{S} \right)^{-1} \right]_{k\ell} r_{\ell i}\right] \left[ 1 - \sum_{k,\ell} r_{jk} \left[ \left( \mathbb{1} - R_\mathcal{S} \right)^{-1} \right]_{k\ell} r_{\ell j} \right]}},
\end{equation}
\normalsize
\end{widetext}
where $R_\mathcal{S}$ contains the partial correlations between nodes in $\mathcal{S}$, and $k$ and $\ell$ run over variables in $\mathcal{S}$. Expanding the matrix inverse using Eq.~\ref{eq:neumann}, and interpreting the resulting terms in terms of paths of partial correlations, we arrive to Eq.~\ref{eq:nesting_transf}. If $\nu(R_\mathcal{S}) \ge 1$, the inverse cannot be expanded in terms of powers of $R_\mathcal{S}$. However, we can employ the trick introduced in Appendix~\ref{app:gen_expansion} and expand the inverse in powers of $R_\mathcal{S}(q)$ (see Eq.~\ref{eq:Rq}), which introduces self-loops in the associated graph.

\section{Conditional mutual information in terms of paths}
\label{app:info}

\begin{figure*}
    \centering
    \includegraphics[scale = 0.35]{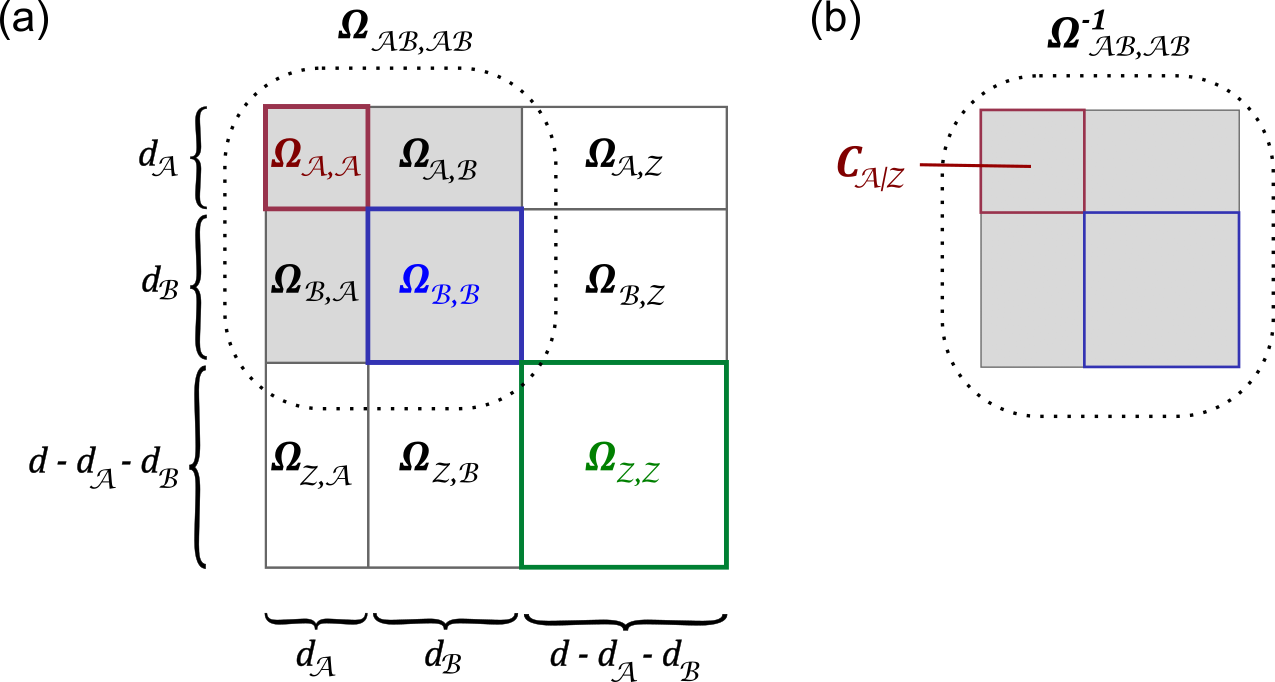}
    \caption{(a): When the variables $X_1, X_2, \dots, X_d$ are ordered so that the first $d_\mathcal{A}$ belong to the subset $\mathcal{A}$, the following $d_\mathcal{B}$ to the subset $\mathcal{B}$, and the remaining ones to the subset $\mathcal{Z}$, the precision matrix is partitioned into blocks. (b): The matrix $C_{\mathcal{A|Z}}$ is the $(\mathcal{A,A})$ block of the inverse of the $(\mathcal{AB,AB})$ block of the precision matrix. }
    \label{fig:cuadrados}
\end{figure*}

Here we derive the path expansion of the conditional mutual information $I(\mathcal{A}; \mathcal{B}| \mathcal{Z})$, where the disjoint subsets $\mathcal{A}$, $\mathcal{B}$ and $\mathcal{Z}$ form a partition of the complete set of Gaussian variables $\{X_1, X_2, \dots, X_d\}$. For definiteness, we order the variables such that the set $\mathcal{A} = \{X_1, \dots, X_{d_A} \}$ contains the first $d_\mathcal{A}$ components, the set $\mathcal{B} = \{X_{d_\mathcal{A} + 1}, \dots X_{d_\mathcal{A} + d_\mathcal{B}} \}$ contains the next $d_\mathcal{B}$, and $\mathcal{Z} = \{X_{d_\mathcal{A} + d_\mathcal{B} + 1}, \dots , X_d\}$ contains the remaining variables. This ordering defines a block structure in the precision matrix (Fig.~\ref{fig:cuadrados}A). For Gaussian distributions, the differential conditional entropies expressed in nats are
\begin{align*}
    h(\mathcal{A}|\mathcal{Z}) &= \frac{1}{2} \ln \left[ \det \left(2\pi e C_{\mathcal{A}|\mathcal{Z}} \right) \right] \\
    h(\mathcal{A}|\mathcal{B},\mathcal{Z}) &= \frac{1}{2} \ln \left[ \det \left( 2\pi e C_{\mathcal{A}|\mathcal{B},\mathcal{Z}} \right) \right]
\end{align*}
where the conditioned covariance matrices are 
\begin{align*}
    C_{\mathcal{A}|\mathcal{Z}} &=\left[\left(\Omega_{\mathcal{AB},\mathcal{AB}}\right)^{-1}\right]_{\mathcal{A},\mathcal{A}}  \\
    C_{\mathcal{A}|\mathcal{B},\mathcal{Z}} &= \left(\Omega_{\mathcal{A},\mathcal{A}}\right)^{-1}.
\end{align*}
As illustrated in Fig.~\ref{fig:cuadrados}, $C_{\mathcal{A}|\mathcal{Z}}$ is the upper left submatrix of the inverse of the matrix that is obtained from $\Omega$ by eliminating the rows and columns of variables belonging to $\mathcal{Z}$. In turn, $C_{\mathcal{A}|\mathcal{B},\mathcal{Z}}$ is the inverse of the matrix that is obtained from $\Omega$ by eliminating the rows and columns of variables belonging to $\mathcal{B} \cup \mathcal{Z}$. 

The conditional mutual information is
\begin{align}
    I(\mathcal{A}; \mathcal{B} | \mathcal{Z})
    &= h(\mathcal{A} | \mathcal{Z}) - h(\mathcal{A}| \mathcal{B}, \mathcal{Z}) \nonumber \\
    &= \frac{1}{2} \ln \left[ \det \left( C_{\mathcal{A}|\mathcal{Z}} \right) \right] - \frac{1}{2} \ln \left[ \det \left( C_{\mathcal{A}|\mathcal{B},\mathcal{Z}} \right) \right] \nonumber \\
    &= - \frac{1}{2} \ln \left[ \det \left( C_{\mathcal{A}|\mathcal{Z}}^{-1} \, C_{\mathcal{A}|\mathcal{B},\mathcal{Z}}  \right) \right] \label{eq:logdet}
\end{align}
Applying the block inversion formula \citep{petersen2008matrix} to the matrix $C_{\mathcal{AB,AB}}$ divided into blocks as in Fig.~\ref{fig:cuadrados}B yields 
\begin{equation*}
C_{\mathcal{A|Z}}^{-1} = \left( \Omega_{\mathcal{A,A}} - \Omega_{\mathcal{A,B}} \, \Omega_{\mathcal{B,B}}^{-1} \, \Omega_{\mathcal{B,A}} \right)^{-1}.
\end{equation*}
After some elementary matrix algebra, Eq.~\ref{eq:logdet} becomes
\begin{equation}
   I(\mathcal{A};\mathcal{B}|\mathcal{Z}) = - \frac{1}{2} \ln \left[ \det \left( \mathbb{1}_{d_\mathcal{A}} - T_{\mathcal{ABA}} \right) \right], \label{eq:taba}
\end{equation}
where
\begin{equation} \label{eq:taba_def}
\begin{split}
     T_\mathcal{ABA} = & R_\mathcal{A,B} \left(\mathbb{1}_{d_\mathcal{B}} - R_\mathcal{B,B}\right)^{-1} \times \\
     & R_\mathcal{B,A} \left(\mathbb{1}_{d_\mathcal{A}} - R_\mathcal{A,A}\right)^{-1}.
\end{split}
\end{equation}
$R_\mathcal{C,D}$ is the $\mathcal{CD}$--block of the partial correlation matrix $R$ (i.e. has the rows of $R$ corresponding to variables in $\mathcal{C}$ and columns corresponding to variables in $\mathcal{D}$). Since $\mathbb{1}_{d_\mathcal{A}} - T_\mathcal{ABA}$ is positive definite, and hence invertible, Jacobi's formula can be applied to Eq.~\ref{eq:taba} yielding
\begin{equation} \label{eq:infomutua_tr_ln}
    I(\mathcal{A; B | Z}) = - \frac{1}{2} {\rm tr} \left[ \ln \left( \mathbb{1}_{d_\mathcal{A}} - T_\mathcal{ABA} \right) \right].
\end{equation}
If the spectral radius $\nu(T_\mathcal{ABA}) < 1$, then \citep{petersen2008matrix}
\begin{equation} \label{eq:ln_matrix_taylor}
    \ln \left( \mathbb{1}_{d_\mathcal{A}} - T_\mathcal{ABA} \right) = -\sum_{n=1}^{\infty} \frac{1}{n} \left(T_\mathcal{ABA}\right)^{n}
\end{equation}
and
\begin{equation} \label{eq:infomutuatrtaba}
    I(\mathcal{A; B | Z}) = \sum_{n=1}^{\infty} \frac{1}{2 n} {\rm tr} \left[\left( T_\mathcal{ABA} \right)^n\right].
\end{equation}
Otherwise, if $\nu(T_\mathcal{ABA}) \ge 1$, a trick similar to the one employed in Appendix~\ref{app:gen_expansion} can be used to define a re-scaled matrix
\begin{equation*}
    T_\mathcal{ABA}(q) = (1-q) \mathbb{1}_{d_\mathcal{A}} + q \, T_\mathcal{ABA}
\end{equation*}
and choosing the free parameter $q$ such that $\nu(T_\mathcal{ABA}(q)) < 1$. The general expansion can then be obtained using that
\begin{equation*}
    \ln (\mathbb{1}_{d_\mathcal{A}} - T_\mathcal{ABA}) = - \ln q + \ln (\mathbb{1}_{d_\mathcal{A}} - T_\mathcal{ABA}(q))
\end{equation*}
in Eq.~\ref{eq:infomutua_tr_ln} and expanding the logarithm as in Eq.~\ref{eq:ln_matrix_taylor}.

If the spectral radii of $R_\mathcal{A,A}$ and $R_\mathcal{B, B}$ are smaller than unity, $\mathbb{1}_{d_\mathcal{A}} - R_\mathcal{A, A}$ and $\mathbb{1}_{d_\mathcal{B}} - R_\mathcal{B, B}$ can be inverted with Eq.~\ref{eq:neumann}. The matrix $T_\mathcal{ABA}$ defined in Eq.~\ref{eq:taba_def} can then be written as 
\begin{equation} \label{eq:taba_interp}
    T_\mathcal{ABA} = \sum_{\ell_\mathcal{B} = 0}^{+\infty} \sum_{\ell_\mathcal{A} = 0}^{+\infty} R_\mathcal{A,B} \, R_\mathcal{B,B}^{\ell_\mathcal{B}} \, R_\mathcal{B, A} \, R_\mathcal{A,A}^{\ell_\mathcal{A}}
\end{equation}
This equation shows that $T_\mathcal{ABA}$ represents paths that start and finish in elements of $\mathcal{A}$, and visit $\mathcal{B}$ somewhere in the middle.  More precisely, the element $\left(T_\mathcal{ABA}\right)_{a_1 a_2}$, where $a_1, a_2 \in \{ 1, \dots, d_\mathcal{A} \}$, is the sum over all paths that are structured as a concatenation of four types of sub-paths in the following way. The first sub-path starts in $a_1$ and immediately leaves the set $\mathcal{A}$ to reach, without passing through $\mathcal{Z}$, a node in $\mathcal{B}$ (factor $R_\mathcal{A, B}$ in Eq.~\ref{eq:taba_interp}). The second sub-path travels within $\mathcal{B}$ for $\ell_{\mathcal{B}}$ steps ($\ell_{\mathcal{B}}$ ranges over all possible numbers of steps within $\mathcal{B}$). The third sub-path returns to $\mathcal{A}$ (factor $R_\mathcal{B,A}$), and the fourth travels within $\mathcal{A}$ for $\ell_\mathcal{A}$ steps ($\ell_\mathcal{A}$ also ranges over all possible number of steps), finishing in $a_2$. Equation~\ref{eq:infomutuatrtaba} implies that the conditional mutual information is a measure of the ability to transmit information back and forth between the subsets $\mathcal{A}$ and $\mathcal{B}$. It includes all the internal reverberations inside each subset, and accounts, through the factor $\sfrac{1}{2n}$, for the progressive forgetfulness of paths that repeatedly circle from $\mathcal{A}$ to $\mathcal{B}$ and back to $\mathcal{A}$ a number of times $n$, as $n$ increases.  The symmetry of the mutual information $I(\mathcal{A;B|Z}) = I(\mathcal{B;A|Z})$ implies that the sum in Eq.~\ref{eq:infomutuatrtaba} does not change if the matrix $T_\mathcal{ABA}$ is exchanged for the analogous matrix $T_\mathcal{BAB}$.

\end{document}